\documentclass[fleqn,usenatbib]{mnras}

\usepackage{textcomp}
\usepackage{graphicx}
\usepackage{mathptmx}

\usepackage[T1]{fontenc}
\usepackage[utf8]{inputenc}
\usepackage{newunicodechar}
\newunicodechar{−}{\textminus}

\DeclareRobustCommand{\VAN}[3]{#2}
\let\VANthebibliography\thebibliography
\def\thebibliography{\DeclareRobustCommand{\VAN}[3]{##3}\VANthebibliography}

\usepackage{graphicx}	
\usepackage{amsmath}	
\usepackage{subfigure} 

\usepackage{soul}

\title[Accretion Geometry of QS Tel]{Accretion Geometry and Variability in QS Telescopii from multi-epoch Observations \thanks{Based on observations made with the Southern African Large Telescope (SALT) under programs 2023-1-SCI-027, 2024-1-SCI-011, and 2025-1-SCI-016 (PI: Z. N. Khangale).}}
\author[L Booi et al]{
L Booi,$^{1,2}$\thanks{E-mail: luthob847@gmail.com}
S.~B. Potter,$^{2,3}$
Z.~N. Khangale$^{1,2}$
V Moloi$^{1,2}$
and P.~A. Woudt$^{1}$
\\
$^{1}$Department of Astronomy, University of Cape Town, Private Bag X1, Rondebosch 7701, South
Africa\\
$^{2}$South African Astronomical Observatory, Observatory Road, Observatory, 7925, Cape Town,
RSA\\
$^{3}$Department of Physics, University of Johannesburg, PO Box 524, Auckland Park 2006, South Africa\\
}

\date{Accepted XXX. Received YYY; in original form \today}

\pubyear{\the\year{}}

\begin{document}
\label{firstpage}
\pagerange{\pageref{firstpage}--\pageref{lastpage}}
\maketitle

\begin{abstract}
We present phase-resolved photometry, spectroscopy, photopolarimetry, and circular spectropolarimetry of the polar QS Telescopii, obtained between 2017 and 2025 using the South African Astronomical Observatory (SAAO) 1.0-m and 1.9-m telescopes, the Transiting Exoplanet Survey Satellite (TESS), Asteroid Terrestrial-impact Last Alert System (ATLAS) and the Southern African Large Telescope (SALT). From ground-based photometry, we report the first detection of quasi-periodic oscillations (QPOs) in QS Tel, with timescales ranging between 2 and 11 minutes. This makes QS Tel one of only seven polars to show QPOs. The average spectra of QS Tel show broad and strong Balmer ($\rm H\delta$, $\rm H\gamma$, $\rm H\beta$ and $\rm H\alpha$) and He\textsc{ii} 4686 \AA\ emission lines, along with weaker emission from He\textsc{i} at (6678\,\AA, 5017\,\AA, 4921\,\AA, 4712\,\AA, 4471\,\AA, and 4387\,\AA)  and the Bowen fluorescence complex C\textsc{iii}/N\textsc{iii} at 4650\,\AA. For the first time detailed Doppler tomography of the $\rm H\alpha$ and He\textsc{ii}\,4686\,\AA\, lines are presented which reveal emission from the irradiated face of the secondary star, the ballistic stream, the threading region and complex magnetically confined flows. The magnetically confined accretion stream splits into two distinct components coming from the same stream. One component appears near the threading region and the other component is seen along the high velocity in the magnetically confined accretion stream closer to the White Dwarf. Broad band photopolarimetry reveals circular polarisation reaching levels of $+8\%$, however more detailed circular spectropolarimetry shows polarisation ranging from $-10\%$ to $+25\%$, associated with two broad, phase-dependent cyclotron humps centred at $\sim$4800~\AA\ and $\sim$6300~\AA, consistent with a magnetic field strength of $\sim$53~MG.
 
\end{abstract}




\section{Introduction}

Polars are a subclass of magnetic cataclysmic variables (mCVs), consisting of a low-mass companion star and a strongly magnetized white dwarf \citep[WD; ][]{1995cvs..book.....W}. The strong magnetic field suppresses the formation of an accretion disc, forcing the accretion flow to follow magnetic field lines onto the WD’s magnetic poles. The two stars are tidally locked, resulting in synchronous rotation \citep{Frank1992,1995cvs..book.....W,2001cvs..book.....H}. Polars can be divided into two subclasses based on their magnetic field strength. The first subclass has magnetic field strengths between 10 and 80 MG e.g VV Pup (B $\sim$ 56 MG; \citealt{1989ApJ...342L..35W}), V834 Cen (B = 23 MG; \citealt{1992MNRAS.256..252F}). In these systems, the accreting material initially follows a ballistic stream trajectory from the inner Lagrangian point, which is not affected by the magnetic field. Once it reaches the threading region, the material becomes ionised and is captured by the magnetic field lines, which channel the flow to the WD’s magnetic poles \citep{1990SSRv...54..195C,1995cvs..book.....W,2001cvs..book.....H}.

The second subclass comprises systems with magnetic field strengths exceeding 80 MG, such as AR UMa (B = 230 MG; \citealt{1996ApJ...473..483S}) and V884 Her (B = 150 MG; \citealt{2001ApJ...553..823S}). In these systems, the material does not form a ballistic stream but instead couples directly to the magnetic field lines and accretes almost immediately onto the magnetic poles. For detailed reviews of accretion processes and magnetic coupling in polars \citep[see][]{1990SSRv...54..195C, 1995cvs..book.....W, 2001cvs..book.....H}. A defining characteristic of cataclysmic variables (CVs) is their ongoing accretion process, which results in a wide range of mass transfer rates. In polars, this process is directly observable due to the absence of an accretion disc. During low accretion states, accretion rate decreases, allowing the stellar components of the binary to be observed more clearly. In contrast, during high accretion states, cyclotron-emitting regions dominate the emission, giving rise to quasi-periodic oscillations (QPOs) and cyclotron humps in optical and infrared regime. In polars, QPOs are attributed to oscillations of the radiative accretion shock formed where the accretion flow impacts the magnetic poles \citep[see e.g.,][]{2015gacv.workE..18B}. These oscillations arise from thermal instabilities in the post-shock region, producing quasi-periodic modulations in both optical and X-ray flux \citep{1995cvs..book.....W}. QPOs in polars were first detected in V834 Cen and AN UMa \citep{1982ApJ...257L..71M}, and have since been observed in other systems such as EF Eri (2--3 s; \citealt{1987A&A...181L..15L}), BL Hyi (1.25--5 s; \citealt{1997ApJ...489..912M}). In longer timescales, QPO's have been seen in sources such as IGR J14536−5522 (5--6 min; \citealt{2010MNRAS.402.1161P}), in this case the QPO's are thought to occur due to instabilities in the threading region, where the ballistic stream couples to the WD’s magnetic field. These oscillations provide a direct probe of shock stability and cooling mechanisms in magnetically confined accretion flows.

QS Tel is a polar that was discovered by \citet{1993MNRAS.262...93B} using data from the ROSAT Extreme Ultraviolet (EUV; \citealt{1982AdSpR...2d.241T,1993MNRAS.260...77P,1995MNRAS.274.1165P}) survey, where it was identified as the brightest of all polars observed by the EUV instrument. Optical photometry and spectroscopy established its classification as a polar, with an orbital period of 2.33 hours \citep{1993MNRAS.262...93B,1995A&A...293..764S,10.1093/mnras/280.4.1121}. This period places QS Tel within the period gap of CVs, orbital period between 2 and 3 hours where magnetic braking is believed to become inefficient as the secondary star becomes fully convective \citep{1993MNRAS.262...93B,2011A&A...529A.116T}. Systems within the period gap are relatively rare, making QS Tel an important target for studying CV evolution. The long-term photometric light curves of QS Tel reveal three distinct accretion states: high, intermediate, and low. The first high accretion state in the optical was reported by \citet{1993MNRAS.262...93B}, who measured a magnitude of V = 15.6 mag in the V band. In follow-up studies, \citet{1995A&A...293..764S} and \citet{10.1093/mnras/280.4.1121} reported mean magnitudes of $V \simeq$ 15.1 mag and $V \simeq$ 15.4 mag, respectively. The intermediate and low states were reported with magnitudes of $V \simeq$ 16.1--16.7 mag and $V \simeq 17.3$ mag, respectively \citep{10.1093/mnras/280.4.1121}. These distinct states reflect significant changes in the mass transfer rate and accretion geometry.

The first optical spectra of QS Tel were reported by \citet{1993MNRAS.262...93B}. These spectra, obtained during a high accretion state, revealed strong Balmer emission lines and a prominent He\textsc{ii}\,4686\,\AA\, line both characteristic of magnetic CVs. Follow-up phased-resolved spectroscopic studies in the high state were conducted by \citet{1995A&A...293..764S} and \citet{10.1093/mnras/280.4.1121} revealing multiple components with radial velocities ranging from 250--1500 $\mathrm{km\,s^{-1}}$. \citet{1995A&A...293..764S} and \citet{2001MNRAS.322..631R} further presented low-state spectra of QS Tel showing weak Balmer emission lines, no trace of the He\textsc{ii}\,4686\,\AA\, line, and clear Zeeman absorption features. Spectropolarimetric observations by \citet{1994MNRAS.268..128F}, conducted during a low accretion state, suggested that QS Tel may operate as a one-pole accreting system under certain conditions. They estimated a magnetic field strength of 56~MG for the active pole. Their polarimetric results showed a wavelength dependence in the degree of circular polarisation, peaking in the blue and sometimes becoming negative toward the red end of the spectrum. A high-state follow-up study by \citet{2001MNRAS.322..631R} revealed that the second cyclotron emission region had become active, indicating that the system had returned to a two-pole accreting configuration. \citet{1995A&A...293..764S} subsequently derived magnetic field strengths of 47 MG for the weaker pole and 70–80 MG for the stronger pole by modelling phase-dependent cyclotron features seen in the optical spectroscopy, suggesting a non-dipolar magnetic field structure \citep{10.1093/mnras/280.4.1121}. Although QS Tel is one of the most extensively observed polars, a comprehensive, phase-resolved analysis integrating Doppler tomography with both photo-polarimetry and circular spectropolarimetry has yet to be conducted.

In this work, we present photometric, phase resolved spectroscopic, and spectropolarimetric observations of QS Tel, aimed at improving our understanding of its accretion geometry and magnetic field structure across accretion states. 
This paper is organised as follows: in Section \ref{section:observations}, we describe the observations and data reduction procedures. Section \ref{section:results} presents the main observational results, including Doppler tomography. In Section \ref{section:discussion}, we interpret these results in the context of accretion geometry and magnetic field structure, comparing with recent studies of similar systems. Finally, Section \ref{section:summary} summarizes our conclusions and outlines prospects for future work.

\section{Observations}\label{section:observations}
\begin{table*}
\caption{Spectroscopic, photometric, polarimetric and circular spectropolarimetric observations log of QS~Tel.}
\label{tab:table_of_obs}
\centering
\resizebox{\linewidth}{!}{%
\begin{tabular}{llllll}
\hline
\textbf{Date of observations} & \textbf{Telescope} & \textbf{Exposure} & \textbf{Wavelength} & \textbf{Observations} & \textbf{Instrument} \\
& & \textbf{time (s)} & \textbf{Range (\AA{})}& \textbf{type} & \textbf{used}\\ \hline
2017/06/01 - 2025/07/25      & ATLAS       & 30        & o-band                   & Photometry (forced)        & ATLAS 0.5‑m Schmidt + 110 Mpix CCD   \\
2017/06/23              & SAAO 1.9-m  & 100       & Red Filter               & Photopolarimetry   & HIPPO   \\
2017/06/24              & SAAO 1.9-m  & 100       & Unfiltered               & Photopolarimetry   & HIPPO   \\
2017/06/25              & SAAO 1.9-m  & 100       & Unfiltered               & Photopolarimetry   & HIPPO   \\
2020/09/05 - 2020/09/25 & TESS            & 120       & 6000 - 1000                         & Photometry         & TESS    \\
2022/09/21              & SAAO 1.9-m  & 250       & 4050 - 5250           & Spectroscopy       & SpUpNIC \\
2022/09/22              & SAAO 1.9-m  & 250       & 6250 - 7250           & Spectroscopy       & SpUpNIC \\
2022/09/23              & SAAO 1.9-m  & 250       & 6250 - 7250           & Spectroscopy       & SpUpNIC \\
2022/09/26              & SAAO 1.9-m  & 250       & 4050 - 5250           & Spectroscopy       & SpUpNIC \\
2023/08/01 - 2023/09/21 & TESS            & 120       & 6000 - 1000                        & Photometry         & TESS    \\
2023/05/24              & SALT            & 300   & 4060 - 7250           & Spectropolarimetry & RSS     \\
2023/07/11              & SALT            & 300   & 4060 - 7250           & Spectropolarimetry & RSS     \\
2023/08/04              & SALT            & 300   & 4060 - 7250           & Spectropolarimetry & RSS     \\
2024/07/05              & SAAO 1-m  & 10        & Unfiltered                 & Photometry         & SHOC    \\
2024/07/21              & SAAO 1-m  & 10        & Unfiltered                 & Photometry         & SHOC   \\
2023/05/13              & SALT            & 300   & 4060 - 7250           & Spectropolarimetry & RSS     \\
2024/10/30              & SALT            & 300   & 4060 - 7250           & Spectropolarimetry & RSS     \\
2025/05/30              & SALT            & 300   & 4060 - 7250          & Spectropolarimetry & RSS      \\
2025/07/18              & SALT            & 300   & 4060 - 7250          & Spectropolarimetry & RSS      \\
2025/07/20              & SALT            & 300   & 4060 - 7250          & Spectropolarimetry & RSS      \\
2025/07/23              & SALT            & 300   & 4060 - 7250          & Spectropolarimetry & RSS      \\
2025/08/19              & SALT            & 300   & 4060 - 7250          & Spectropolarimetry & RSS      \\
2025/08/20              & SALT            & 300   & 4060 - 7250           & Spectropolarimetry & RSS      \\
2025/06/29 - 2025/07/25 & TESS            & 120       & 6000 - 1000                        & Photometry         & TESS    \\
\hline
\end{tabular}%
}
\end{table*}

\subsection{ATLAS Observations}

We obtained photometry of QS~Tel using the Asteroid Terrestrial-impact Last Alert System (ATLAS; \citealt{2018PASP..130f4505T, 2018ApJ...867..105T}) covering the period from 2017 June 01 to 2025 July 25. During this interval, observations were acquired in two broad passbands: the o--band (5582–8249 \AA) and the c--band (4157–6556 \AA). The data were retrieved from the interactive ATLAS forced-photometry database\footnote{ATLAS interactive forced-photometry service available at \url{https://fallingstar-data.com/forcedphot/}}, which provides position-fixed PSF photometry for all survey epochs. The resulting long-term light curve is presented in  Figure~\ref{fig:ATLAS} and discussed in Section~\ref{Altas results}.

\subsection{SHOC photometry}

High-speed photometric observations were obtained on 2024 July 5 and 2024 July 21 using the South African Astronomical Observatory (SAAO) 1-m telescope (Table~\ref{tab:table_of_obs}). The data were acquired under photometric conditions with the Sutherland High-speed Optical Camera (SHOC; \citealt{Coppejans_2013}). The observations were unfiltered, and the SHOC detector was operated in frame-transfer mode with 4$\times$4 binning. Each run covered approximately one full orbital cycle of the binary. Data reduction was performed using the IO\_Phot software\footnote{IO\_Phot is a Python-based pipeline for aperture photometry, developed at SAAO, and available at \url{https://www.saao.ac.za/~sbp/IO_phot/IO_phot.py}}, which is specifically designed for processing SHOC data cubes. The pipeline applies flat-field corrections, background subtraction, and performs both aperture and differential photometry. All timestamps were converted to the barycentric reference frame of the Solar System (BJD) using the corrections described by \citet{2010PASP..122..935E}. 

\subsection{TESS Observations}

We obtained space-based photometric observations from the archival datasets of the Transiting Exoplanet Survey Satellite (TESS; \citealt{2015JATIS...1a4003R}), accessed via the Mikulski Archive for Space Telescopes (MAST; \citealt{2015AAS...22533659F}). TESS observed the system at a 120-second cadence in three sectors: Sector~27 (2020 July 5–30), Sector~67 (2023 July 1–28), and Sector~94 (2025 June 29–July 25). TESS light curves are available in two forms: Simple Aperture Photometry (SAP; \citealt{Kinemuchi_2012}), and Pre-search Data Conditioning Simple Aperture Photometry (PDCSAP), the latter applying corrections for instrumental systematics, removing isolated outliers, and adjusting source flux for crowding. For our analysis, we adopted the PDCSAP light curves, which are presented in Figure~\ref{fig:tess} and will be discussed in detail in Section~\ref{Phot_res}.

\subsection{Spectroscopy}

Phase-resolved spectroscopic observations were obtained over four nights between 2022 September 21 and 26, as listed in Table~\ref{tab:table_of_obs}. The data were collected with the SAAO 1.9-m telescope, using the SpUpNIC spectrograph \citep{2019JATIS...5b4007C}. Two gratings were employed: grating~4, covering 4250–5250 \AA, and grating~5, covering 6250–7250 \AA. Each observing session spanned approximately two orbital cycles of the binary. Arc lamp exposures (CuAr for grating~4 and CuNe for grating~5) were taken frequently after every eight science frames to ensure accurate wavelength calibration. A 1.5~arcsec slit was used, with an exposure time of 250~seconds per spectrum, corresponding to an orbital phase interval of $\Delta\phi \approx 0.03$ ($\approx 3\%$ of QS~Tel's 2.33-hr orbital period. Data reduction followed standard \texttt{IRAF}\footnote{\texttt{IRAF} (Image Reduction and Analysis Facility) is a software suite for the reduction and analysis of astronomical data \citep{1986SPIE..627..733T, 1993ASPC...52..173T}.} procedures, including overscan correction, bias subtraction, and gain correction. One-dimensional spectra were extracted using the \texttt{apall} task within the \texttt{twodspec} package. All timestamps were converted to BJD reference frame.
The resulting average spectrum is shown in Figure~\ref{fig:QS_tel_spectrums} and is discussed in Section~\ref{Res_Spectroscopy}.

\subsection{Photopolarimetry}

Photopolarimetric observations were obtained over three nights, from 2017 June 23 to 25 (see Table~\ref{tab:table_of_obs}). All data were collected using the HI-speed Photo-POlarimeter (HIPPO; \citealt{2010MNRAS.402.1161P}) mounted on the SAAO 1.9-m telescope in Sutherland, South Africa. HIPPO was operated in both linear and circular polarimetry modes, as well as in photometry mode. An unfiltered configuration, defined by the response of the RCA31034A GaAs photomultiplier, was used on two nights, while a broadband red OG57 filter was used on the remaining night. Several polarised and unpolarised standard stars were observed to calibrate the instrument, allowing the determination of wave-plate position angle offsets and efficiency factors. All observations were synchronized to GPS with sub-millisecond precision, and timestamps were converted to BJD reference frame. The resulting polarimetric light curves were phase-folded on the orbital period and are presented in Figure~\ref{fig:unfiltered} in Section~\ref{Photopol_results} and described in Section \ref{section:discussion}.

\subsection{Spectropolarimetry}

Circular spectropolarimetric observations were obtained with the Southern African Large Telescope (SALT; \citealt{2006SPIE.6267E..0ZB}) using the Robert Stobie Spectrograph (RSS; \citealt{10.1117/12.460312,10.1117/12.460315}) in circular spectropolarimetry mode \citep{2003SPIE.4843..170N}. As summarized in Table~\ref{tab:table_of_obs}, observations were carried out on multiple occasions between 2023 May 24 and 2025 July 20, with individual exposure times of 300~s. The spectropolarimetric setup employed rotating quarter- and half-wave plates near the focal plane, together with a calcite mosaic beam splitter ahead of the camera. The PG0900 grating was used, providing wavelength coverage from 4250--7250\,\AA. The CCD data were reduced using the \texttt{Polsalt-beta}\footnote{For details, see \url{https://github.com/saltastro/polsalt/}} pipeline, which is based on the \textsc{pysalt} package. Standard reduction steps included bias subtraction, gain correction, and overscan correction. The ordinary (O) and extraordinary (E) beams were extracted with the \texttt{Polsalt} software and written as separate extensions within a single output file using a custom Python script. All timestamps were corrected to BJD reference frame. The resulting total-flux spectra, circular polarisation, and polarised-flux measurements are presented in Figure~\ref{fig:Spectropolarimetry} and described in Section~\ref{Spectropol_res}.

\section{Results}\label{section:results}

\subsection{Photometric Lightcurves}

\subsubsection{ATLAS Long-term Photometry}
\label{Altas results}

\begin{figure*}
    \centering
    \includegraphics[width=\linewidth]{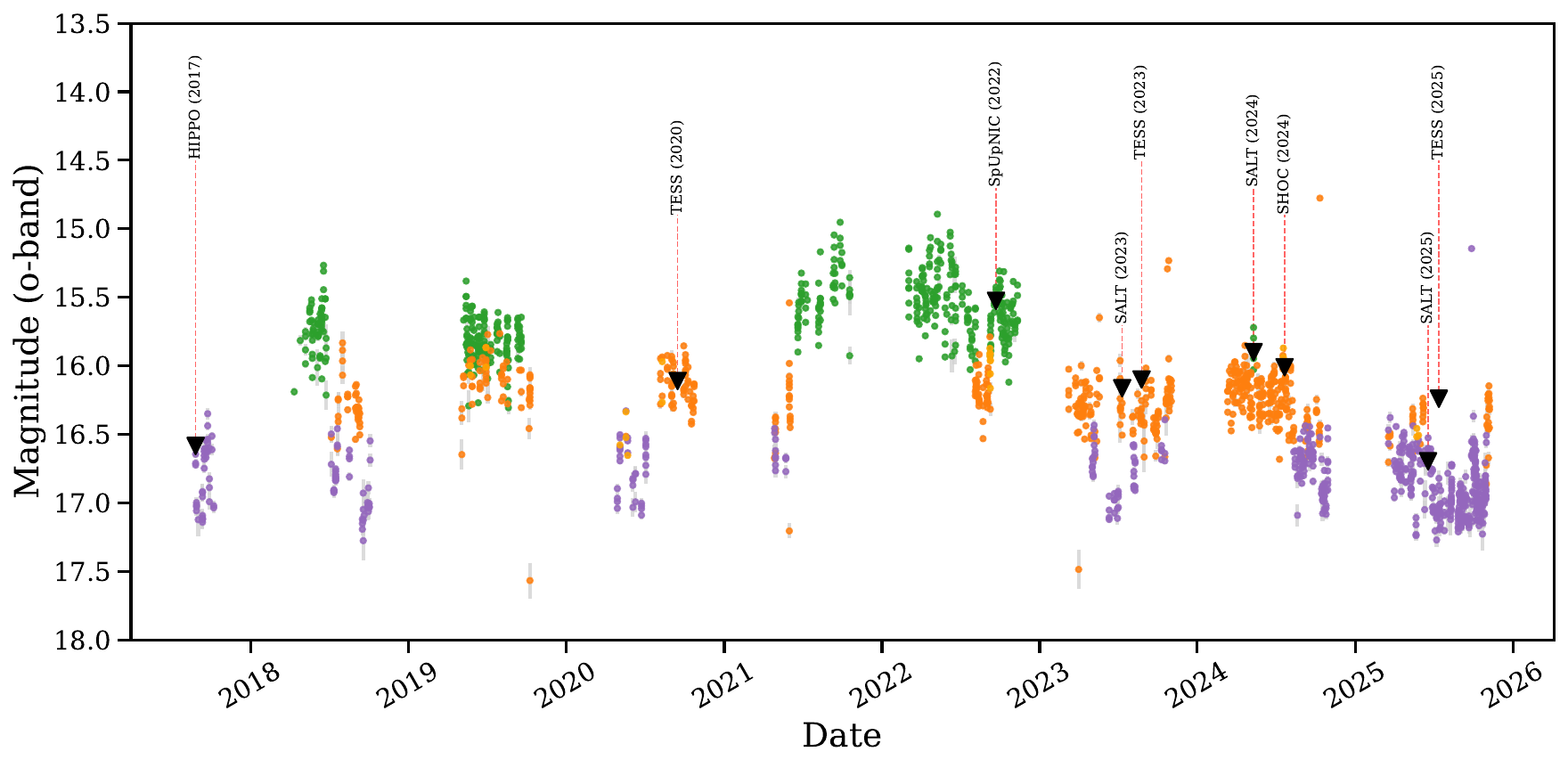}    
    \caption{ATLAS ($o$-band) long-term photometry of QS~Tel spanning 2017--2025 August. The three colours indicate the classified accretion states: green corresponds to the high state, orange to the intermediate state, and purple to the low state.}
    \label{fig:ATLAS}
\end{figure*}

Figure \ref{fig:ATLAS} shows the ATLAS $o$-band long-term photometric light curve  of QS~Tel spanning 2017 June--2025 August. The red vertical dashed lines marks the dates of our observations. The $o$-band light curve was classified into three accretion states using a Gaussian Mixture Model (GMM\footnote{Implemented with the \texttt{scikit-learn} Python library: \url{https://scikit-learn.org/stable/modules/generated/sklearn.mixture.GaussianMixture.html}}). The optimal number of components was determined by minimizing the Bayesian Information Criterion \citep{1978AnSta...6..461S}, with three components providing the best representation corresponding to high, intermediate, and low accretion states. Each observation was assigned to the most likely component, while classifications with probabilities below 65$\%$ were flagged as uncertain to avoid isolated outliers. The components were ordered by brightness and labeled accordingly, and a rolling majority filter was applied to stabilize the final-state sequence, and the resulting light curve is shown in Figure \ref{fig:ATLAS}. We find mean magnitudes of $15.653 \pm 0.013$\,mag for the high state (green), $16.201 \pm 0.022$\,mag for the intermediate state, and $16.784 \pm 0.034$\,mag for the low state consistent with the accretion states presented by \citet{10.1093/mnras/280.4.1121,1995A&A...293..764S,10.1093/mnras/280.4.1121,2001MNRAS.322..631R} (to be discussed in section 4.1) .

\subsubsection{TESS lightcurves}
\label{Phot_res}

\begin{figure}
  \centering
  \includegraphics[width=\linewidth]{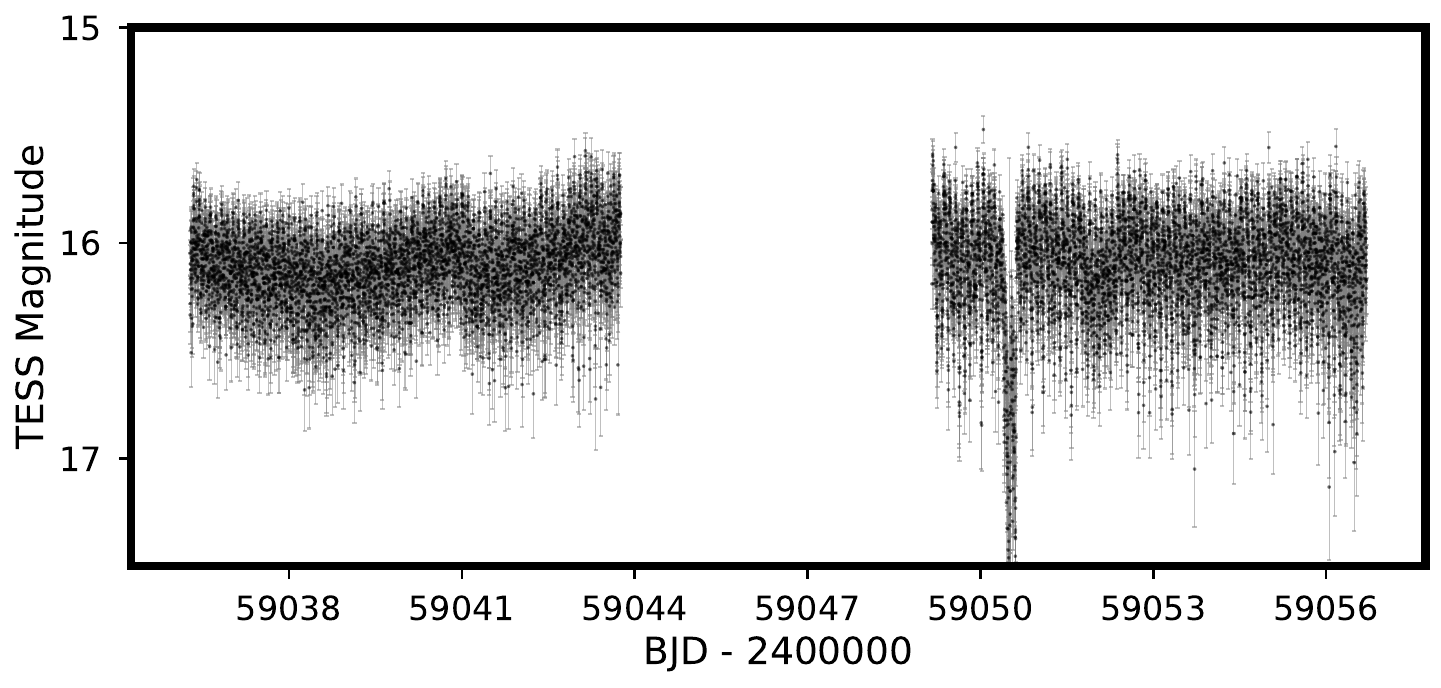}
  \includegraphics[width=\linewidth]{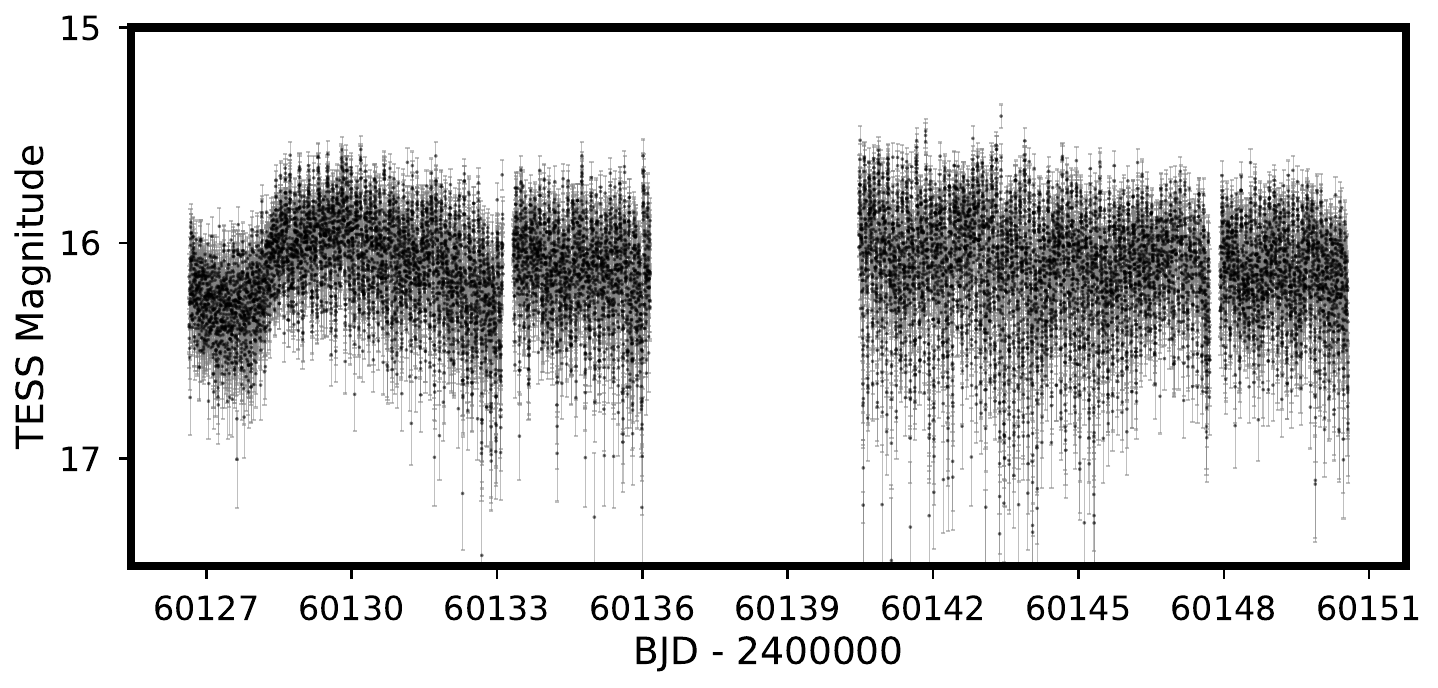}
  \includegraphics[width=\linewidth]{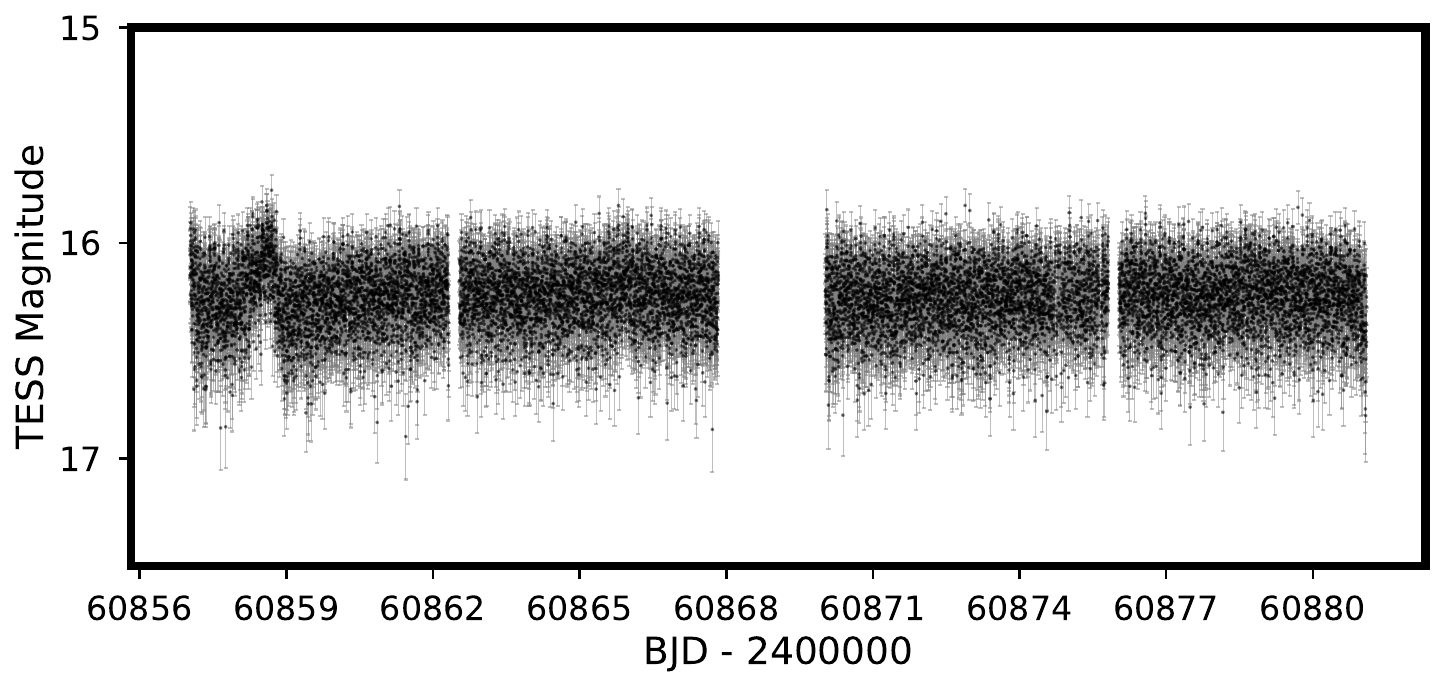}
  \caption{Photometric light curves of QS~Tel obtained with the TESS satellite.
The top, middle, and bottom panels show data from Sectors 27 (2020 July), 67 (2023 July), and 94 (2025 July), respectively. Error bars indicate 1$\sigma$ uncertainties.}
  \label{fig:tess}
\end{figure}

Figure~\ref{fig:tess} shows the TESS light curves of QS~Tel. These datasets were obtained while the system was in an intermediate accretion state, as indicated by the long-term ATLAS ($o$-band) light curve shown in Figure~\ref{fig:ATLAS}. Across all three observation sectors, the system maintains a mean optical brightness centered near 16th magnitude. The top panel ($\mathrm{BJD}$ 59038--59057) displays a steady baseline interrupted by a single deep flux dip dropping past 17.5 mag at $\mathrm{BJD} \sim 59050.5$. The middle panel ($\mathrm{BJD}$ 60127--60151) exhibits recurring flux dips down to 17.0--17.3 mag across consecutive orbital cycles. The bottom panel shows continuous, smooth orbital modulation bounded between 15.8 and 16.7 mag, with a complete absence of sharp flux drops or deep localized dips. The TESS light curves from all sectors were combined and analysed using Lomb–Scargle periodograms \citep[see][]{1976Ap&SS..39..447L, 1982ApJ...263..835S}. A strong periodic signal was detected at a frequency of $10.289399340 \pm 0.000002366~\mathrm{cycles\,d^{-1}}$. The associated uncertainty was estimated from the full width at half maximum (FWHM) of a Gaussian fit to the primary peak in the power spectrum. The corresponding orbital period, obtained directly from the inverse of the measured frequency, is $2.332498 \pm 0.000005~\mathrm{h}$, where the uncertainty was propagated from the frequency error. This value is consistent with the orbital period reported by \citet{1993MNRAS.262...93B} and \citet{1995A&A...293..764S}, but with a significantly reduced uncertainty.

\subsubsection{Orbital Ephemeris}
\label{orbital_ephem}

The first orbital ephemeris of QS~Tel was published by \citet{1993MNRAS.262...93B} and later refined by \citet{1995A&A...293..764S}. As these studies are now more than three decades old, an updated determination of the ephemeris of the system is needed. We adopted the orbital period derived from the TESS observations presented in the previous section. To determine a new epoch, we analysed the trailed spectra of the He\textsc{ii} 4686 \AA\, emission line shown in Figure~\ref{fig:HeII}. Multiple emission components are present (to be discussed later), but we identified a faint, narrow, red-shifted, low-velocity component between orbital phases 0.3 and 0.6. In polars, such a narrow component typically originates from the irradiated inner hemisphere of the secondary star. Consequently, phase 0 corresponds to the inferior conjunction of the secondary, marked by the blue-to-red crossing of this feature.
We therefore determined the epoch such that the observed red-to-blue crossing occurs at orbital phase 0.5. The uncertainty in the epoch was determined by examining the range of phase offsets derived from the position of the secondary star in the Doppler tomograms. The resulting orbital ephemeris is:

\begin{equation}
T(\text{BJD}) = 2459844.14193(4) + 0.097187340(2) \cdot E
\label{eq2}
\end{equation}
where $E$ is the cycle number, the numbers in brackets show the uncertainties of the values in the ephemeris. This ephemeris was used to phase all observations presented in this work. Comparing this updated ephemeris with that of \citet{1995A&A...293..764S}, our 31-year baseline spans $E = 108\,493$ orbital cycles ($\Delta T \approx 11\,044$~d). Accounting for the HJD-to-BJD coordinate conversion ($\le 4$~s), error propagation yields an accumulated timing uncertainty of $\approx 0.00022$~d ($\approx 19$~s). This corresponds to a fractional phase uncertainty of $\Delta \phi < 0.003$ ($<0.3\%$ of the $8397$~s orbit), confirming exceptional long-term phase coherence across three decades. We are unable to detect the period derivative ($\dot{P}$) previously suggested by \citet{1995A&A...293..764S}, neither within our continuous 5-year dataset nor over the extended 31-year baseline.

\subsubsection{Folded Lightcurves}

\begin{figure*}
  \centering
  \includegraphics[width=\linewidth]{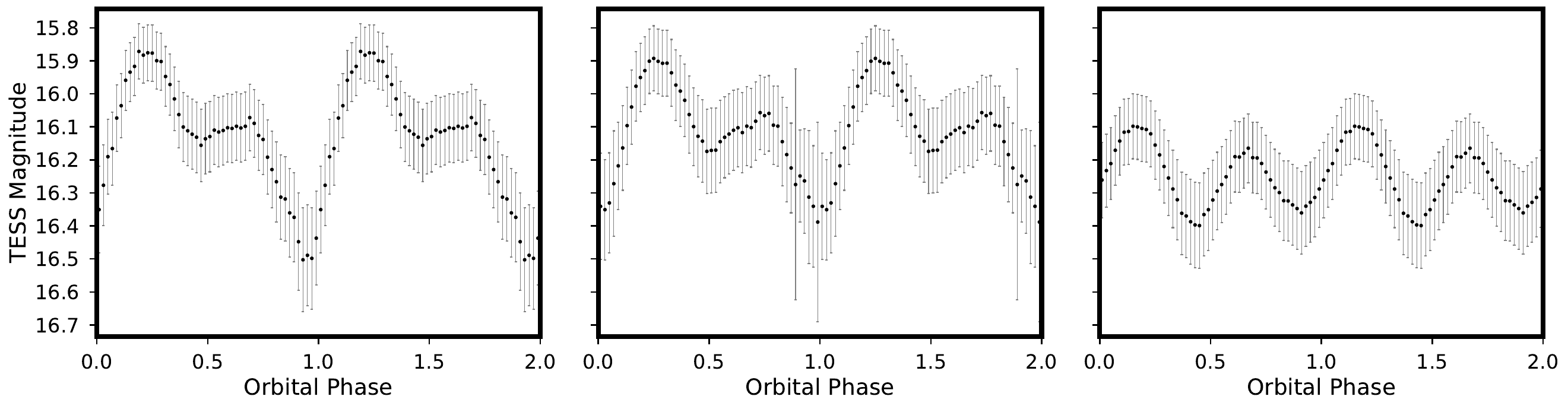}
  \caption{Phase-folded TESS light curves of QS~Tel. The left panel shows data from 2020 July, the middle panel from 2023 July, and the right panel from 2025 July. Each light curve is binned into 50 phase bins and folded on the system’s orbital period to emphasize the periodic modulation. Error bars represent 1$\sigma$ uncertainties.}
  \label{fig:folded}
\end{figure*}

Figure~\ref{fig:folded} shows the TESS phase-folded light curves of QS~Tel from the three TESS datasets. The data were binned into 50 phase bins. All datasets are phase-folded on the orbital period and plotted over two consecutive cycles ($0 \le \phi \le 2$) for visual clarity, revealing a distinct orbital modulation. The left and central panels of Figure~\ref{fig:folded} retain a distinct double-humped structure, with a well-defined orbital minimum at phases 0.9--1.0. In contrast, the right-hand panel of Figure~\ref{fig:folded} exhibits a smoother, more sinusoidal orbital modulation with reduced contrast between the two humps. The orbital minimum is also shallower and less clearly defined, lacking the pronounced deep minimum evident in both the left and central panels of Figure~\ref{fig:folded}.

\subsubsection{Flickering and Quasi-Periodic Oscillations}

\begin{figure*}
    \centering
    \includegraphics[width=0.9\linewidth]{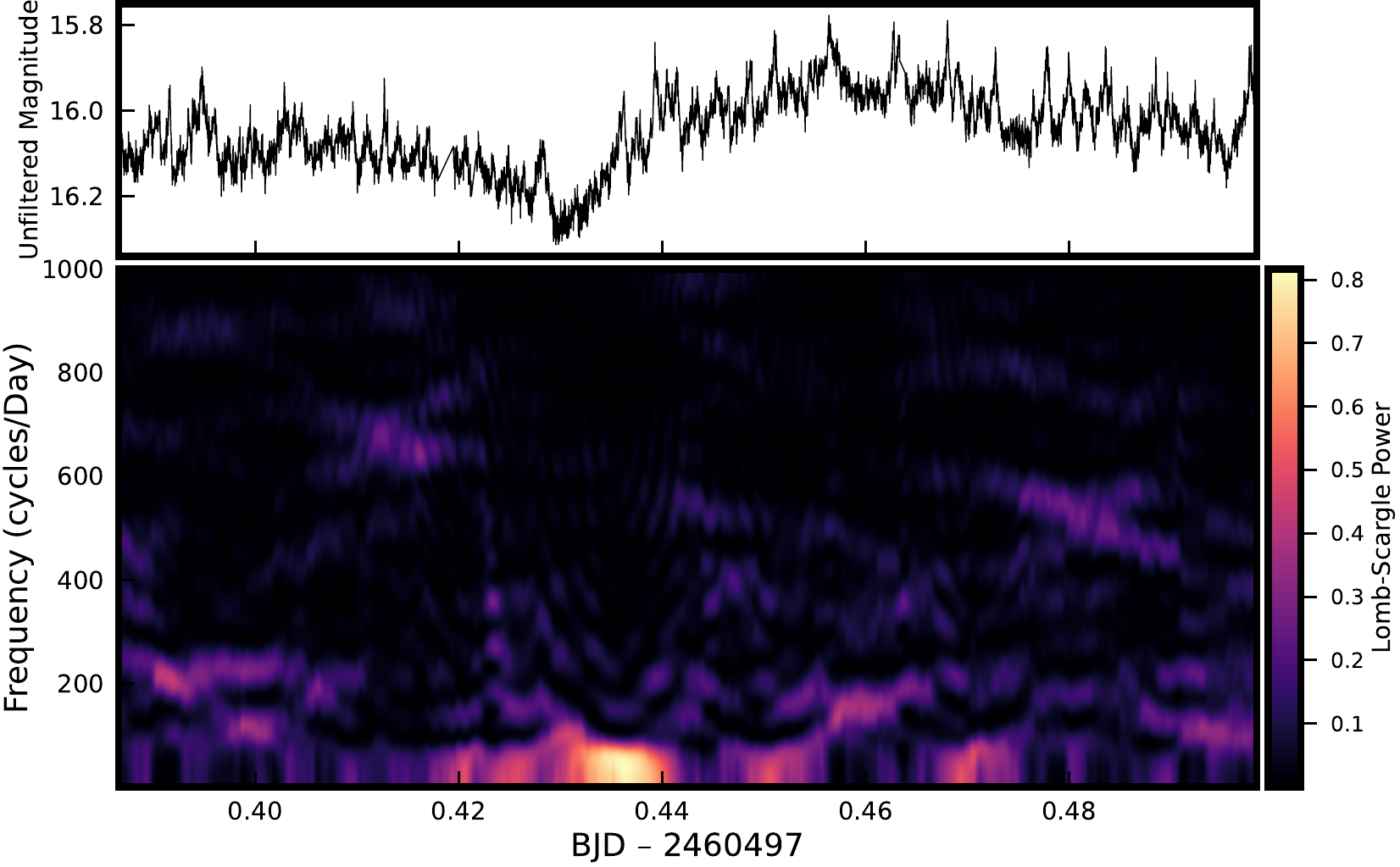}    
    \caption{Trailed Lomb–Scargle periodogram (dynamic power spectrum) of QS~Tel from high-speed photometry on 2024 July 05. The top panel shows the light curve, and the bottom panel displays the trailed Lomb–Scargle power. The colour scale represents normalized power, tracing the evolution of quasi-periodic oscillations (QPOs) throughout the night.}
    \label{fig:QPO_Trailed_july_05}
\end{figure*}

\begin{figure*}
    \centering
    \includegraphics[width=0.9\linewidth]{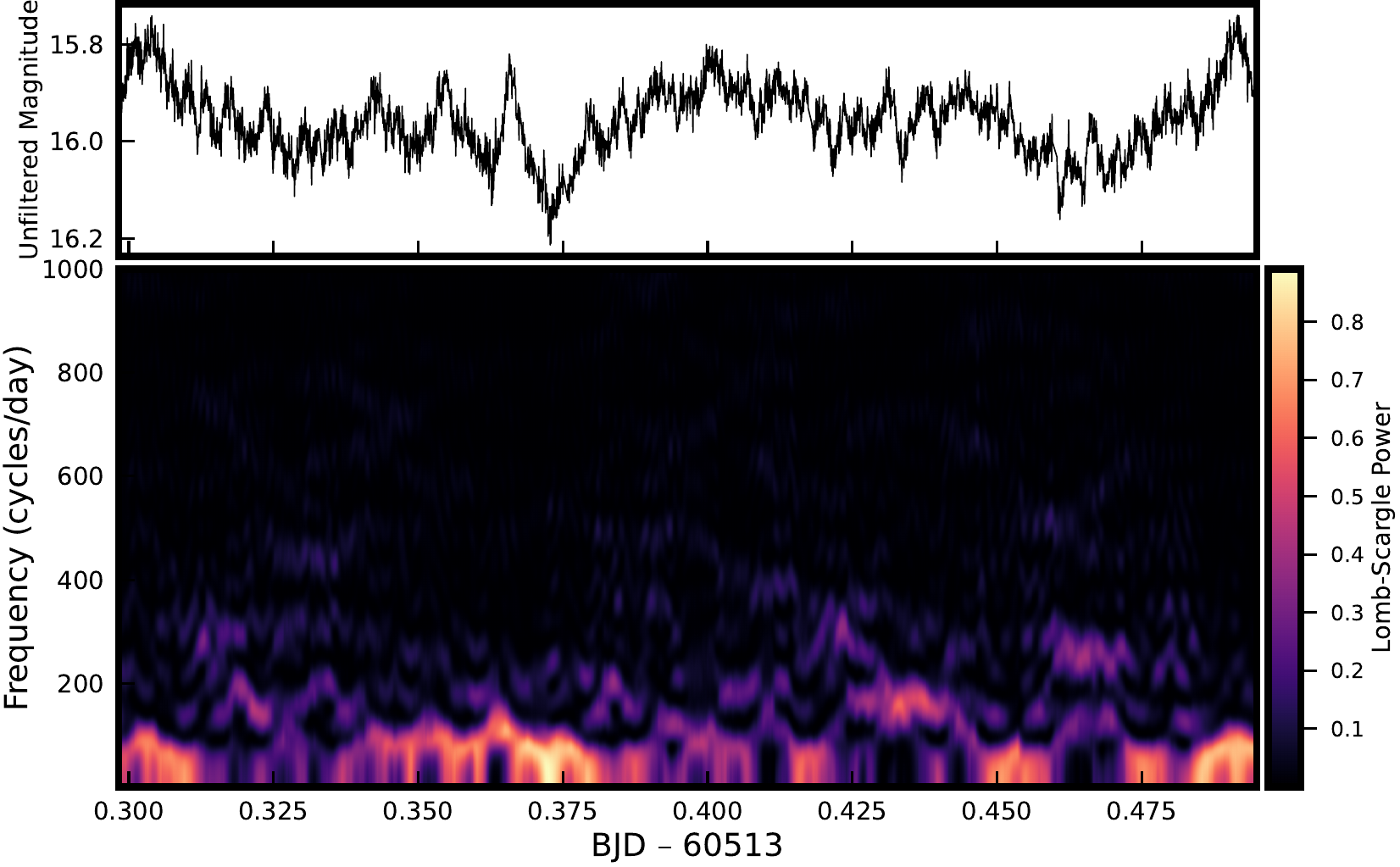}    
    \caption{Trailed Lomb–Scargle periodogram (dynamic power spectrum) of QS~Tel from high-speed photometry on 2024 July 21. The top panel shows the light curve, and the bottom panel displays the trailed Lomb–Scargle power. The colour scale represents normalized power, tracing the evolution of quasi-periodic oscillations (QPOs) throughout the night.}
    \label{fig:QPO_Trailed_july_21}
\end{figure*}

The top panels of Figures~\ref{fig:QPO_Trailed_july_05} and \ref{fig:QPO_Trailed_july_21} show the high-cadence, unfiltered SHOC light curves of QS Tel, each covering approximately one orbital cycle. In addition to the general double-humped modulation, the light curves display flickering and possible quasi-periodic oscillations (QPOs). To investigate the presence of QPOs, we computed dynamic power spectra by calculating sliding Lomb–Scargle periodograms over 20-minute segments of each light curve. 
The resulting power maps are shown in the bottom panels of Figures~\ref{fig:QPO_Trailed_july_05} and \ref{fig:QPO_Trailed_july_21}. For the most part, the dynamic spectra are consistent with stochastic flickering. However, there are short intervals where apparent periodic oscillations in the light curves produce significant power at well-defined frequencies. Specifically, QPOs centred near $\sim$200 cycles d$^{-1}$ (7–8 minutes) appear at BJD = 2460497.40 and 2460497.46 (Figure~\ref{fig:QPO_Trailed_july_05}), and at BJD = 2460513.44 and 2460513.46 (Figure~\ref{fig:QPO_Trailed_july_21}). In addition, the oscillations visible near BJD = 2460497.48 in Figure~\ref{fig:QPO_Trailed_july_05} exhibit an evolving QPO component at higher frequencies, around 500–600 cycles d$^{-1}$ (2–3 minutes). To our knowledge, this represents the first reported detection of quasi-periodic oscillations in QS~Tel (see Section~\ref{section:discussion_phot}).

\subsection{Spectroscopy}
\label{Res_Spectroscopy}

\begin{figure*}
  \centering
  \includegraphics[width=0.9\linewidth]{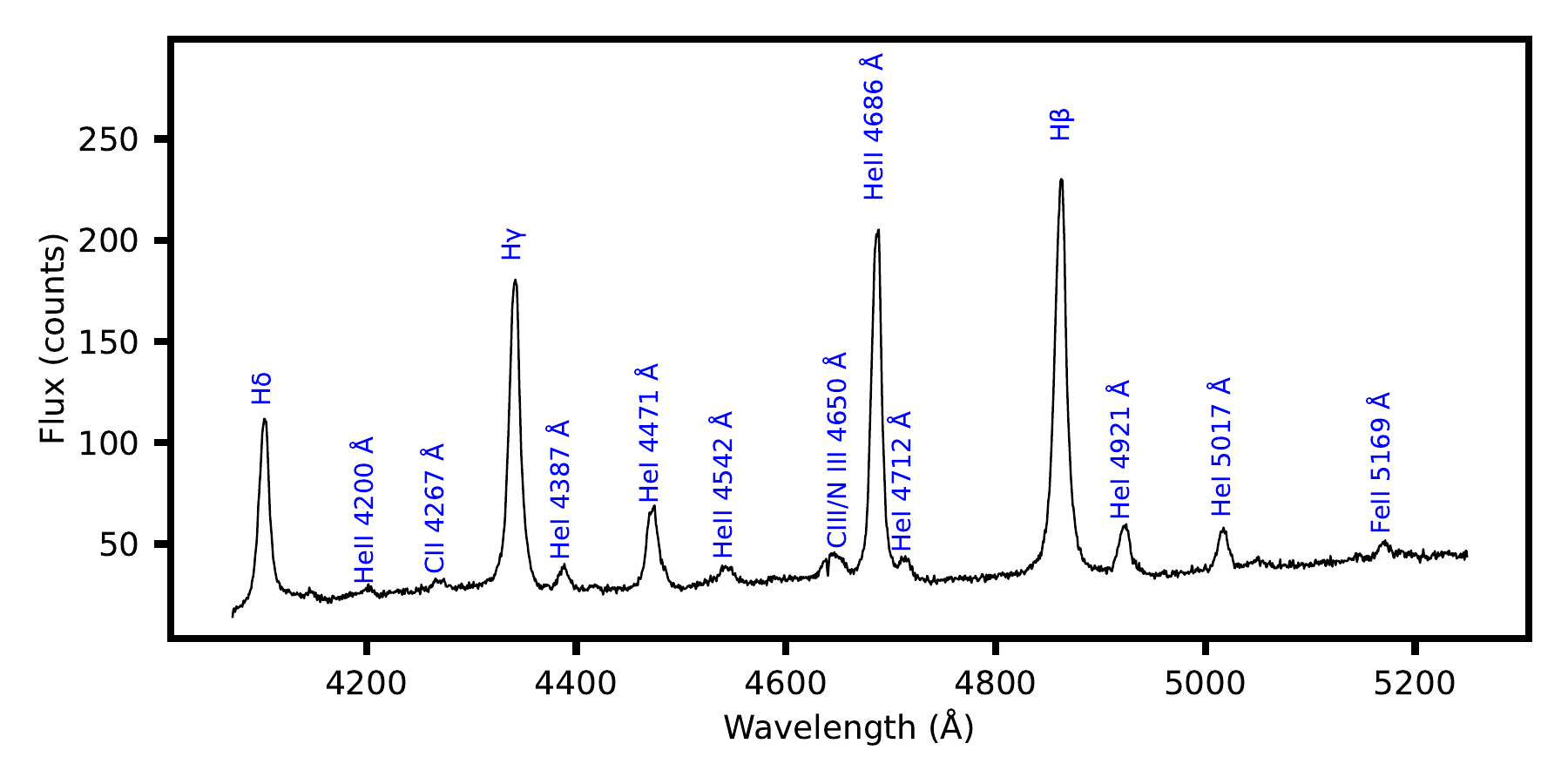}

  \vspace{1em} 

  \includegraphics[width=0.9\linewidth]{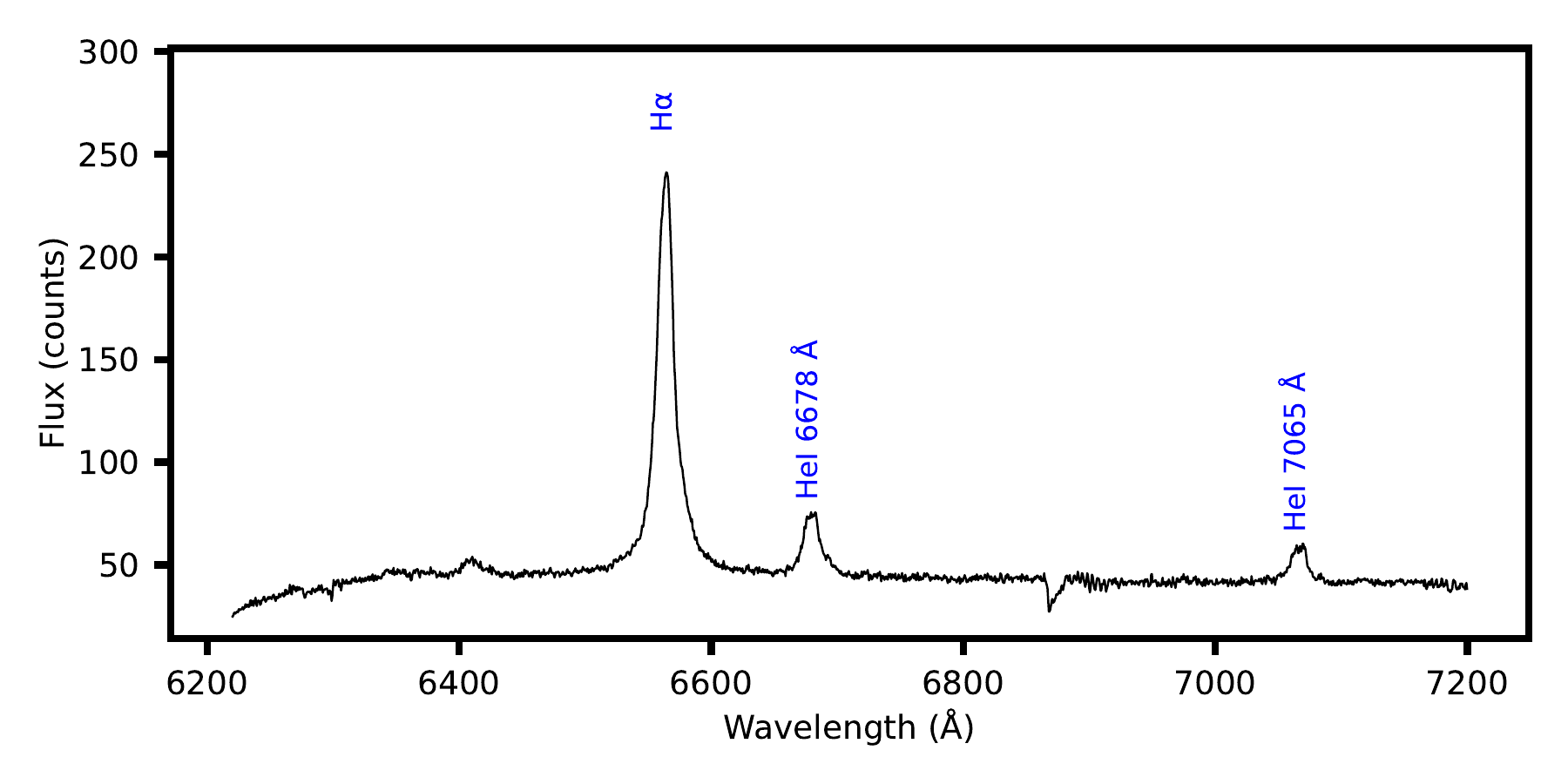}

  \caption{Average spectra of QS Telescopii obtained using the SAAO 1.9-m telescope. The top panel shows the bluer spectrum obtained on 2022 September 21 with Grating 4, and the bottom panel shows the redder spectrum obtained on 2022 September 23  with Grating 5. The observed absorption line at 6867\,\AA\, is a telluric sky absorption feature.}
  \label{fig:QS_tel_spectrums}
\end{figure*}

Figure~\ref{fig:QS_tel_spectrums} presents the average spectra of QS~Tel. The top panel shows the blue part of the spectra. The spectra displays both broad single-peaked and double-peaked Balmer emission lines. Strong He\textsc{ii} emission is present at 4686\,\AA, while weaker He\textsc{ii} lines are observed at 4542\,\AA\ and 4200\,\AA. Several He\textsc{i} lines are also detected at 5017\,\AA, 4921\,\AA, 4712\,\AA, 4471\,\AA, and 4387\,\AA. The Bowen (C\textsc{iii}/N\textsc{iii}) blend at 4650\,\AA\ and the C\textsc{ii} line at 4267\,\AA\ appear in weak emission. The bottom panel shows the redder part of the spectra. This spectra is dominated by a strong, single-peaked H$\alpha$ emission line, together with weak He\textsc{i} emission at 6678\,\AA\, and 7065\,\AA\, which appears either single- or double-peaked. The absorption feature observed at 6867\,\AA\ is identified as a telluric atmospheric absorption line. The prominence of strong Balmer and He\textsc{ii}\,4686\,\AA\, emission lines indicates that QS~Tel was in a high accretion state at the time of these observations, consistent with the long-term ATLAS photometry presented in Figure~\ref{fig:ATLAS} and discussed in Section~\ref{Altas results}.

\subsection{Doppler Tomography}\label{section:doppler_tomography}
The strong emission lines, particularly the He\textsc{ii}\,4686\,\AA\, and $\rm H\alpha$,  were used to construct Doppler maps using the standard and inside-out Doppler map techniques described in \citep{2015A&A...579A..77K, 2016A&A...595A..47K} and the results are shown in Figures \ref{fig:HeII} and \ref{fig:Ha}. The Doppler maps and the trailed spectra were normalized by the maximum flux in the input spectra of each spectrum. To aid in interpretation, we overlaid a binary model on the Doppler maps, initially adopting a WD mass of $M_1 = 0.5\,M_\odot$ and a donor mass of $M_2 = 0.23\,M_\odot$, corresponding to a mass ratio of $q \sim 0.46$ and an inclination angle of $i = 73^\circ$ \citep{2005ASPC..330..429R}. While this model captures the ballistic and magnetically confined accretion streams, we slightly adjusted the parameters to better match our data, resulting in an overlay with $M_1 \sim 0.55\,M_\odot$, $q \sim 0.38$, and $i = 73^\circ$. The centres of mass of the WD and secondary star are marked with ‘$\times$’, and their Roche lobes are outlined by dashed and solid black lines, respectively. The ballistic accretion stream is shown as a solid black line from the Roche lobe of the secondary star extending up to about $65^\circ$ around the WD. Magnetically confined accretion stream follow magnetic dipole trajectories are depicted as dotted black lines at $7^\circ$ azimuthal intervals around the WD.

\subsubsection{Observed and reconstructed trailed spectra}

The bottom panels of Figures~\ref{fig:HeII} and \ref{fig:Ha} present the trailed spectra of He\textsc{ii}\,4686\,\AA\ and $\rm H\alpha$ emission lines. The standard, observed and inside-out reconstruction are shown from left to right respectively. The observed trailed spectra of both emission lines reveal multiple emission components typical of polars. A dominant component (blue) is present throughout the orbital cycle and splits into two between $\phi\sim$0.3--0.6. The broader component ($\phi\sim$0.0--0.55) crosses zero velocity at $\phi\sim$0.3 and reaches maximum blueshift at $\phi\sim$0.5, while the narrower component ($\phi\sim$0.3--0.6) crosses zero velocity at $\phi\sim$0.5. These components merge again between $\phi\sim$0.6--1.0. A redshifted wing is present at $\phi\sim$0.9--1.0, merging with the dominant component near $\phi\sim$1.0. A weaker component (yellow) is visible throughout the orbital cycle, reaching maximum blueshift at $\phi\sim$0.3 and maximum redshift at $\phi\sim$0.8. The reconstructed spectra of He\textsc{ii},$\lambda 4686$ and H$\alpha$ reproduce the main structure of the observed trailed spectra but, as expected, do not capture the variations in line intensity. Both the standard (left) and inside-out (right) reconstructions recover the overall morphology with minimal differences between them, yet neither reproduces the splitting of the dominant component at $\phi \sim 0.3\text{--}0.6$ nor the redshifted wing at $\phi \sim 0.9\text{--}1.0$. Trailed spectra of QS~Tel have previously been presented by \citet{1993MNRAS.262...93B} and \citet{1995A&A...293..764S}, showing features consistent with those typically observed in other polars, such as V834~Cen \citep{2016A&A...595A..47K}.

\subsubsection{Standard and Inside-Out Doppler Maps}

\begin{figure}
\begin{center}$
\begin{array}{ccc}
\includegraphics[height=4cm,width=4cm]{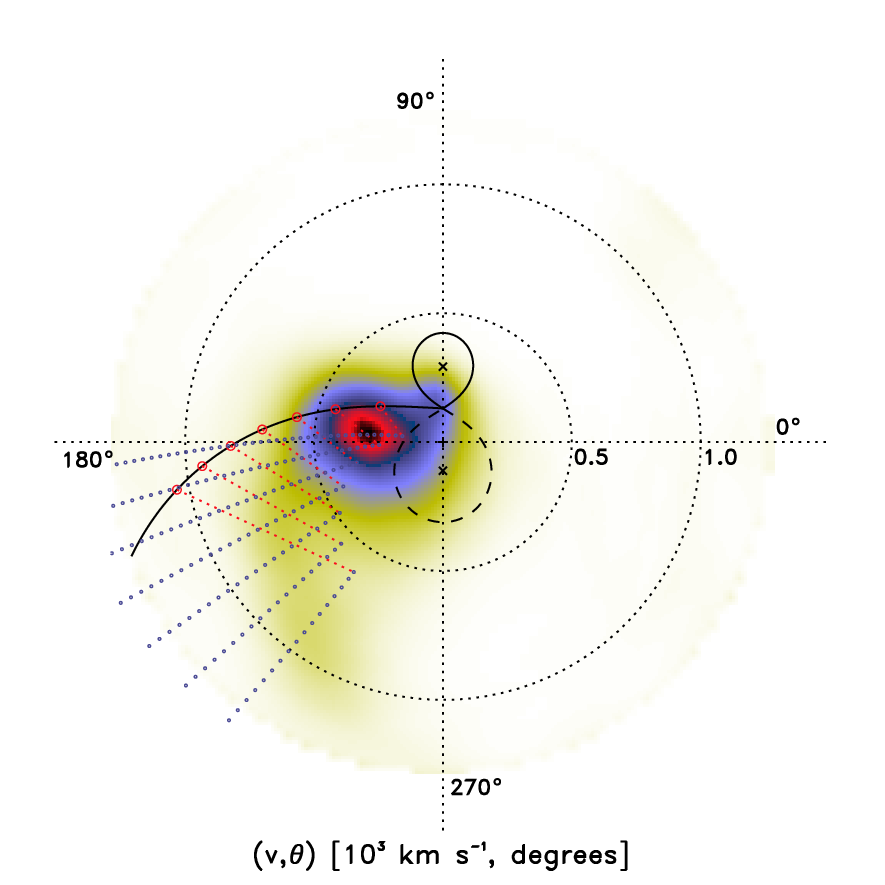}
 \hspace{-0.25cm}
\includegraphics[height=4cm,width=4cm]{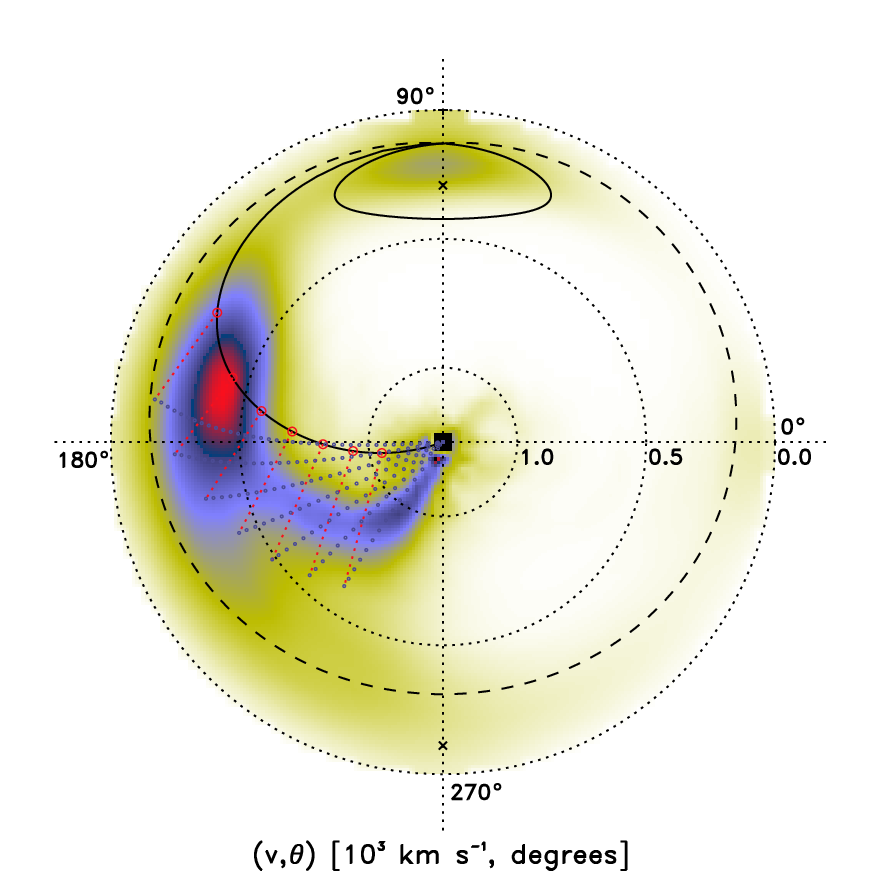}
 \hspace{-0.15cm}
\raisebox{-0.15cm}{\includegraphics[height=4.cm]{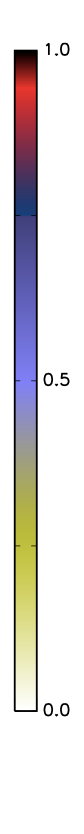}} \\ 
\end{array}$
\end{center}
\vspace{-0.15cm}
\begin{center}$
\begin{array}{cccc}
\includegraphics[width=0.15\textwidth,height=4.5cm]{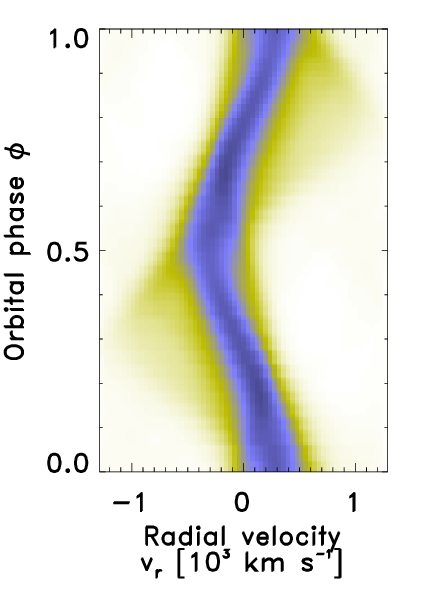} 
 \hspace{-0.10cm}
\includegraphics[width=0.15\textwidth,height=4.5cm]{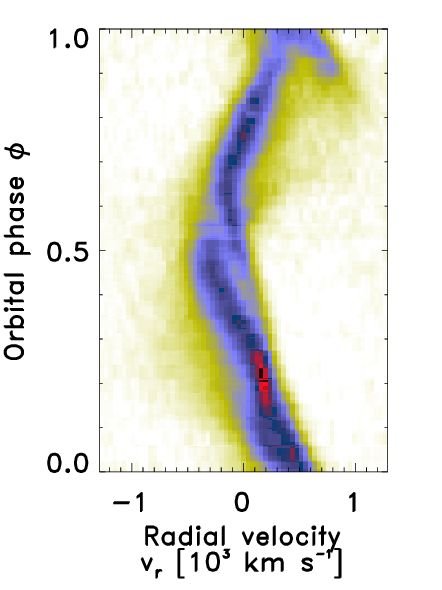} 
 \hspace{-0.10cm}
\includegraphics[width=0.15\textwidth,height=4.5cm]{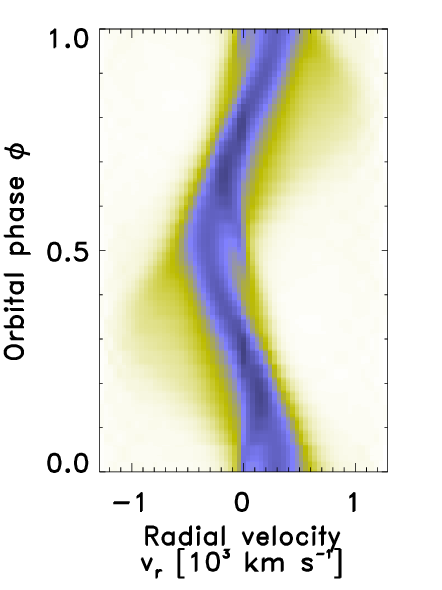} 
 \hspace{-0.10cm}
\raisebox{0.29cm}{\includegraphics[ height= 4.2cm]{Spectra_colorbar_final_a.png}} \\ \\
\end{array}$
\end{center}
\caption{Standard and inside-out Doppler tomograms  based on the He\textsc{ii}\,4686\,\AA\, emission line. Top row: the standard (left) and inside-out (right) Doppler tomograms. Bottom row: the input trailed spectra (centre) with the reconstructed trailed spectra for the standard (left) and inside-out (right) tomograms, respectively.}
\label{fig:HeII}
\end{figure}

\begin{figure}
\begin{center}$
\begin{array}{ccc}
\includegraphics[height=4cm,width=4cm]{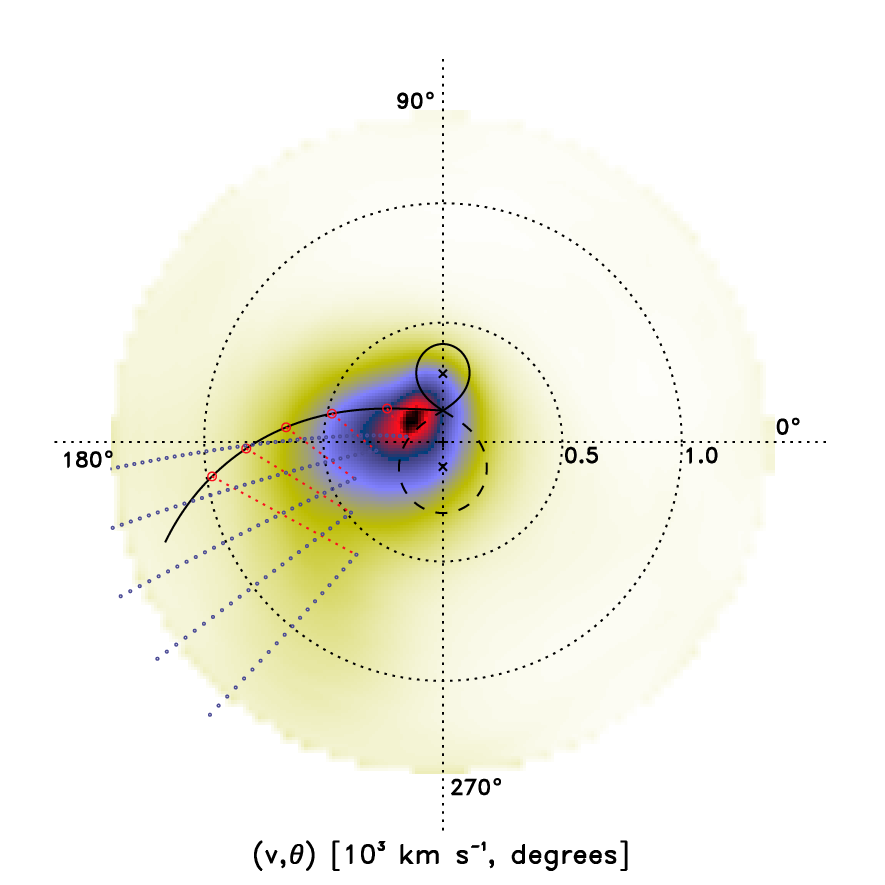}
 \hspace{-0.25cm}
\includegraphics[height=4cm,width=4cm]{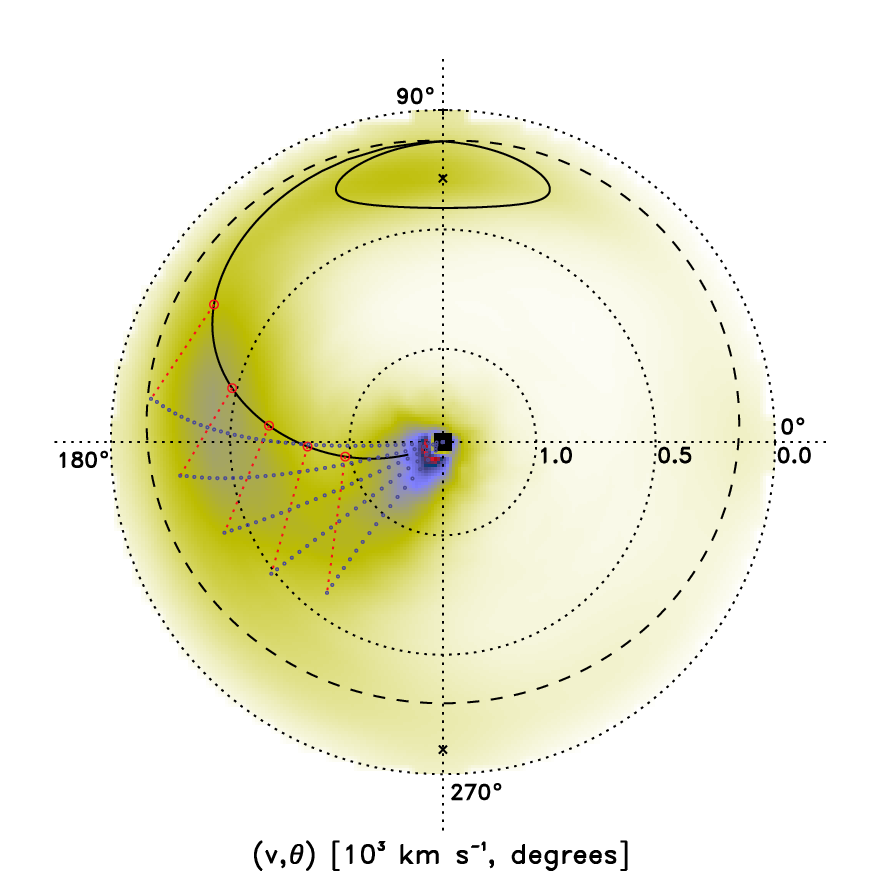}
 \hspace{-0.15cm}
\raisebox{-0.15cm}{\includegraphics[height=4cm]{Spectra_colorbar_final_a.png}} \\ 
\end{array}$
\end{center}
\vspace{-0.15cm}
\begin{center}$
\begin{array}{cccc}
\includegraphics[width=0.15\textwidth,height=4.5cm]{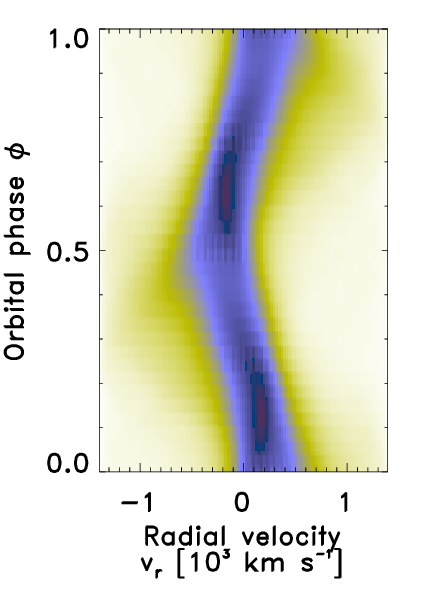} 
 \hspace{-0.10cm}
\includegraphics[width=0.15\textwidth,height=4.5cm]{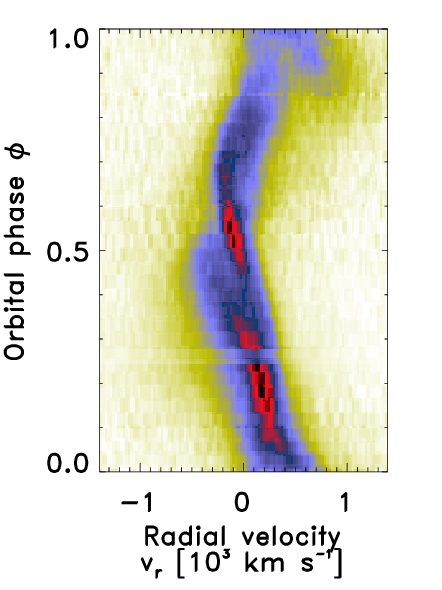}  
 \hspace{-0.10cm}
\includegraphics[width=0.15\textwidth,height=4.5cm]{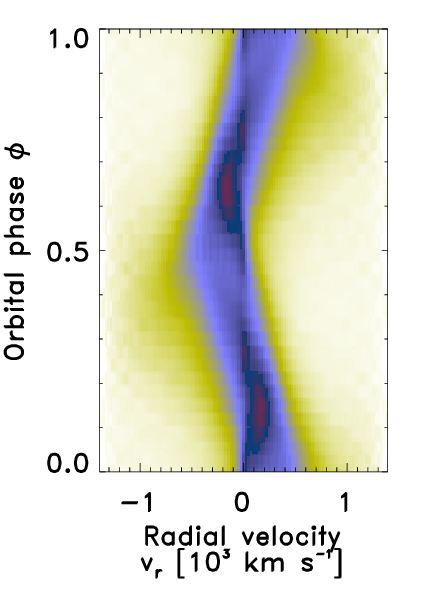} 
 \hspace{-0.30cm}
\raisebox{0.29cm}{\includegraphics[ height= 4.25cm]{Spectra_colorbar_final_a.png}} \\ \\
\end{array}$
\end{center}
\caption{As in Figure~\ref{fig:HeII}, but for the $\rm H\alpha$ emission line. }
\label{fig:Ha}
\end{figure}

The standard Doppler maps of QS Tel are shown in the top-left panels of Figures \ref{fig:HeII} and \ref{fig:Ha}. For both the He\textsc{ii}\,4686\,\AA\ and $\rm H\alpha$ emission lines, the maps show compact, blended emission associated with the ballistic accretion stream and the threading region, with additional contribution from the irradiated face of the secondary star. In the He\textsc{ii}\,4686\,\AA\ tomogram, this emission is split between the upper- and lower-left regions, with the brightest components (black, red, and blue) tracing the ballistic stream and the early stages of magnetically confined accretion stream. Clear evidence of this magnetically confined flow (yellow) appears as an arc extending through the lower-left quadrant between $\theta \approx 180^\circ$ and $270^\circ$, with diagonal dotted blue lines tracing the motion toward higher radial velocities. In contrast, the $\rm H\alpha$ show emission which is more centrally concentrated, with its brightest component following the ballistic stream near the threading region and blending with emission from the irradiated face of the secondary star. The magnetically confined accretion stream is less distinct in the $\rm H\alpha$ map, appearing more diffuse in the lower-left region compared to the well-defined structure seen in He\textsc{ii}\,4686\,\AA\, tomogram.

The inside-out Doppler maps of the He\textsc{ii}\,4686\,\AA\ and $\rm H\alpha$ emission lines (top right panels of Figures~\ref{fig:HeII} and \ref{fig:Ha}) reveal emission from the irradiated face of the secondary star, the ballistic stream, and the threading region, and the magnetically confined accretion flow. In the He\textsc{ii}\,4686\,\AA\ tomogram, emission from the secondary is identified as diffuse structure (yellow) within the Roche lobe at $\theta \sim 90^\circ$. The ballistic stream appears as an arc extending from the L1 point toward the WD, increasing in brightness and reaching a maximum at velocities of $\sim$200--500~$\mathrm{km~s^{-1}}$ and azimuthal angles of $\theta \sim 160^\circ$--$180^\circ$, consistent with the threading region (blue and red). The magnetically confined accretion stream in He\textsc{ii}\,4686\,\AA\, is clearly defined in the lower-left region, where an arc-like blue feature traces the accelerating flow as it follows magnetic field lines toward the WD. In contrast, the $\rm H\alpha$ map shows more diffuse emission overall, with the same components present but appearing fainter and more extended. The magnetically confined flow is less distinct and brightens only closer to the WD, lacking the well-defined arc-like structure seen in the He\textsc{ii}\,4686\,\AA\, tomogram, indicating that the Balmer emission is more broadly distributed throughout the system.

\subsubsection{Flux Modulation Amplitude  Doppler Maps}

\begin{figure}
\begin{center}$
\begin{array}{ccc}
\includegraphics[height=4cm,width=4cm]{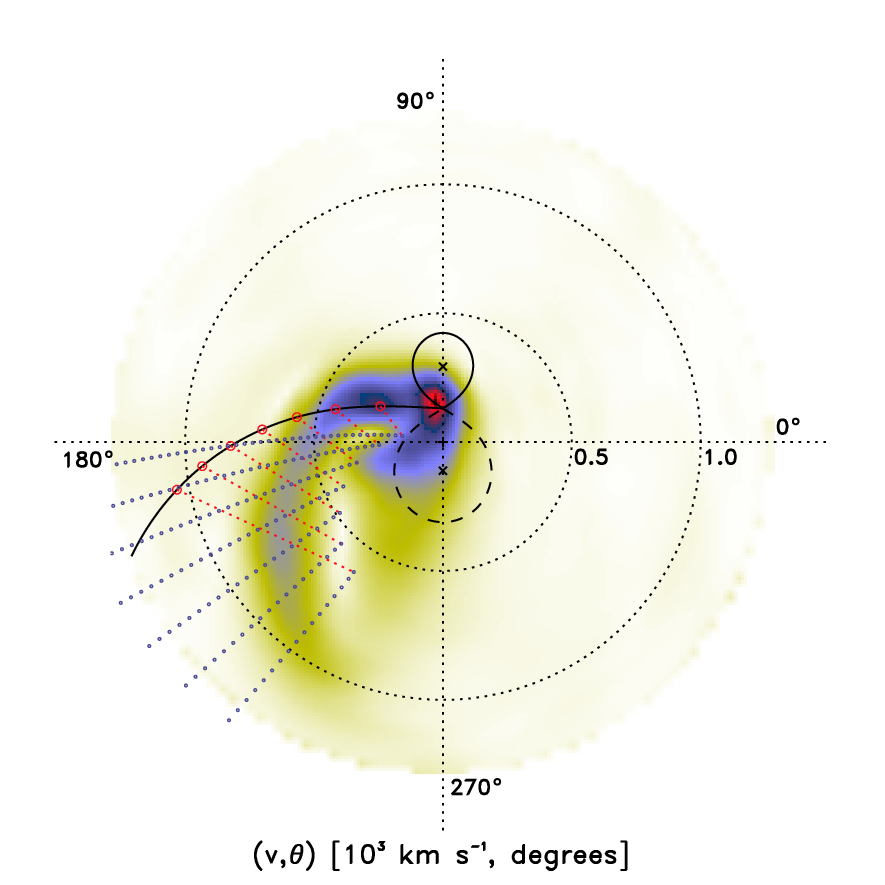}
 \hspace{-0.25cm}
\includegraphics[height=4cm,width=4cm]{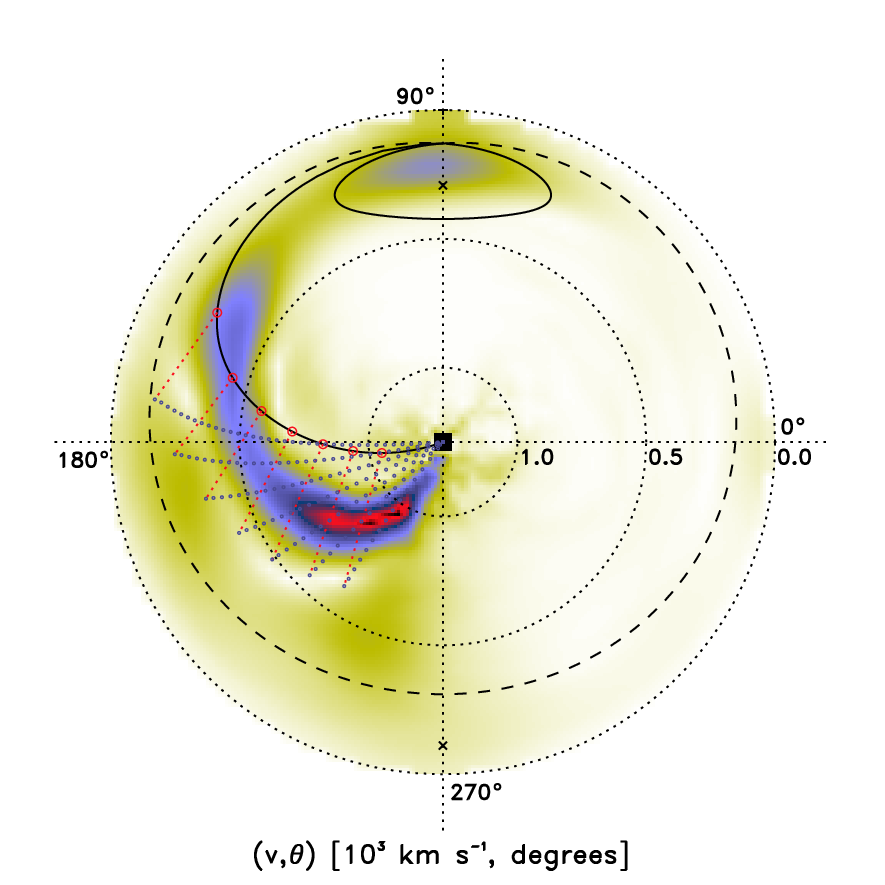}
 \hspace{-0.15cm}
\raisebox{-0.15cm}{\includegraphics[height=4cm]{Spectra_colorbar_final_a.png}} \\
\end{array}$
\end{center}

\vspace{-0.35cm}

\begin{center}$
\begin{array}{ccc}
\includegraphics[height=4cm,width=4cm]{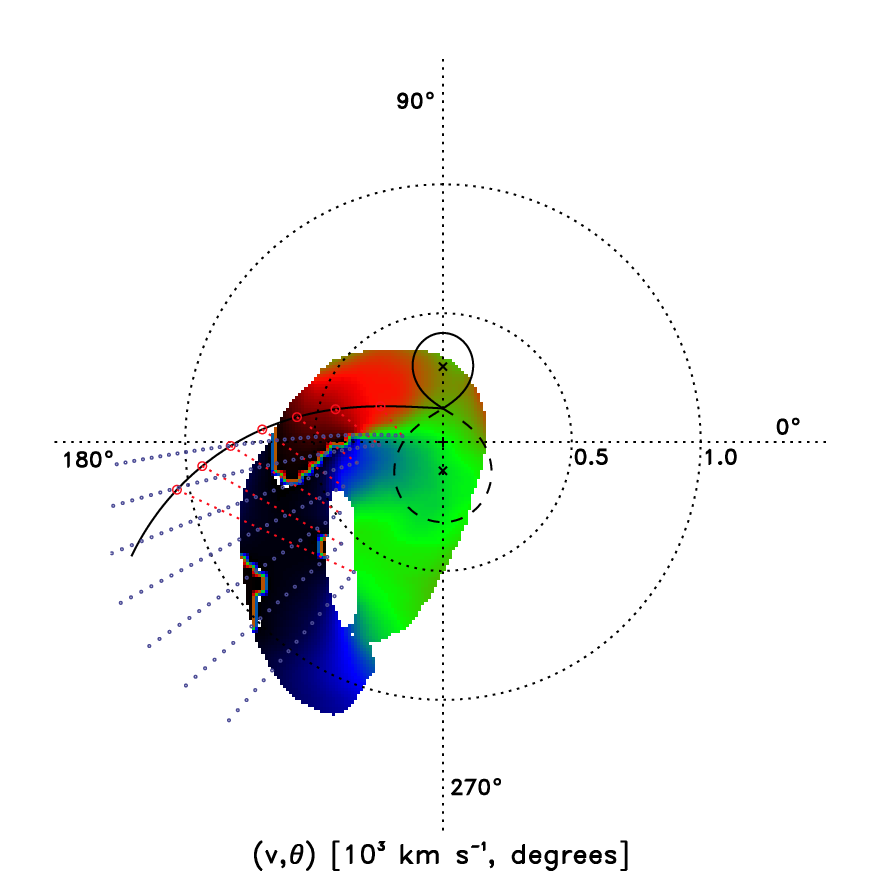}
 \hspace{-0.25cm}
\includegraphics[height=4cm,width=4cm]{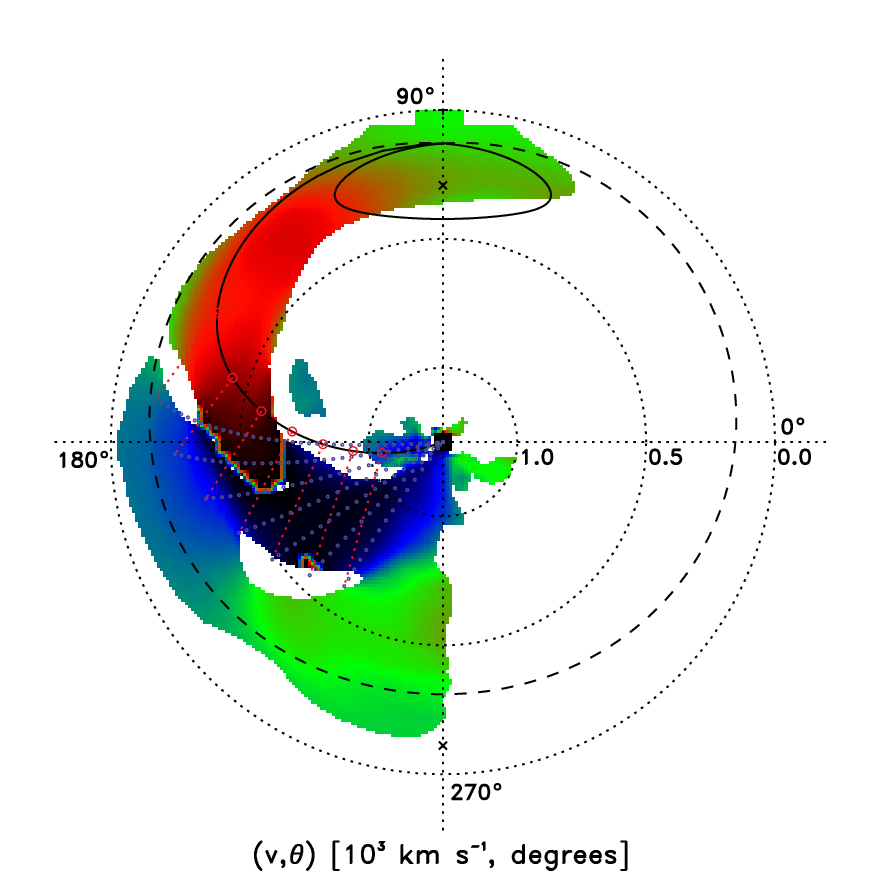}
 \hspace{-0.15cm}
\raisebox{-0.15cm}{\includegraphics[height=4.2cm]{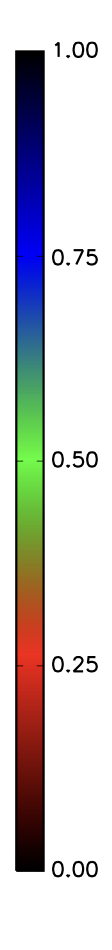}} \\
\end{array}$
\end{center}

\vspace{-0.15cm}

\begin{center}$
\begin{array}{cccc}
\includegraphics[width=0.15\textwidth,height=4.5cm]{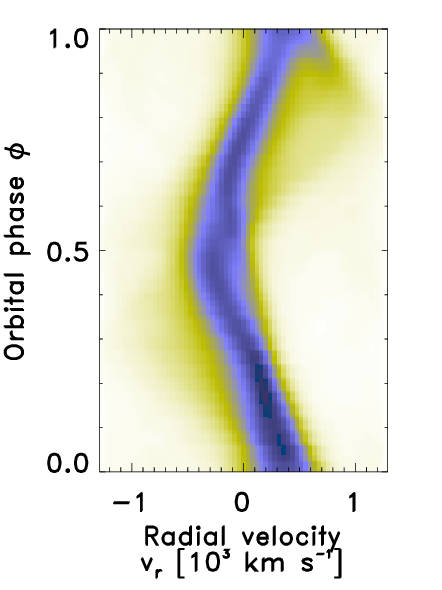}
 \hspace{-0.10cm}
\includegraphics[width=0.15\textwidth,height=4.5cm]{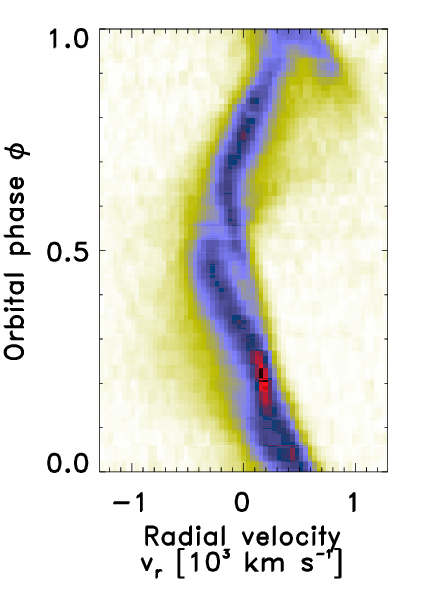}
 \hspace{-0.10cm}
\includegraphics[width=0.15\textwidth,height=4.5cm]{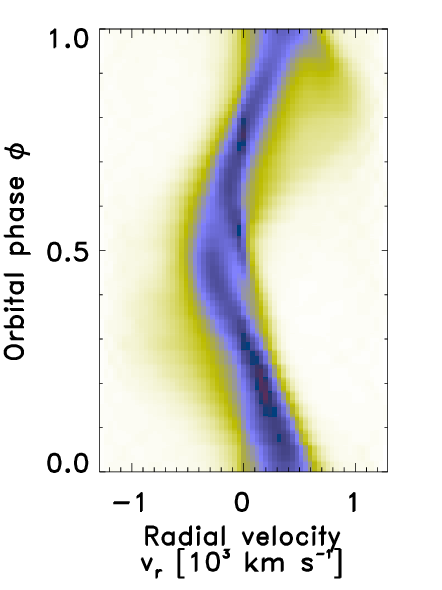} 
\hspace{-0.15cm}
\raisebox{0.29cm}{\includegraphics[ height= 4.25cm]{Spectra_colorbar_final_a.png}} \\ \\
\end{array}$
\end{center}

\vspace{-0.15cm}

\caption{Flux-modulated standard and inside-out Doppler maps of the He\textsc{ii}\,4686\,\AA\, emission line. Top row: flux-modulated standard and inside-out amplitude maps. Middle row: standard and inside-out phase-of-maximum-flux maps. Bottom row: the observed trailed spectra (centre) with the corresponding reconstructed trailed spectra.}
\label{fig:HeM}
\end{figure}

\begin{figure}
\begin{center}$
\begin{array}{ccc}
\includegraphics[height=4cm,width=4cm]{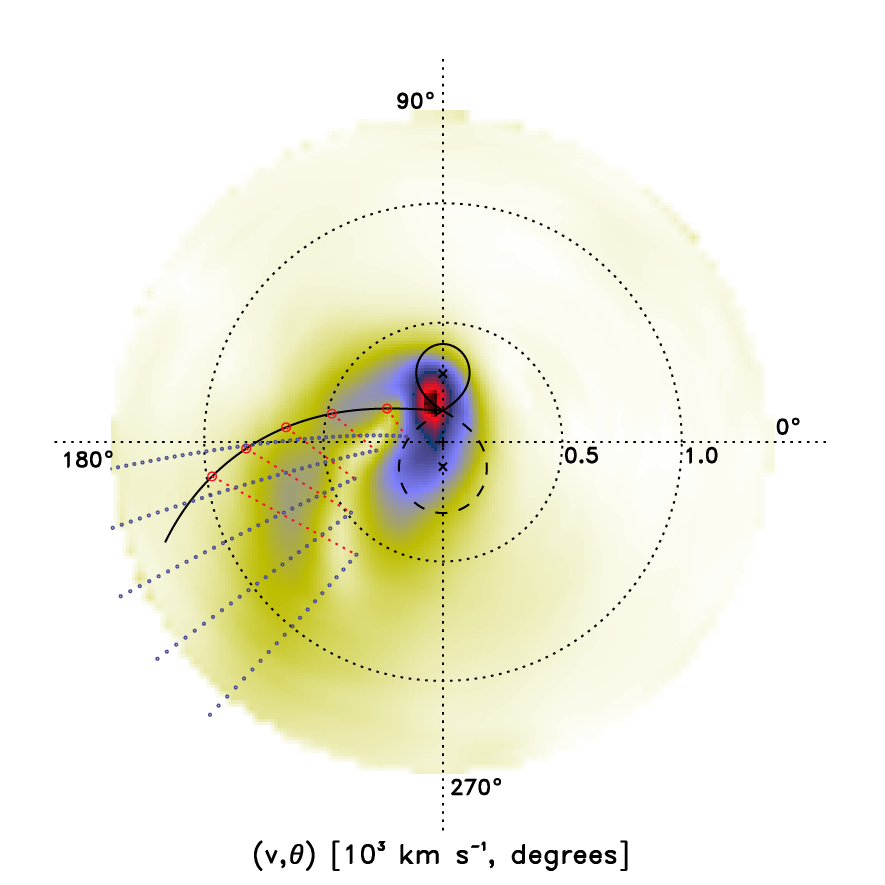}
 \hspace{-0.25cm}
\includegraphics[height=4cm,width=4cm]{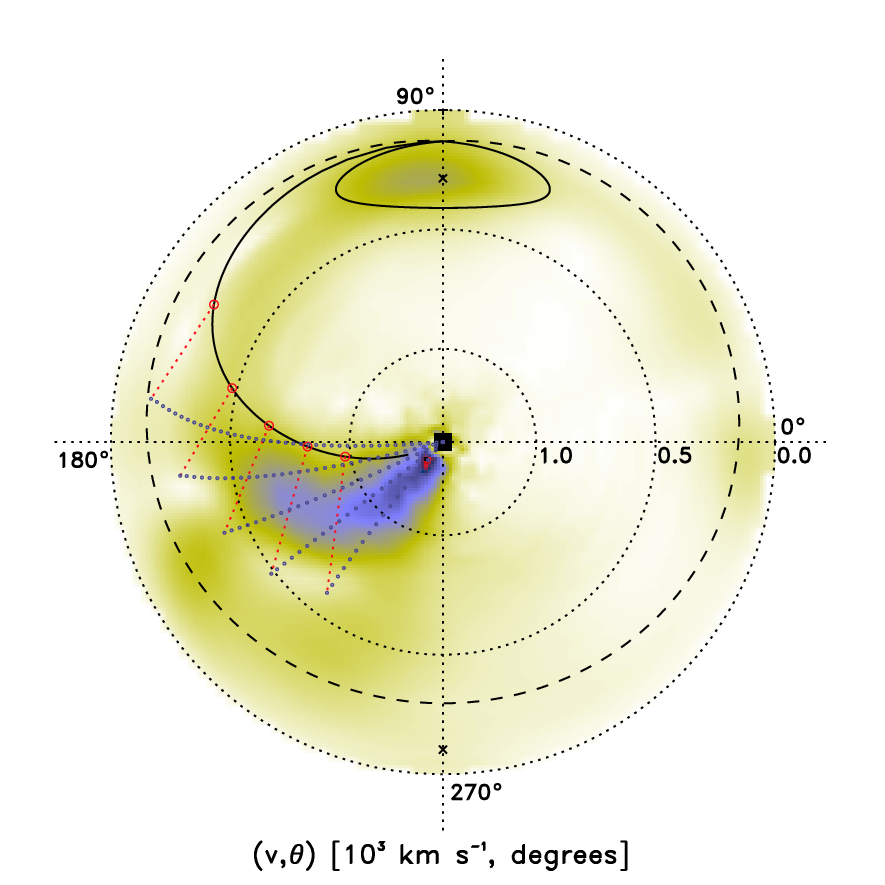}
 \hspace{-0.15cm}
\raisebox{-0.15cm}{\includegraphics[height=4cm]{Spectra_colorbar_final_a.png}} \\
\end{array}$
\end{center}

\vspace{-0.35cm}

\begin{center}$
\begin{array}{ccc}
\includegraphics[height=4cm,width=4cm]{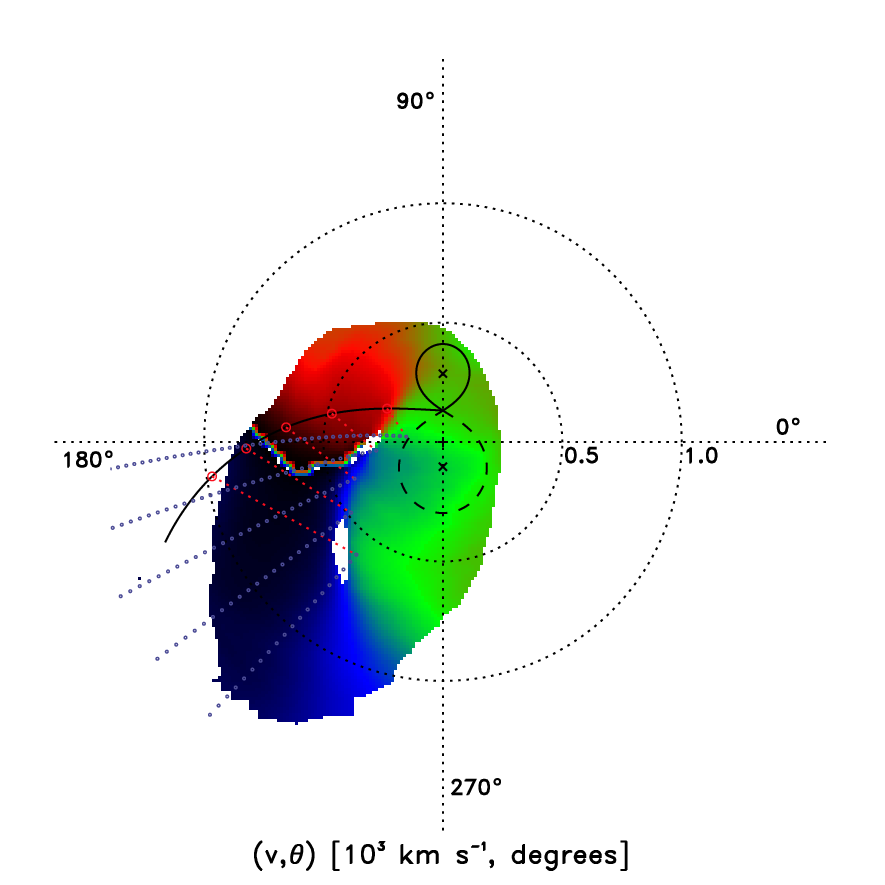}
 \hspace{-0.25cm}
\includegraphics[height=4cm,width=4cm]{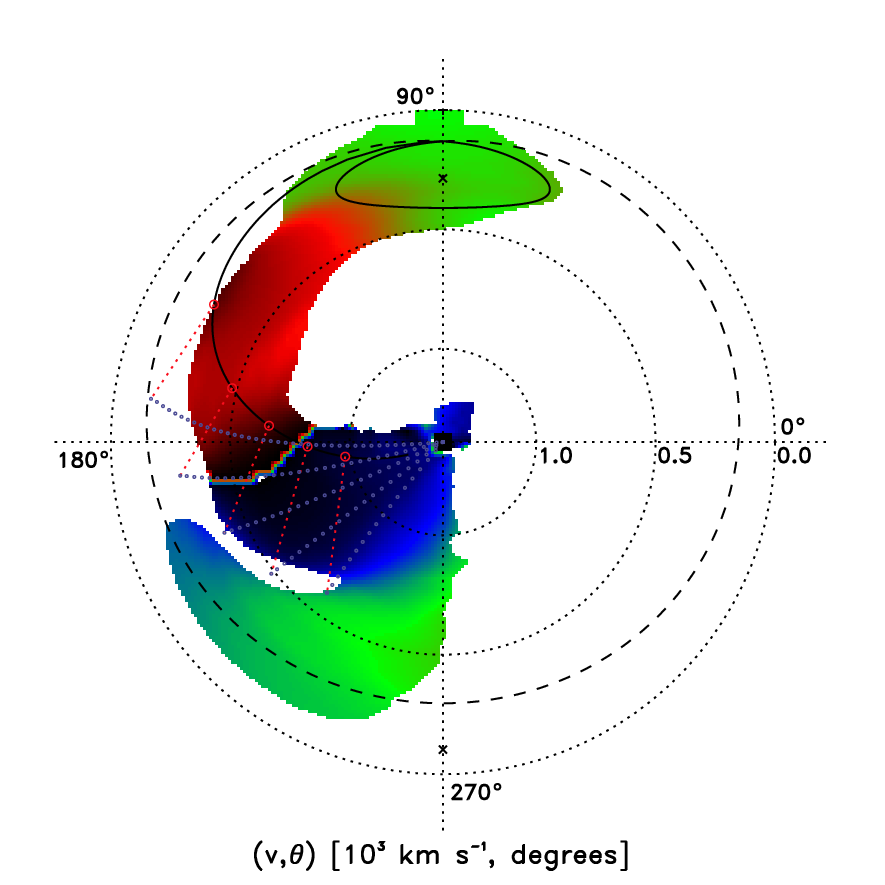}
 \hspace{-0.15cm}
\raisebox{-0.15cm}{\includegraphics[height=4.2cm]{Phase_colorbar_final.png}} \\
\end{array}$
\end{center}

\vspace{-0.15cm}

\begin{center}$
\begin{array}{ccc}
\includegraphics[width=0.15\textwidth,height=4.5cm]{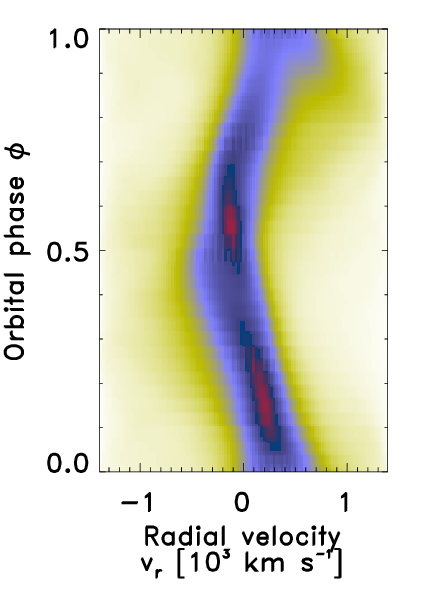}
 \hspace{-0.10cm}
\includegraphics[width=0.15\textwidth,height=4.5cm]{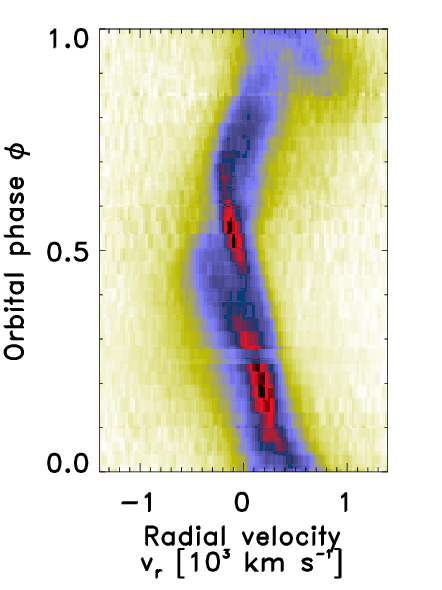}
 \hspace{-0.10cm}
\includegraphics[width=0.15\textwidth,height=4.5cm]{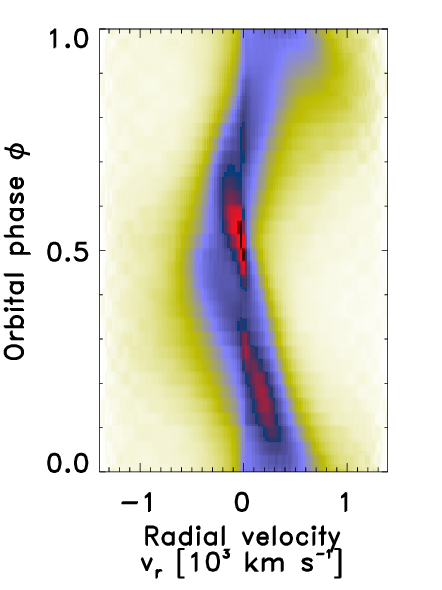}
\hspace{-0.15cm}
\raisebox{0.29cm}{\includegraphics[ height= 4.25cm]{Spectra_colorbar_final_a.png}} \\ \\
\end{array}$
\end{center}
\vspace{-0.15cm}
\caption{ As in Figure~\ref{fig:HeM}, but for the $\rm H\alpha$ emission line.}
\label{fig:HaM}
\end{figure}

We also present Doppler maps constructed using the flux modulation technique described in  technique \citet{2016A&A...595A..47K}. This method extends traditional Doppler tomography by decomposing the emission into three components: the average flux, the modulation amplitude, and the phase of maximum flux. It exploits the fact that key emitting regions such as the irradiated face of the secondary star, the ballistic stream, and the magnetically confined accretion flow undergo systematic variations in line-of-sight visibility, allowing their contributions to be more effectively isolated and visualised. Figures~\ref{fig:HeM} and \ref{fig:HaM} show the standard and inside-out flux-modulated Doppler maps of the He\textsc{ii}\,4686\,\AA\ and $\rm H\alpha$ emission lines, respectively. Compared to the traditional maps, the flux-modulated tomograms reveal more clearly defined emission structures. In both He\textsc{ii}\,4686\,\AA\ and the $\rm H\alpha$ maps, emission is associated with the irradiated face of the secondary star, the ballistic stream, the threading region, and the magnetically confined accretion stream. These components are particularly well resolved in He\textsc{ii}\,4686\,\AA\ , where the high-velocity magnetic stream and threading region stand out with superior brightness and clarity. In particular, both lines display a distinct splitting in the lower-left velocity quadrant—interpreted as separate emission from the threading region and the high-velocity section of the magnetic stream—though it appears substantially fainter in $\mathrm{H}\alpha$ than in He\textsc{ii}\,4686\,\AA\. While this feature is less prominent in $\mathrm{H}\alpha$, the $\mathrm{H}\alpha$ map clearly highlights the overall arc-like trajectory of both the ballistic and magnetically confined streams. Notably, this splitting is also present in the $\rm H\gamma$ and $\rm H\beta$ maps which are not shown in this paper, supporting the interpretation that it arises from the threading region and the high-velocity magnetic flow.

\label{Fluxmodulated}

\subsubsection{Phase of Maximum flux maps}

The middle rows of Figures~\ref{fig:HeM} and \ref{fig:HaM} present the standard (left) and inside-out (right) phase-of-maximum-flux maps for the He\textsc{ii}\,4686\,\AA\ and $\rm H\alpha$ emission lines. These maps show the orbital phase at which each emission component reaches its peak brightness. Only pixels with modulation amplitudes exceeding $10\%$ of the maximum are displayed. The colour scale represents orbital phase, with 0.0 in black, 0.25 in red, 0.5 in green, and 0.75 in blue. Similar maps have been reported for other polars, including HU Aqr and V834 Cen (\citealt{2016A&A...595A..47K}), UZ For (\citealt{2020MNRAS.492.4298K,2022PhDT........26K}), and EF Eri (\citealt{2025MNRAS.544..309K}).

In the He\textsc{ii}\,4686\,\AA\, maps, the irradiated face of the secondary star appears predominantly green, indicating maximum flux at $\phi \sim 0.5$, when the irradiated face of the secondary star is pointing towards the observer \citep{2016A&A...595A..47K}. The ballistic stream is mainly red, with some green contributions, suggesting peak visibility around $\phi \sim 0.25$ and $\phi \sim 0.5$. At $\phi \sim 0.25$, the inner stream near the L1 point is most directly visible, while at $\phi \sim 0.5$ it is viewed more side-on. The magnetically confined accretion stream exhibits a mix of black, green, and traces of blue, indicating visibility across $\phi \sim 0.5$, 0.75, and 1.0. These phases correspond to views of different stream regions, from the outer surface to the inner edge.

The $\rm H\alpha$ phase-of-maximum-flux maps show a similar behaviour for the irradiated secondary, with peak emission at $\phi \sim 0.5$. The ballistic stream is predominantly red, indicating a maximum near $\phi \sim 0.25$ with traces of black, suggesting emission extending to $\phi \sim 1.0$. The magnetically confined stream is largely characterised by black and green, with consistent blue contributions, indicating emission spanning phases $\sim 0.5$ to 1.0, including $\phi \sim 0.75$.

\subsection{Photopolarimetry}
\label{Photopol_results}

\begin{figure}
  \centering
  \includegraphics[width=1.0\linewidth]{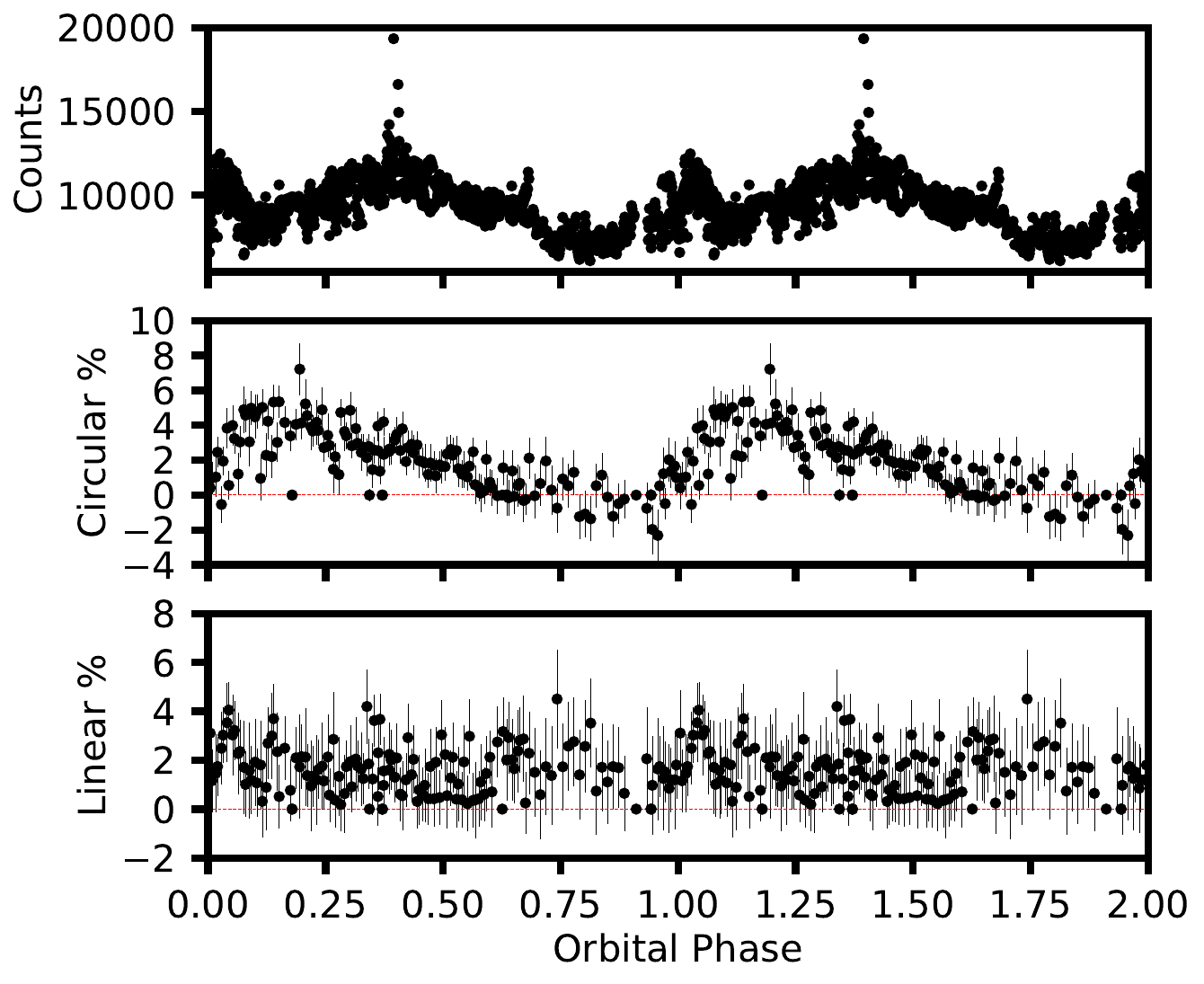}
  \caption{Phased photopolarimetric observations obtained with HIPPO instrument from 2017 June 24 to 2017 June 25. From top to bottom, the panels show the photometric flux, circular polarisation (\%), and linear polarisation (\%). The red dotted line denotes zero polarisation.
  }
  \label{fig:unfiltered}
\end{figure}

Figure~\ref{fig:unfiltered} shows the photopolarimetric observations, phased using the orbital ephemeris given in Equation~\ref{eq2}. The circular polarisation varies between approximately $-2\%$ and $+5\%$, peaking near orbital phase 0.25 and reaching a minimum around phase 0.8, in close agreement with the behaviour reported in the discovery paper by \citet{1993MNRAS.262...93B}. The linear polarisation exhibits only weak, seemingly random variability at $\sim 1\%$ level, although there is evidence for a brief linear polarisation pulse just after orbital phase $\sim 0.0$.

\subsection{Circular Spectropolarimetry}
\label{Spectropol_res}

\begin{figure*}
  \centering
  \includegraphics[width=0.75\linewidth]{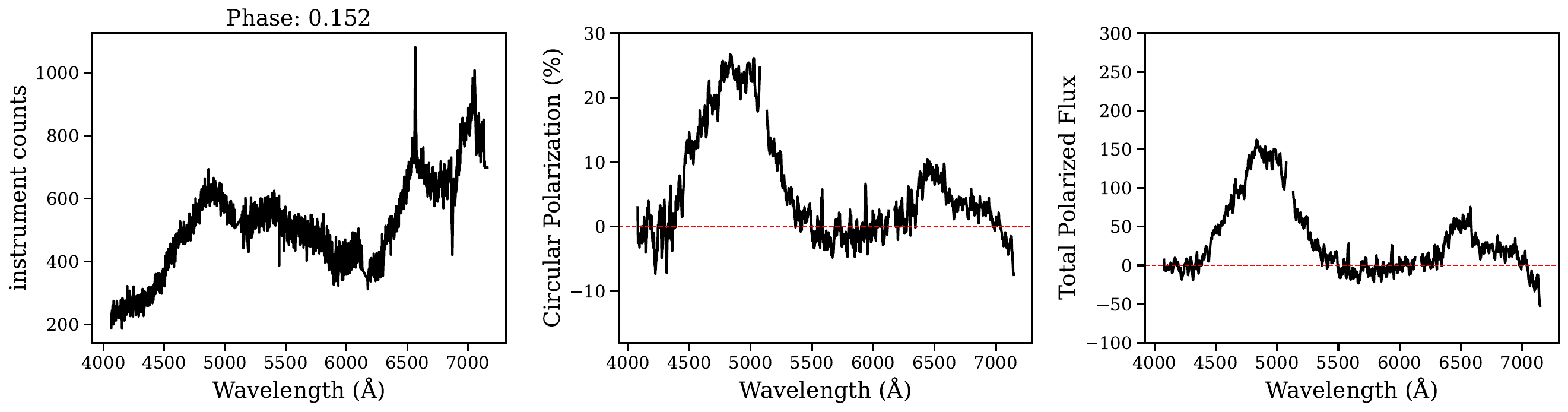}
  \includegraphics[width=0.75\linewidth]{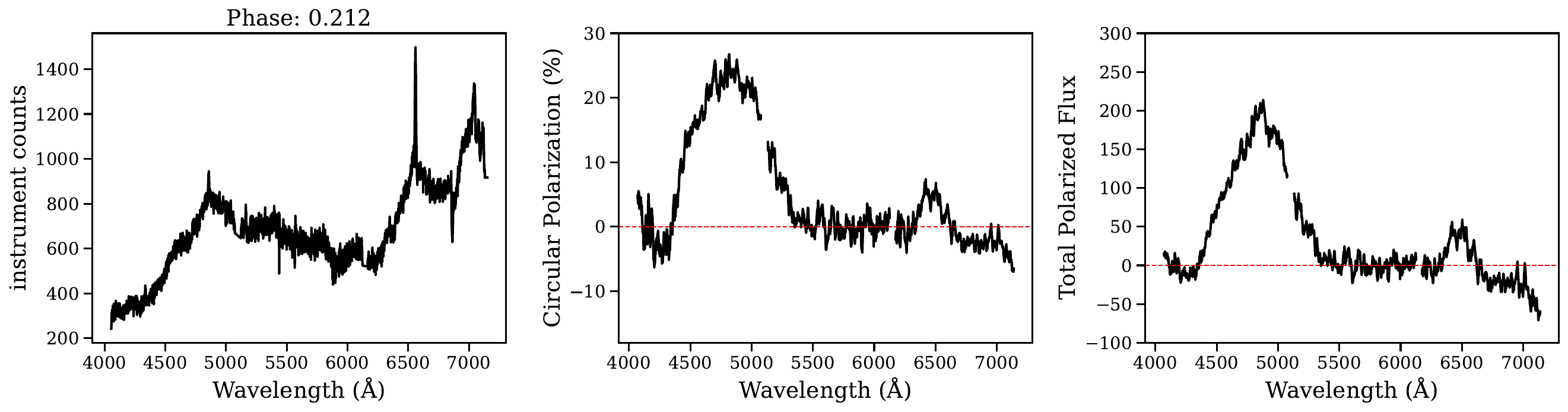}
  \includegraphics[width=0.75\linewidth]{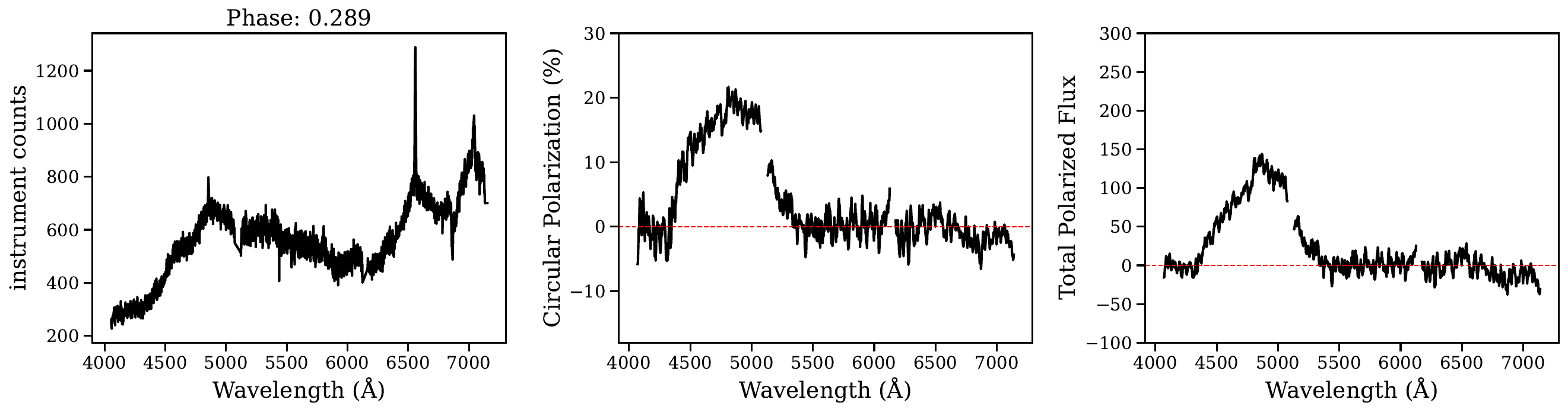}
  \includegraphics[width=0.75\linewidth]{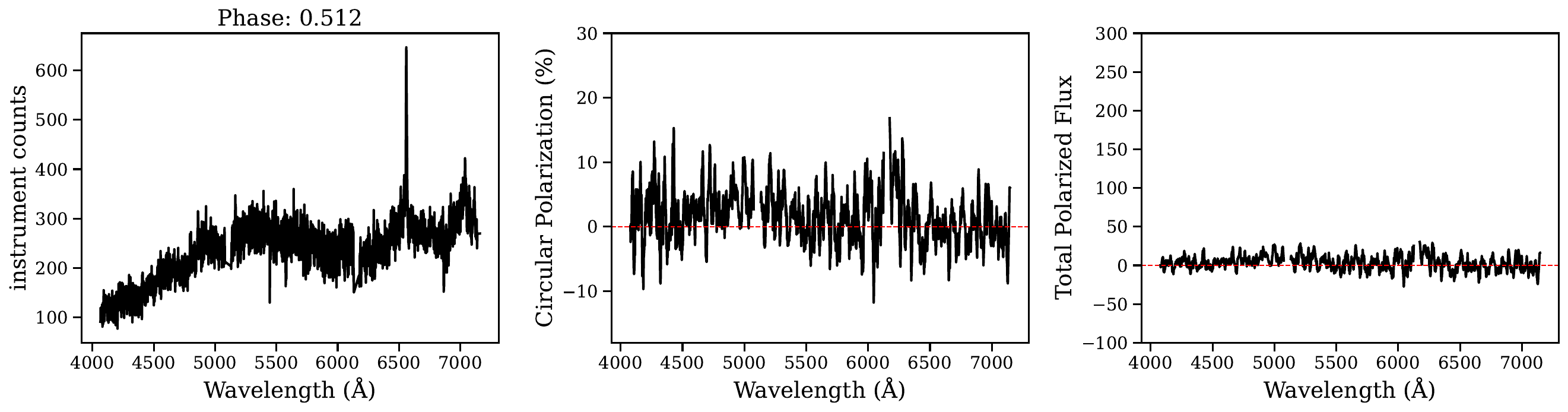}
  \includegraphics[width=0.75\linewidth]{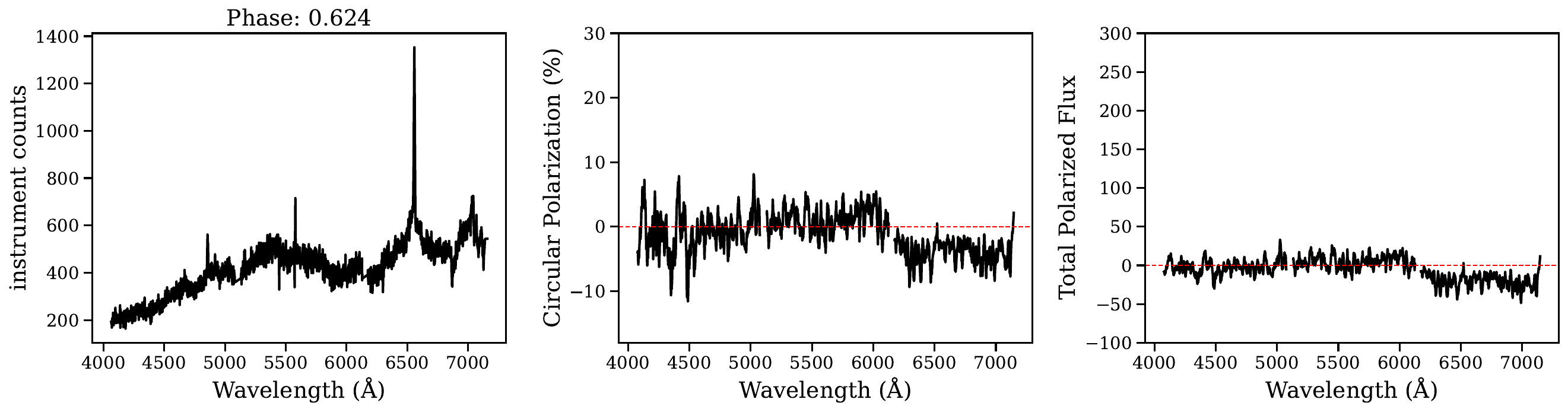}
  \caption{Spectropolarimetric data of QS Tel obtained during the low accretion state between 2025 May and August, folded over the orbital phase. The left panel shows the total flux, the middle panel displays the circular polarisation percentage, and the right panel presents the total intensity polarised flux}
  \label{fig:low_Spectropolarimetry}
\end{figure*}

\begin{figure*}
  \centering
  \includegraphics[width=0.75\linewidth]{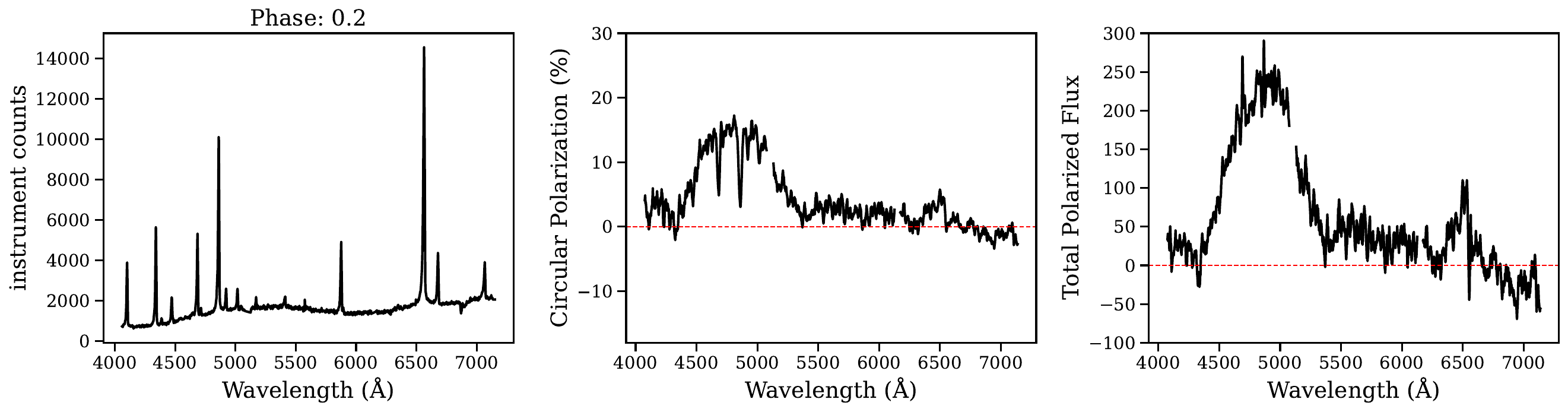}
  \includegraphics[width=0.75\linewidth]{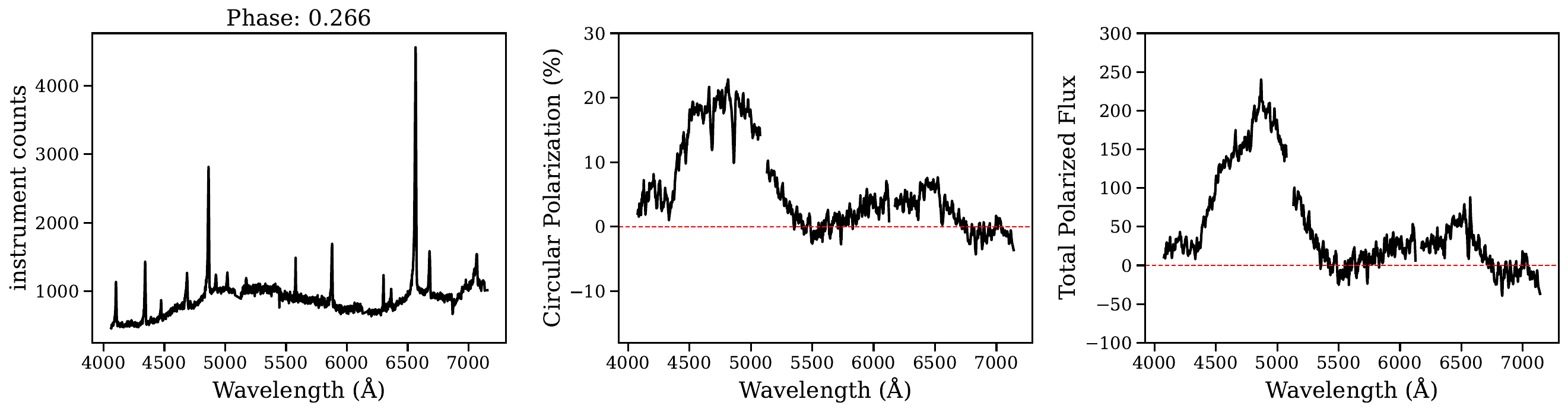}
  \includegraphics[width=0.75\linewidth]{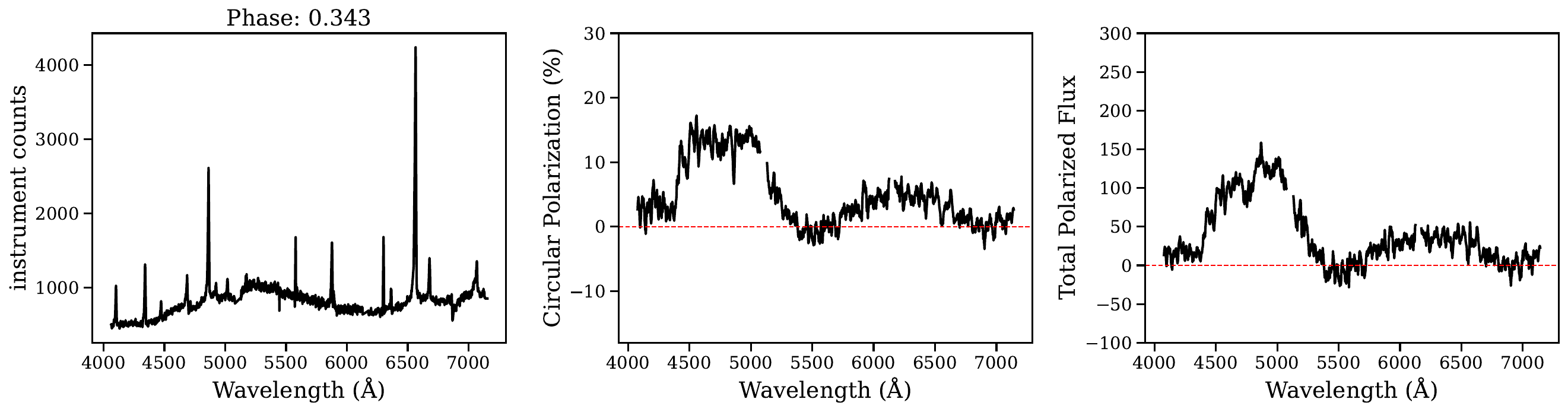}
  \includegraphics[width=0.75\linewidth]{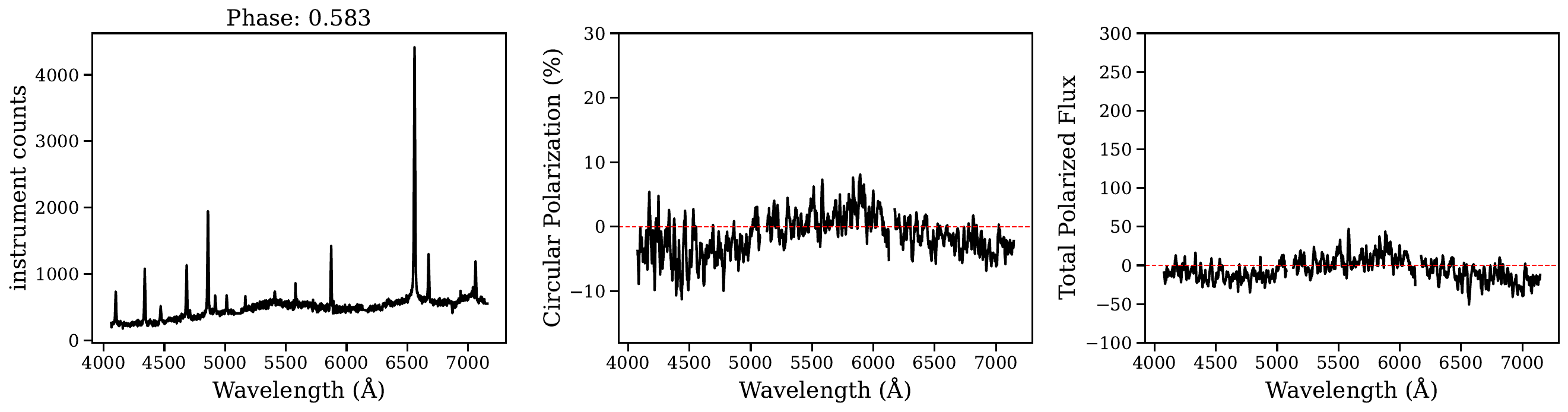}
  \includegraphics[width=0.75\linewidth]{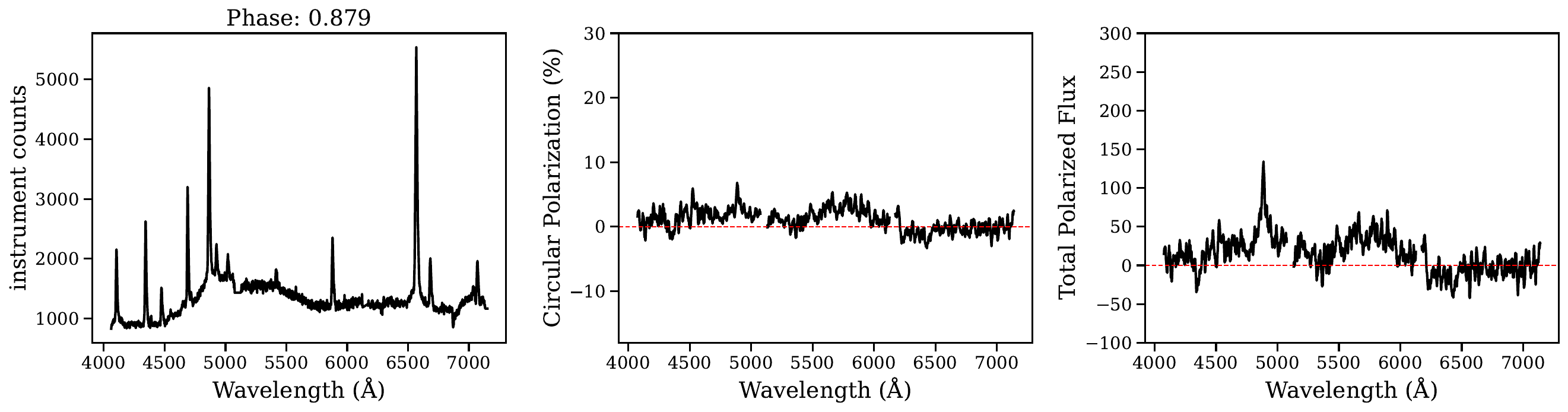}
  \includegraphics[width=0.75\linewidth]{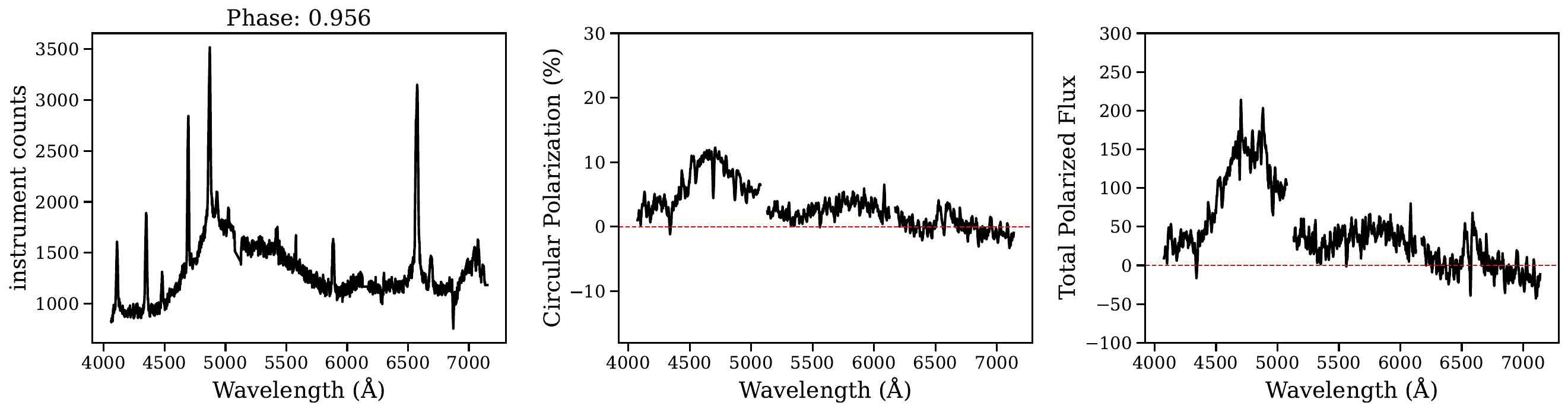}
  \caption{Spectropolarimetric data of QS Tel obtained during the intermediate accretion state between 2023 May and 2024 October, folded over the orbital phase. The left panel shows the spectroscopic flux, the middle panel displays the circular polarisation percentage, and the right panel presents the total polarised flux.}
  \label{fig:Spectropolarimetry}
\end{figure*}

Figures \ref{fig:low_Spectropolarimetry} and \ref{fig:Spectropolarimetry} show circular spectropolarimetry results of QS Tel, plotted as a function of orbital phase in ascending order. The left panels of these figures show the total instrument counts (total flux), the middle panels show the degree of circular polarisation and the right panels show the total polarised flux. The left panels of Figure~\ref{fig:low_Spectropolarimetry} display no prominent emission lines; the Balmer series is absent apart from weak $\mathrm{H}\alpha$ emission, and there is no trace of He\textsc{ii}\,$\lambda 4686$. This is consistent with QS~Tel being in a low accretion state during our observations. In contrast, the left panels of Figure~\ref{fig:Spectropolarimetry} show prominent Balmer and He\textsc{ii}\,4686\,\AA\, emission lines, together with weak He\textsc{i} features. The presence of strong high-excitation He\textsc{ii} emission and enhanced Balmer line strengths indicates that the system was in a higher accretion state compared to the low-state observations shown in Figure~\ref{fig:low_Spectropolarimetry}. 

The middle panels of Figures~\ref{fig:low_Spectropolarimetry} and~\ref{fig:Spectropolarimetry} show the percentage of circular polarisation, which varies between $-8\%$ and $+25\%$ over the covered orbital phases. The circular polarisation shows the presence of two phase-dependent cyclotron humps centred at $\approx$ 4800\,\AA\, and 6400\,\AA\,, respectively. The cyclotron hump at the bluer end of the spectrum is more dominant than the one in the red end of the spectrum. At certain orbital phases, for example, at phase 0.289 in Figure~\ref{fig:low_Spectropolarimetry}, the cyclotron hump on the blue side remains prominent, while the hump on the red side of the spectrum disappears. Polarity reversals is evident at phase 0.583 (see Figure~\ref{fig:Spectropolarimetry}), where the cyclotron hump on the blue side (4500–5000 \AA) of the spectrum exhibits negative circular polarisation, while the cyclotron emission in the 5000–6000 \AA\, remains positively polarised. Further evidence of polarity reversals is observed at phase 0.624 seen in Figure \ref{fig:low_Spectropolarimetry}: the circular polarisation is positive at bluer wavelengths, but becomes negative beyond 6000 \AA. The circular spectropolarimetry in Figure \ref{fig:Spectropolarimetry} shows depolarisation at the positions of the emission lines, as expected, since the emission lines are not intrinsically polarised. This behaviour is not observed in Figure \ref{fig:low_Spectropolarimetry}, where no prominent emission lines are present.

The right panels of Figures~\ref{fig:low_Spectropolarimetry} and~\ref{fig:Spectropolarimetry} show the total polarised flux. This was derived by multiplying the total flux spectra by the corresponding circular polarisation percentage. The key difference between the circular polarisation percentage and the total polarised flux is that the non polarised regions, which are visible in the circular polarisation percentage, are absent in the total polarised flux. Consequently, the polarised flux represents emission of pure cyclotron origin, largely free from any contribution by the secondary star or other unpolarised emission components \citep{1990SSRv...54..195C, 1989MNRAS.236...31M}. To estimate the magnetic field strength, we used the spacing between the cyclotron harmonics in the spectrum, which provides a direct measure of the field strength in the accretion region \citep{2005A&A...439..823K}. We fitted Gaussian profiles to the cyclotron humps at orbital phases 0.229, 0.266, and 0.343, as these phases display broad, well-defined harmonics suitable for reliable profile fitting. From these fits, we measured harmonic central wavelengths of 4800\,\text{\AA} and 6300\,\text{\AA}. Using the cyclotron fundamental spacing relation \citep[Eq.~1]{2005A&A...439..823K}, the magnetic field strength is given by:
\begin{equation}
    B = \frac{2\pi m_e c^2}{e} \left( \frac{1}{\lambda_{n+1}} - \frac{1}{\lambda_{n}} \right)
    \label{eq3}
\end{equation}
where $c$ is the speed of light, $m_e$ is the electron mass, $e$ is the elementary charge, and $\lambda_n$ is the central wavelength of the $n$-th harmonic. This relation yields an average magnetic field strength of $B \approx 53\,\text{MG}$. 

To determine the corresponding harmonic numbers, we applied the inclination-dependent cyclotron relation \citep[Eq.~1]{1989MNRAS.236...31M}:
\begin{equation}
    n = \frac{10710}{\lambda_n} \left( \frac{10^8\,\text{G}}{B} \right) \sin\theta
    \label{eq4}
\end{equation}
where $\lambda_n$ is the harmonic wavelength in Angstroms, $B$ is the magnetic field strength in Gauss, and $\theta$ is the viewing angle relative to the magnetic field vector. Assuming a viewing angle of $\theta = 90^\circ$, Equation~\ref{eq4} yields harmonic numbers of $n = 4$ (at 4800\,\text{\AA}) and $n = 3$ (at 6300\,\text{\AA}). These values agree with those reported by \citet{1995A&A...293..764S}, who identified harmonics $n = 3\text{--}7$ for a magnetic field strength of 47\,\text{MG}.

\subsubsection{Modelling of cyclotron humps}

\begin{figure}
    \centering
    \includegraphics[width=1.0\linewidth]{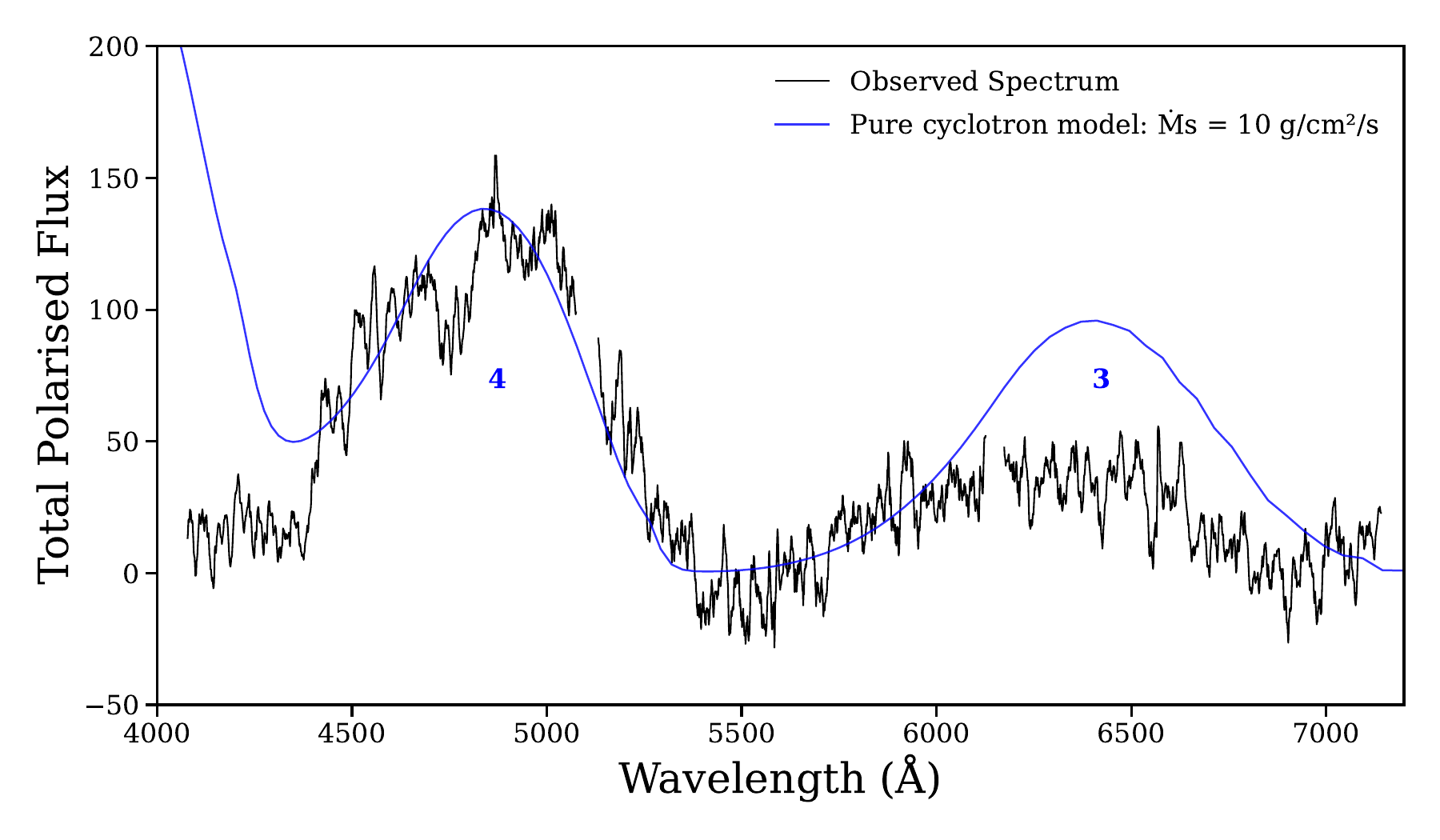}
    \caption{Total circularly polarised flux at orbital phase 0.343. Overlaid is a pure cyclotron model (blue) with a magnetic field strength of $\sim$53\,MG and a viewing angle of $73^{\circ}$, the mass transfer rate in the shock ($\dot{M}_s$) is 10~g\,cm$^{-2}$\,s$^{-1}$. The numbers 4 and 3 in blue mark the theoretical position of cyclotron harmonics for each the overlayed model.}
    \label{fig:modelling}
\end{figure}

To constrain the magnetic field strength of QS Tel and investigate the observed cyclotron features, we compared the total polarised flux with the pure cyclotron model of stratified accretion shock described in \citet{2000MNRAS.314..672P} and \citet{potter1998multiwavelenght}. The comparison was performed using a pure cyclotron model calculated for a mass-transfer rate of 10~g\,cm$^{-2}$\,s$^{-1}$ in the shock region. Figure~\ref{fig:modelling} presents the model overlaid on the total polarised flux at orbital phase 0.343, where the cyclotron humps are broad and well defined. The best-fitting overlay yields a magnetic field strength of 53\,MG and a viewing angle of $73^\circ$ between the line of sight and the magnetic field direction, with the model shown in blue.

\medskip
\noindent

The pure cyclotron model showed two cyclotron humps centred at 4850\,\AA\, and 6400\,\AA\, with harmonic numbers $n=4$ and $n=3$ respectively. While the model provides a reasonable match to the observed features, it slightly overestimates the amplitude of the 6400\,\AA\ hump. To place these results in context, we used Equation~\ref{eq4} to calculate the magnetic field strength expected for a viewing angle of $\theta = 73^\circ$, consistent with the modelling above. This yields an average field strength of 53\,MG for the pure cyclotron model, in excellent agreement with the value obtained from the overlay. This consistency suggests that the pure cyclotron model provides a physically plausible description of the cyclotron emission in QS~Tel, and that the inferred magnetic field strength is robust.

\section{Discussion}\label{section:discussion}

\subsection{Photometric Light Curves}\label{section:discussion_phot}

The ATLAS long-term photometric light curve of QS~Tel, spanning 2017–2025, reveals transitions on monthly to yearly timescales between three distinct brightness levels corresponding to high, intermediate, and low accretion states. The mean magnitudes of these high, intermediate, and low states (shown in Figure~\ref{fig:ATLAS}), derived using the GMM model, are consistent with previous values reported in the literature \citep[see e.g.,][]{1993MNRAS.262...93B, 1995A&A...293..764S, 10.1093/mnras/280.4.1121, 2001MNRAS.322..631R}. The high-cadence SHOC observations reveal short-term, irregular brightness fluctuations in QS Tel on minute timescales, consistent with QPOs \citep{1982ApJ...257L..71M}.We detected QPOs occurring on timescales of $\sim$2--8~minutes across two observing nights separated by two weeks. In mCVs, such minute-timescale QPOs are commonly interpreted as signatures of instabilities in the threading region, where the ballistic accretion stream couples to the WD’s magnetic field \citep{1981ApJ...245..618P}. The presence of these QPOs supports an accretion scenario that is intrinsically inhomogeneous, characterised by intermittent, structured accretion and regulated by magneto-hydrodynamic processes \citep{1987MNRAS.226..867W}. Minute-timescale QPOs have been detected in systems such as IGR~J14536$-$5522 which exhibits optical QPOs on characteristic timescales of 5--6~min \citep{2010MNRAS.402.1161P}.
In contrast, a number of polars exhibit much faster optical QPOs on timescales of seconds, including systems such as AN~UMa, V834~Cen, EF~Eri, BL~Hyi, VV~Pup and CTCV J1928–5001 \citep[e.g.,][]{1982ApJ...257L..71M,1987A&A...181L..15L,2015gacv.workE..18B,2017A&A...600A..53M,2020A&A...635A.204B,breytenbach2026study}. In these systems, the oscillations are generally interpreted as arising from instabilities in the post-shock accretion column near the WD surface. The detection of minute-scale QPOs in QS~Tel constitutes the first report of QPO behaviour in this system and places it among only a handful of polars (of order $\leq$10 systems) known to exhibit QPO variability, highlighting the rarity of this phenomenon. This suggests that the variability is dominated by processes occurring far from the WD surface, consistent with an origin in the threading region rather than the shock region. This makes QS Tel one of the few polars that shows QPOs with longer periods.

\medskip
\noindent

The orbital modulation observed in the TESS folded light curves shown in Figure \ref{fig:folded} is characterised by a double-humped profile and a pronounced minimum near orbital phase 0.9–1.0. Both of these features have previously been reported for QS Tel by different authors \citep[see][]{1993MNRAS.262...93B, 2001MNRAS.322..631R, 10.1093/mnras/280.4.1121}. The double-humped morphology is typical of high-field polars in a high and intermediate accretion states, in which cyclotron emission dominates the optical flux. The unequal amplitudes of the humps suggest asymmetric accretion, possibly one pole accreting more efficiently, or having a more favourable viewing geometry \citep{1990SSRv...54..195C}. The persistent orbital minimum is most plausibly caused by self-occultation of the primary accretion region as it rotates behind the white dwarf limb. Similar orbital morphologies have been observed in other mCV systems during high accretion states, including AM Her \citep{2022MNRAS.516.3144D,2020A&A...642A.134S, 2001A&A...372..557G}. 
In these systems, the optical high-state emission can be modelled with a cyclotron-emitting region that becomes progressively obscured as it rotates out of view. However, partial obscuration by the secondary star cannot be excluded \citep{1990SSRv...54..195C}. As shown in the left and middle panels of Figure~\ref{fig:folded}, a change in the shape of the folded light curve is observed within the intermediate accretion state. The amplitude of the first peak decreases, leading to a more pronounced second peak, suggesting a redistribution of accretion between the two magnetic poles as mass-transfer rate decreases slightly but remains within the intermediate-state regime \citep{1990SSRv...54..195C,1995A&A...293..764S,2001cvs..book.....H}. Consequently, the orbital minimum near phases 0.9--1.0 becomes shallower, consistent with a reduced dominance of the primary accretion region rather than a transition to a low accretion state.

\medskip
\noindent 

\subsection{Spectroscopy and Doppler Tomography}

We present phase-resolved optical spectroscopy of QS Tel obtained between 2022 September 21 and 23. The average spectra display broad and strong Balmer and He\textsc{ii}\,4686\,\AA\, emission lines. The prominence of the He\textsc{ii}\,4686\,\AA\, and its comparable intensity to the $\rm H\beta$ emission line, indicates a highly ionised accretion environment and confirms that the system was observed in a high accretion state. This is in agreement with our ATLAS $o$-band photometry seen in Figure \ref{fig:ATLAS}, which also placed QS Tel in high state during the period of our observations. Similar high-state optical spectra of QS~Tel, characterized by prominent Balmer and He\textsc{ii},$\lambda 4686$ emission alongside weaker He\textsc{i} lines, have been reported previously in the literature \citep[e.g.,][]{1993MNRAS.262...93B, 1995A&A...293..764S, 2001MNRAS.322..631R}. Comparable high accretion state spectra, exhibiting strong and broad Balmer and He\textsc{ii}\,4686\,\AA\ emission lines, have also been observed in other polars such as HU~Aqr, RX J1846.9+5538, BY Cam, SRGA J213151.5+491400 and EF Eri \citep{1997A&A...319..894S,2002A&A...392..505S,2005A&A...444..213S,2024A&A...684A.190B,2025MNRAS.544..309K}. 
 
We also presented the first detailed Doppler maps of QS Tel using both the standard and flux-modulated techniques, together with the corresponding observed and reconstructed trailed spectra of the He\textsc{ii}\,4686\,\AA\, and $\rm H\alpha$ emission lines. The input trailed spectra of QS~Tel reported here closely resemble those presented by \citet{1993MNRAS.262...93B} and \citet{1995A&A...293..764S}, showing similar features, including a high-velocity redshifted wing extending over orbital phases $\phi \sim 0.9$–1.1. Both the standard and inside-out Doppler maps of QS~Tel reveal multiple emission components reflecting the complex accretion geometry typical of high-field polars. 
Specifically, the tomograms of He\textsc{ii}\,4686\,\AA\, and the $\rm H\alpha$ emission lines show contributions from the irradiated face of the secondary star, the ballistic stream, the threading region, and the magnetically confined accretion stream. These results highlight spatial and kinematic differences between the emission lines, reflecting their distinct excitation and ionisation conditions. 

For the He\textsc{ii}\,4686\,\AA\, line, the dominant emission components are the ballistic stream, the threading region, and the magnetically confined accretion stream. This is particularly evident in the inside-out projection, which redistributes velocity-space features that appear compacted in the standard projection \citep{2016A&A...595A..47K}. Similar high-state Doppler tomograms based on the He\textsc{ii}\,4686\,\AA\, emission line have been presented in the literature for other polars, e.g. V834 Cen \citep{2004MNRAS.348..316P,2016A&A...595A..47K}, UZ For \citep{2020MNRAS.492.4298K}, and SRGA J213151.5+491400 \citep{2024A&A...684A.190B}. For the $\rm H\alpha$ line, the dominant emission components are the ballistic stream, and the magnetically confined accretion stream. The ballistic stream dominates most of the emission in the standard projection, whereas in the inside-out projection the high-velocity portion of the magnetically confined stream becomes the most prominent component. Similar high-state Doppler maps based on the $\rm H\alpha$ line have been presented by \citet{2025MNRAS.544..309K} for EF Eri in a high accretion state.

The flux-modulated maps of He\textsc{ii}\,4686\,\AA\, and $\rm H\alpha$ lines reveal contributions from the irradiated face of the secondary star, the ballistic stream, and the magnetically confined accretion stream. This behaviour is expected for a high-inclination system, where flux from different emitting components is modulated with orbital phase due to changing viewing angles. The flux-modulated maps of the He\textsc{ii}\,4686\,\AA\, line show modulation from all three emission components; however, the dominant modulated component depends on the projection used. In the standard projection, the irradiated face of the secondary star and the ballistic stream appear most strongly modulated, while the magnetically confined stream shows weaker modulation. In contrast, the inside-out projection shows that the magnetically confined stream exhibits the strongest modulation, followed by the ballistic stream. According to \citet{2016A&A...595A..47K}, this behaviour reflects the redistribution of flux contrast in velocity space produced by the inside-out projection, which suppresses extended low-contrast structures relative to strongly modulated components. Similarly, the $\rm H\alpha$ flux-modulated maps show that the most strongly modulated component corresponds to the magnetically confined stream. The ballistic stream appears faint or absent in these maps, while the irradiated face of the secondary star represents the second most strongly modulated component.

As mentioned in Section \ref{Fluxmodulated}, both the standard and inside-out flux-modulated tomograms shown in Figures \ref{fig:HeM} and \ref{fig:HaM}, exhibit evidence for the splitting of the magnetically confined accretion stream into two distinct components. The same behaviour is observed in the flux-modulated maps of the $\rm H\gamma$ and $\rm H\beta$ emission lines (not shown here), confirming that these components are real. Further support is provided by the phase-of-maximum-flux maps shown in the bottom middle panels of the same figures, where different colours highlight distinct emission components. The presence of these components allows two possible interpretations. One possibility is that QS~Tel accretes onto both magnetic poles of the WD, consistent with earlier studies showing that the system can switch between one-pole and two-pole accretion during high accretion states \citep{1995A&A...293..764S, 2001MNRAS.322..631R}. Alternatively, the two components may originate from different regions within a single accretion stream. In this scenario, the high-velocity component represents a direct magnetically confined flow close to the WD surface, while the lower-velocity component corresponds to a more gradually accelerated, magnetically guided flow, possibly associated with the threading region where material couples to the magnetic field lines. This behaviour may be influenced by the system inclination and magnetic field configuration, which can produce projection effects and partial visibility of different stream segments at different orbital phases \citep{1990SSRv...54..195C}.

Our results favour the latter interpretation, as there is no clear evidence for a second independent magnetically confined stream. The emission therefore likely arises from physically distinct regions along a single accretion flow, with differences in velocity, temperature, density, and modulation strength reflecting changing physical conditions along the accretion column. This interpretation is consistent with the observed continuity of the stream in velocity space and with expectations for flux-modulated emission originating from regions with significantly different physical properties along the flow. Such splitting of the magnetically confined accretion stream is not commonly observed in other high-field polars. Systems such as HU Aqr and V834 Cen, studied by \citet{2016A&A...595A..47K}, as well as SRGA J213151.5+491400 studied by \citet{2024A&A...684A.190B}, do not show evidence for two distinct components in their flux-modulated maps; instead, most emission arises from the ballistic and magnetically confined streams. However, recent observations of EF Eri in a high accretion state presented by \citet{2025MNRAS.544..309K} show a splitting of the magnetically confined accretion stream into two components, which was interpreted as evidence for two-pole accretion (see their Figures 6 and 7). QS~Tel may therefore represent only the second known non-eclipsing polar exhibiting this behaviour.

\subsection{Photopolarimetry}
\label{Disc_Phtotopolorimetry}
The photopolarimetric results shown in Figure~\ref{fig:unfiltered} reveal clear phase-dependent modulation in the circular polarisation, consistent with cyclotron emission originating from one or more magnetic accretion regions on the surface of the WD. The circular polarisation reaches a maximum of about $ +8\%$ near orbital phase 0.25 and a minimum of approximately $-2\%$ around phase 0.8, indicating that the dominant accretion region becomes most visible at phase 0.25 and is either self-eclipsed or viewed at a less favourable angle near phase 0.8. Such behaviour is typical of polars, where the observed polarisation is governed by the geometry of the magnetic field and the changing line of sight to the accretion column \citep{1995cvs..book.....W}. The weakly modulated linear polarisation (ranging between $0$–$5\%$) suggests that the linearly polarised component either arises from a more extended emission region or is significantly affected by depolarising processes such as Faraday rotation or scattering within the accretion flow \citep{2001cvs..book.....H}. The low level of linear polarisation further implies that the cyclotron emission is predominantly circularly polarised, consistent with an optically thick emission regime. These results are consistent with earlier studies of QS Tel presented by \citet{1993MNRAS.262...93B}. These authors also reported strong phase-dependent circular polarisation variations with minimal changes in linear polarisation. Overall, the observed polarimetric behaviour supports a model in which a single dominant magnetic pole contributes most of the cyclotron emission, with its visibility modulated by the orbital geometry. This is consistent with the system being in a low state, as indicated by the ATLAS $o$-band long-term photometry shown in Figure~\ref{fig:ATLAS}, and is in agreement with previous studies \citep{2001MNRAS.322..631R,1993MNRAS.262...93B,1995A&A...293..764S}.

\subsection{Circular Spectropolarimetry}
Our circular spectropolarimetry observations of QS~Tel reveals two distinct accretion states, reflected in the intensity differences between the left panels of Figures \ref{fig:low_Spectropolarimetry} and \ref{fig:Spectropolarimetry}. The spectra obtained between 2025 May and August (Figure~\ref{fig:low_Spectropolarimetry}) correspond to a low accretion state. This is evidenced by the absence of the Balmer emission lines and He\textsc{ii}\,4686\,\AA\, emission, with the exception of $\rm H \alpha$. Such spectral characteristics reflect a significant reduction in the mass-transfer rate from the donor star, as expected for low-accretion states in mCVs \citep{1990SSRv...54..195C}. The ATLAS $o$-band photometry (Figure~\ref{fig:ATLAS}) further confirms that the system remained in a low state during this period. In contrast, the spectra presented in Figure~\ref{fig:Spectropolarimetry} show prominent Balmer emission lines and He\textsc{ii}\,4686\,\AA\ emission, indicating an increase in the mass-transfer rate from the donor star. The relatively low He\textsc{ii}/H$\beta$ ratio suggests that the ionising flux was lower than expected for a high accretion state (see Figure~\ref{fig:QS_tel_spectrums}, Section~\ref{Res_Spectroscopy}), implying that the accretion rate was insufficient to produce the high-energy photons typical of high states. This places the spectra in an intermediate accretion state, consistent with the ATLAS $o$-band photometry (Figure~\ref{fig:ATLAS}). Such behaviour is characteristic of transitional states in mCVs \citep{1990SSRv...54..195C}. These states are likely driven by short-term variations in mass transfer or magnetic coupling at the L1 point.

The circular polarisation varies between $-10\%$ and $+25\%$, displaying significant phase-dependent variations. These include polarity reversals around phases 0.583 and 0.624. Such polarity reversals are commonly interpreted as arising from two accreting regions with opposite magnetic polarities that alternate over the orbital cycle. The stronger cyclotron hump in the bluer wavelengths which is visible across most phases, suggests that one accretion region dominates the observed emission. This behaviour is consistent with previous studies of QS Tel and other systems that show two-pole accretion. In these systems, the cyclotron emission geometry is governed by the changing viewing angle of the magnetic field lines during the orbit \citep{1993MNRAS.265..605F,1995A&A...293..764S}. The similarity of the polarisation behaviour to that observed in systems such as AR UMa \citep{2003MNRAS.338..340F} and VV Pup \citep{2008A&A...490..279M} further supports the presence of a strong magnetic field. Because HIPPO is most sensitive toward the red optical wavelengths, the broadband photopolarimetric measurements are expected to be weighted toward emission from the redder, lower-order cyclotron harmonics. As a result, the strong circular polarisation observed in the HIPPO data is consistent with cyclotron emission dominated by the red cyclotron humps identified in the spectropolarimetric observations (Figures~\ref{fig:low_Spectropolarimetry} and~\ref{fig:Spectropolarimetry}). The similarity in both the amplitude and phase dependence of the circular polarisation between the broadband photopolarimetry and the spectropolarimetric data therefore supports a common cyclotron origin for the polarised emission \citep{1990SSRv...54..195C,2010MNRAS.402.1161P}.

The total circularly polarised flux spectra show smooth cyclotron humps with no significant contribution from unpolarised components. This indicates that the observed polarised flux is dominated by cyclotron radiation rather than by unpolarised emission from other components in the binary system. Our results therefore support an origin of the polarised emission in the magnetically confined post-shock accretion plasma, rather than in unpolarised components such as the ballistic accretion stream or the secondary star \citep{1989MNRAS.236...31M,1990SSRv...54..195C}. The two phase-dependent cyclotron features identified in our spectra correspond to the $n = 4$ and $n = 3$ harmonics reported by \citet{1995A&A...293..764S} in the top panel of their Figure~11. Their modulation with orbital phase provides further evidence of varying optical depth and viewing angle geometry within the accretion column. Our derived average magnetic field strength of 53 MG and the corresponding harmonic numbers for the phase-dependent cyclotron features demonstrate excellent agreement between the overlaid model and theoretical calculations. This magnetic field strength is in good agreement with the 50--60\,MG range reported for QS~Tel for the pole that is dominant in the optical region \citep[see][]{1995A&A...293..764S,10.1093/mnras/280.4.1121,2001MNRAS.322..631R}. The viewing angle from the overlaid model is consistent with expectations for systems in which parts of the accretion stream remain visible throughout the orbital cycle \citep[e.g.][]{1995A&A...293..764S,10.1093/mnras/280.4.1121}. Notably, \citet{2005ASPC..330..429R} adopted a similar inclination of 73$^\circ$ for their Doppler tomography modelling of QS~Tel and we used the same inclination angle for our Doppler tomograms seen in section \ref{section:doppler_tomography}. It should be noted that only two cyclotron harmonics are observed in the polarised flux spectra. The limited wavelength coverage restricts our ability to access the lower-order harmonics, which contain information about the optical depth and electron temperature \citep[e.g.,][]{1989MNRAS.236...31M,1998MNRAS.301..767S,2001A&A...373..211F}, and therefore introduces some uncertainty in these parameters. Determination of the temperature, optical depth, and other related quantities would require observations of the lower-order cyclotron harmonics, which are not available in the current data.

\section{Conclusion}\label{section:summary}

We have presented a comprehensive multi-epoch, multi-wavelength study of QS~Tel, combining photometry, spectroscopy and polarimetry to investigate its accretion geometry and variability. Long-term ATLAS photometry confirms that QS~Tel transitions between low, intermediate, and high accretion states on timescales of months to years. High-speed SHOC and TESS photometry obtained during an intermediate state reveal a double-humped orbital modulation with a deep minimum near phase 0.9–1.0, consistent with a grazing self-eclipse of the primary accretion region or partial obscuration by the secondary star. We report the first detection of short-timescale QPOs in QS~Tel, revealing previously unobserved rapid variability within the accretion flow, likely associated with the threading region.

Our phase-resolved spectroscopic observations obtained during a high accretion state shows strong Balmer and He\textsc{ii}\,4686\,\AA\, emission, indicative of a highly ionised accretion environment. We also present the first detailed Doppler tomography study of QS~Tel, using both standard and flux-modulated techniques. These reveal emission from the irradiated face of the secondary star, the ballistic stream, and the magnetically confined accretion flow. 
Our flux-modulated maps reveal two distinct components within the magnetically confined accretion stream. We interpret these components as arising from different regions of a single accretion flow, characterised by distinct velocities and physical conditions, rather than as unambiguous evidence for two independent accretion poles.

Our SALT circular spectropolarimetry observations reveal two prominent phase-dependent cyclotron humps centred near 4800\,\AA\ and 6400\,\AA. Modelling of these features yields a magnetic field strength of ~53 MG and viewing angle of ~73$^\circ$, consistent with cyclotron emission originating in strongly magnetically confined regions near the WD's surface. The observed polarisation behaviour supports a geometry in which one dominant pole contributes most of the optical cyclotron emission, with possible contributions from a second pole at certain phases.

Overall, QS~Tel exhibits a complex and state-dependent accretion structure governed by the interplay between mass-transfer rate, magnetic confinement, and viewing geometry. This work provides the first detection of QPOs in the system. It also presents the first detailed Doppler tomographic analysis, offering new constraints on the large-scale accretion geometry. These results highlight the importance of coordinated, multi-wavelength observations in advancing our understanding of magnetic accretion in polars.
 
\section*{Acknowledgements}
This paper uses spectroscopic and photometric observations taken using the South African Astronomical Observatory (SAAO) facilities\footnote{\url{https://www.saao.ac.za/explore/our-telescopes/}}, the SAAO 1.0-m and 1.9-m telescopes. Spectropolarimetry observations were taken using the Southern African Large Telescope (SALT) under programs 2023-1-SCI-027, 2024-1-SCI-011, and 2025-1-SCI-016 (PI: Z.~N. Khangale). This paper also includes data from the Transiting Exoplanet Survey Satellite (TESS)\footnote{\citep{2015JATIS...1a4003R}}, which are publicly accessible via the Mikulski Archive for Space Telescopes (MAST). The financial assistance of the South African Radio Astronomy Observatory (SARAO) towards this research is hereby acknowledged (\url{www.sarao.ac.za}). We acknowledge the use of data from the Asteroid Terrestrial-impact Last Alert System\footnote{data available at:\url{https://fallingstar-data.com/forcedphot/queue/}} (ATLAS), a project funded by NASA and operated by the University of Hawaii Institute for Astronomy.

\section{Data Availability}

The TESS data used in this study are publicly available through the Mikulski Archive for Space Telescopes (MAST)\footnote{https://mast.stsci.edu/portal/Mashup/Clients/Mast/Portal.html}. The ATLAS dataset is also publicly available online through the ATLAS Forced Photometry service\footnote{https://fallingstar-data.com/forcedphot/queue/}. The SAAO data used in this project are available from the authors upon reasonable request.

\bibliographystyle{mnras}
\bibliography{Reference} 

@mscthesis{breytenbach2026study,
  author       = {Breytenbach, Hannes},
  title        = {{A study of quasi-periodic oscillations in magnetic cataclysmic variable stars}},
  school       = {University of Cape Town},
  year         = {2026},
  type         = {{masters} thesis},
  url          = {http://hdl.handle.net/11427/43323}
}

@inproceedings{10.1117/12.460312,
author = {Eric B. Burgh and Kenneth H. Nordsieck and Henry A. Kobulnicky and Ted B. Williams and Darragh O'Donoghue and Michael P. Smith and Jeffrey W. Percival},
title = {{Prime focus imaging spectrograph for the Southern African Large Telescope: optical design}},
volume = {4841},
booktitle = {Instrument Design and Performance for Optical/Infrared Ground-based Telescopes},
editor = {Masanori Iye and Alan F. M. Moorwood},
organization = {International Society for Optics and Photonics},
publisher = {SPIE},
series    = {{Society of Photo-Optical Instrumentation Engineers (SPIE) Conference Series}},
pages = {1463 -- 1471},
year = {2003},
doi = {10.1117/12.460312},
URL = {https://doi.org/10.1117/12.460312}
}

@inproceedings{10.1117/12.460315,
author = {Henry A. Kobulnicky and Kenneth H. Nordsieck and Eric B. Burgh and Michael P. Smith and Jeffrey W. Percival and Ted B. Williams and Darragh O'Donoghue},
title = {{Prime Focus Imaging Spectrograph for the Southern African Large Telescope: operational modes}},
volume = {4841},
booktitle = {Instrument Design and Performance for Optical/Infrared Ground-based Telescopes},
editor = {Masanori Iye and Alan F. M. Moorwood},
organization = {International Society for Optics and Photonics},
publisher = {SPIE},
series    = {{Society of Photo-Optical Instrumentation Engineers (SPIE) Conference Series}},
pages = {1634 -- 1644},
year = {2003},
doi = {10.1117/12.460315},
URL = {https://doi.org/10.1117/12.460315}
}

@ARTICLE{1981ApJ...245..618P,
       author = {{Patterson}, J. and {Williams}, G. and {Hiltner}, W.~A.},
        title = "{The amazing X-ray light curve of 2A 0311-22.7.}",
      journal = {\apj},
         year = 1981,
        month = apr,
       volume = {245},
        pages = {618-623},
          doi = {10.1086/158837},
       adsurl = {https://ui.adsabs.harvard.edu/abs/1981ApJ...245..618P}
}

@ARTICLE{1987MNRAS.226..867W,
       author = {{Watson}, M.~G. and {King}, A.~R. and {Williams}, G.~A.},
        title = "{X-ray and optical flickering in EF Eri.}",
      journal = {\mnras},
         year = 1987,
        month = jun,
       volume = {226},
        pages = {867-878},
          doi = {10.1093/mnras/226.4.867},
       adsurl = {https://ui.adsabs.harvard.edu/abs/1987MNRAS.226..867W}
}

@ARTICLE{2010MNRAS.402.1161P,
       author = {{Potter}, Stephen B. and {Buckley}, David A.~H. and {O'Donoghue}, Darragh and {Romero-Colmenero}, Encarni and {O'Connor}, James and {Fourie}, Piet and {Evans}, Geoff and {Sass}, Craig and {Crause}, Lisa and {Still}, Martin and {Butters}, O.~W. and {Norton}, A.~J. and {Mukai}, Koji},
        title = "{Polarized QPOs from the INTEGRAL polar IGRJ14536-5522 (=Swift J1453.4-5524)}",
      journal = {\mnras},
         year = 2010,
        month = feb,
       volume = {402},
       number = {2},
        pages = {1161-1170},
          doi = {10.1111/j.1365-2966.2009.15944.x},
archivePrefix = {arXiv},
       eprint = {0911.1198},
 primaryClass = {astro-ph.SR},
       adsurl = {https://ui.adsabs.harvard.edu/abs/2010MNRAS.402.1161P}
}

@ARTICLE{2001A&A...373..211F,
       author = {{Fischer}, A. and {Beuermann}, K.},
        title = "{Accretion physics of AM Herculis binaries. I. Results from one-dimensional stationary radiation hydrodynamics}",
      journal = {\aap},
         year = 2001,
        month = jul,
       volume = {373},
        pages = {211-221},
          doi = {10.1051/0004-6361:20010600},
archivePrefix = {arXiv},
       eprint = {astro-ph/0105190},
 primaryClass = {astro-ph},
       adsurl = {https://ui.adsabs.harvard.edu/abs/2001A&A...373..211F}
}

@ARTICLE{2005A&A...444..213S,
       author = {{Schwarz}, R. and {Schwope}, A.~D. and {Staude}, A. and {Remillard}, R.~A.},
        title = "{Doppler tomography of the asynchronous polar BY Camelopardalis}",
      journal = {\aap},
         year = 2005,
        month = dec,
       volume = {444},
       number = {1},
        pages = {213-220},
          doi = {10.1051/0004-6361:20053711},
       adsurl = {https://ui.adsabs.harvard.edu/abs/2005A&A...444..213S}
}

@ARTICLE{10.1093/mnras/280.4.1121,
       author = {{Rosen}, S.~R. and {Mittaz}, J.~P.~D. and {Buckley}, D.~A. and {Layden}, A.~C. and {Clayton}, K.~L. and {McCain}, C. and {Wynn}, G.~A. and {Sirk}, M.~M. and {Osborne}, J.~P. and {Watson}, M.~G.},
        title = "{Accretion mode changes in QS Tel (RE 1938-461): EUVE, ROSAT and optical observations}",
      journal = {\mnras},
         year = 1996,
        month = jun,
       volume = {280},
       number = {4},
        pages = {1121-1142},
          doi = {10.1093/mnras/280.4.1121},
       adsurl = {https://ui.adsabs.harvard.edu/abs/1996MNRAS.280.1121R}
}

@ARTICLE{2004MNRAS.348..316P,
       author = {{Potter}, Stephen B. and {Romero-Colmenero}, E. and {Watson}, C.~A. and {Buckley}, D.~A.~H. and {Phillips}, A.},
        title = "{Stokes imaging, Doppler mapping and Roche tomography of the AM Herculis system V834 Cen}",
      journal = {\mnras},
         year = 2004,
        month = feb,
       volume = {348},
       number = {1},
        pages = {316-324},
          doi = {10.1111/j.1365-2966.2004.07379.x},
archivePrefix = {arXiv},
       eprint = {astro-ph/0311151},
 primaryClass = {astro-ph},
       adsurl = {https://ui.adsabs.harvard.edu/abs/2004MNRAS.348..316P}
}

@ARTICLE{1982AdSpR...2d.241T,
       author = {{Truemper}, J.},
        title = "{The ROSAT mission}",
      journal = {Advances in Space Research},
         year = 1982,
        month = jan,
       volume = {2},
       number = {4},
        pages = {241-249},
          doi = {10.1016/0273-1177(82)90070-9},
       adsurl = {https://ui.adsabs.harvard.edu/abs/1982AdSpR...2d.241T}
}

@ARTICLE{1993MNRAS.260...77P,
       author = {{Pounds}, K.~A. and {Allan}, D.~J. and {Barber}, C. and {Barstow}, M.~A. and {Bertram}, D. and {Branduardi-Raymont}, G. and {Brebner}, G.~E.~C. and {Buckley}, D. and {Bromage}, G.~E. and {Cole}, R.~E. and {Courtier}, M. and {Cruise}, A.~M. and {Culhane}, J.~L. and {Denby}, M. and {Donoghue}, D.~O. and {Dunford}, E. and {Georgantopoulos}, I. and {Goodall}, C.~V. and {Gondhalekar}, P.~M. and {Gourlay}, J.~A. and {Harris}, A.~W. and {Hassall}, B.~J.~M. and {Hellier}, C. and {Hodgkin}, S. and {Jeffries}, R.~D. and {Kellett}, B.~J. and {Kent}, B.~J. and {Lieu}, R. and {Lloyd}, C. and {McGale}, P. and {Mason}, K.~O. and {Matthews}, L. and {Mittaz}, J.~P.~D. and {Page}, C.~G. and {Pankiewicz}, G.~S. and {Pike}, C.~D. and {Ponman}, T.~J. and {Puchnarewicz}, E.~M. and {Pye}, J.~P. and {Quenby}, J.~J. and {Ricketts}, M.~J. and {Rosen}, S.~R. and {Sansom}, A.~E. and {Sembay}, S. and {Sidher}, S. and {Sims}, M.~R. and {Stewart}, B.~C. and {Sumner}, T.~J. and {Vallance}, R.~J. and {Watson}, M.~G. and {Warwick}, R.~S. and {Wells}, A.~A. and {Willingale}, R. and {Willmore}, A.~P. and {Willoughby}, G.~A. and {Wonnacott}, D.},
        title = "{The ROSAT Wide Field Camera all-sky survey of extreme-ultraviolet sources. I. The bright source catalogue.}",
      journal = {\mnras},
         year = 1993,
        month = jan,
       volume = {260},
        pages = {77-102},
          doi = {10.1093/mnras/260.1.77},
       adsurl = {https://ui.adsabs.harvard.edu/abs/1993MNRAS.260...77P}
}

@ARTICLE{1995MNRAS.274.1165P,
       author = {{Pye}, J.~P. and {McGale}, P.~A. and {Allan}, D.~J. and {Barber}, C.~R. and {Bertram}, D. and {Denby}, M. and {Page}, C.~G. and {Ricketts}, M.~J. and {Stewart}, B.~C. and {West}, R.~G.},
        title = "{The ROSAT Wide Field Camera all-sky survey of extreme-ultraviolet sources - II. The 2RE Source Catalogue}",
      journal = {\mnras},
         year = 1995,
        month = jun,
       volume = {274},
       number = {4},
        pages = {1165-1193},
          doi = {10.1093/mnras/274.4.1165},
       adsurl = {https://ui.adsabs.harvard.edu/abs/1995MNRAS.274.1165P}
}

@ARTICLE{2019JATIS...5b4007C,
       author = {{Crause}, Lisa A. and {Gilbank}, David and {Gend}, Carel van and {Worters}, Hannah L. and {Sass}, Craig and {Kotze}, Enrico J. and {Potter}, Stephen and {Sickafoose}, Amanda and {Sefako}, Ramotholo and {Southworth}, John and {Macri}, Lucas and {Thorstensen}, John and {Galan}, Cezary and {Skelton}, Patricia and {Engelbrecht}, Chris and {Braker}, Ian and {Winkler}, Hartmut and {Pie{\'n}kowski}, Daniel and {S{\"u}rgit}, Derya and {Erdem}, Ahmet and {Burleigh}, Matt},
        title = "{SpUpNIC (Spectrograph Upgrade: Newly Improved Cassegrain): a versatile and efficient low- to medium-resolution, long-slit spectrograph on the South African Astronomical Observatory's 1.9-m telescope}",
      journal = {Journal of Astronomical Telescopes, Instruments, and Systems},
         year = 2019,
        month = apr,
       volume = {5},
          eid = {024007},
        pages = {024007},
          doi = {10.1117/1.JATIS.5.2.024007},
       adsurl = {https://ui.adsabs.harvard.edu/abs/2019JATIS...5b4007C}
}

@ARTICLE{Kinemuchi_2012,
       author = {{Kinemuchi}, K. and {Barclay}, T. and {Fanelli}, M. and {Pepper}, J. and {Still}, M. and {Howell}, Steve B.},
        title = "{Demystifying Kepler Data: A Primer for Systematic Artifact Mitigation}",
      journal = {\pasp},
         year = 2012,
        month = sep,
       volume = {124},
       number = {919},
        pages = {963},
          doi = {10.1086/667603},
archivePrefix = {arXiv},
       eprint = {1207.3093},
 primaryClass = {astro-ph.IM},
       adsurl = {https://ui.adsabs.harvard.edu/abs/2012PASP..124..963K}
}

@INPROCEEDINGS{2003SPIE.4843..170N,
       author = {{Nordsieck}, Kenneth H. and {Jaehnig}, Kurt P. and {Burgh}, Eric B. and {Kobulnicky}, Henry A. and {Percival}, Jeffrey W. and {Smith}, Michael P.},
        title = "{Instrumentation for high-resolution spectropolarimetry in the visible and far-ultraviolet}",
    booktitle = {Polarimetry in Astronomy},
         year = 2003,
       editor = {{Fineschi}, Silvano},
       series = {Society of Photo-Optical Instrumentation Engineers (SPIE) Conference Series},
       volume = {4843},
        month = feb,
        pages = {170-179},
          doi = {10.1117/12.459288},
       adsurl = {https://ui.adsabs.harvard.edu/abs/2003SPIE.4843..170N}
}

@INPROCEEDINGS{2006SPIE.6267E..0ZB,
       author = {{Buckley}, David A.~H. and {Swart}, Gerhard P. and {Meiring}, Jacobus G.},
        title = "{Completion and commissioning of the Southern African Large Telescope}",
    booktitle = {Ground-based and Airborne Telescopes},
         year = 2006,
       editor = {{Stepp}, Larry M.},
       series = {Society of Photo-Optical Instrumentation Engineers (SPIE) Conference Series},
       volume = {6267},
        month = jun,
          eid = {62670Z},
        pages = {62670Z},
          doi = {10.1117/12.673750},
       adsurl = {https://ui.adsabs.harvard.edu/abs/2006SPIE.6267E..0ZB}
}

@ARTICLE{1978AnSta...6..461S,
       author = {{Schwarz}, Gideon},
        title = "{Estimating the Dimension of a Model}",
      journal = {Annals of Statistics},
         year = 1978,
        month = jul,
       volume = {6},
       number = {2},
        pages = {461-464},
       adsurl = {https://ui.adsabs.harvard.edu/abs/1978AnSta...6..461S}
}

@ARTICLE{2001A&A...372..557G,
       author = {{G{\"a}nsicke}, B.~T. and {Fischer}, A. and {Silvotti}, R. and {de Martino}, D.},
        title = "{A model for the optical high state light curve of AM Herculis.}",
      journal = {\aap},
         year = 2001,
        month = jun,
       volume = {372},
        pages = {557-562},
          doi = {10.1051/0004-6361:20010522},
archivePrefix = {arXiv},
       eprint = {astro-ph/0104167},
 primaryClass = {astro-ph},
       adsurl = {https://ui.adsabs.harvard.edu/abs/2001A&A...372..557G}
}

@ARTICLE{2020MNRAS.492.4298K,
       author = {{Khangale}, Z.~N. and {Potter}, S.~B. and {Woudt}, P.~A. and {Buckley}, D.~A.~H. and {Semena}, A.~N. and {Kotze}, E.~J. and {Groenewald}, D.~N. and {Hewitt}, D.~M. and {Pretorius}, M.~L. and {Fender}, R.~P. and {Groot}, P. and {Bloemen}, S. and {Klein-Wolt}, M. and {K{\"o}rding}, E. and {Le Poole}, R. and {McBride}, V.~A. and {Townsend}, L. and {Paterson}, K. and {Pieterse}, D.~L.~A. and {Vreeswijk}, P.},
        title = "{A spectroscopic, photometric, polarimetric, and radio study of the eclipsing polar UZ Fornacis: the first simultaneous SALT and MeerKAT observations}",
      journal = {\mnras},
         year = 2020,
        month = mar,
       volume = {492},
       number = {3},
        pages = {4298-4312},
          doi = {10.1093/mnras/staa080},
archivePrefix = {arXiv},
       eprint = {2001.06036},
 primaryClass = {astro-ph.SR},
       adsurl = {https://ui.adsabs.harvard.edu/abs/2020MNRAS.492.4298K}
}

@ARTICLE{2022MNRAS.516.3144D,
       author = {{Duffy}, C. and {Ramsay}, G. and {Wu}, Kinwah and {Mason}, Paul A. and {Hakala}, P. and {Steeghs}, D. and {Wood}, M.~A.},
        title = "{Short-duration accretion states of Polars as seen in TESS and ZTF data}",
      journal = {\mnras},
         year = 2022,
        month = nov,
       volume = {516},
       number = {3},
        pages = {3144-3158},
          doi = {10.1093/mnras/stac2482},
archivePrefix = {arXiv},
       eprint = {2208.14855},
 primaryClass = {astro-ph.SR},
       adsurl = {https://ui.adsabs.harvard.edu/abs/2022MNRAS.516.3144D}
}

@ARTICLE{Coppejans_2013,
       author = {{Coppejans}, R. and {Gulbis}, A.~A.~S. and {Kotze}, M.~M. and {Coppejans}, D.~L. and {Worters}, H.~L. and {Woudt}, P.~A. and {Whittal}, H. and {Cloete}, J. and {Fourie}, P.},
        title = "{Characterizing and Commissioning the Sutherland High-Speed Optical Cameras (SHOC)}",
      journal = {\pasp},
         year = 2013,
        month = aug,
       volume = {125},
       number = {930},
        pages = {976},
          doi = {10.1086/672156},
       adsurl = {https://ui.adsabs.harvard.edu/abs/2013PASP..125..976C}
}

@INPROCEEDINGS{2015AAS...22533659F,
       author = {{Fleming}, Scott W. and {Abney}, Faith and {Donaldson}, Tom and {Dower}, Theresa and {Fraquelli}, Dorothy A. and {Koekemoer}, Anton M. and {Levay}, Karen and {Matuskey}, Jacob and {McLean}, Brian and {Quick}, Lee and {Rogers}, Anthony and {Shiao}, Bernie and {Thompson}, Randy and {Tseng}, Shui-Ay and {Wallace}, Geoff and {White}, Richard L.},
        title = "{Beyond The Prime Directive: The MAST Discovery Portal and High Level Science Products}",
    booktitle = {American Astronomical Society Meeting Abstracts \#225},
         year = 2015,
       series = {American Astronomical Society Meeting Abstracts},
       volume = {225},
        month = jan,
          eid = {336.59},
        pages = {336.59},
       adsurl = {https://ui.adsabs.harvard.edu/abs/2015AAS...22533659F}
}

@BOOK{1995cvs..book.....W,
       author = {{Warner}, Brian},
        title = "{Cataclysmic variable stars}",
        series={Cambridge Astrophysics},
        publisher={Cambridge University Press},
         year = 1995,
       volume = {28},
       adsurl = {https://ui.adsabs.harvard.edu/abs/1995cvs..book.....W}
}

@ARTICLE{2017A&A...600A..53M,
       author = {{Mouchet}, M. and {Bonnet-Bidaud}, J. -M. and {Van Box Som}, L. and {Falize}, E. and {Buckley}, D.~A.~H. and {Breytenbach}, H. and {Ashley}, R.~P. and {Marsh}, T.~R. and {Dhillon}, V.~S.},
        title = "{A VLT-ULTRACAM study of the fast optical quasi-periodic oscillations in the polar V834 Centauri}",
      journal = {\aap},
         year = 2017,
        month = apr,
       volume = {600},
          eid = {A53},
        pages = {A53},
          doi = {10.1051/0004-6361/201630166},
archivePrefix = {arXiv},
       eprint = {1701.02677},
 primaryClass = {astro-ph.HE},
       adsurl = {https://ui.adsabs.harvard.edu/abs/2017A&A...600A..53M}
}

@ARTICLE{2003MNRAS.338..340F,
       author = {{Ferrario}, Lilia and {Wickramasinghe}, D.~T. and {Schmidt}, Gary},
        title = "{Analysis of new spectropolarimetric data of AR UMa}",
      journal = {\mnras},
         year = 2003,
        month = jan,
       volume = {338},
       number = {2},
        pages = {340-346},
          doi = {10.1046/j.1365-8711.2003.06044.x},
       adsurl = {https://ui.adsabs.harvard.edu/abs/2003MNRAS.338..340F}
}

@ARTICLE{2002A&A...392..505S,
       author = {{Schwarz}, R. and {Greiner}, J. and {Tovmassian}, G.~H. and {Zharikov}, S.~V. and {Wenzel}, W.},
        title = "{A new two-pole accretion polar: RX J1846.9+5538}",
      journal = {\aap},
         year = 2002,
        month = sep,
       volume = {392},
        pages = {505-514},
          doi = {10.1051/0004-6361:20021193},
archivePrefix = {arXiv},
       eprint = {astro-ph/0208545},
 primaryClass = {astro-ph},
       adsurl = {https://ui.adsabs.harvard.edu/abs/2002A&A...392..505S}
}

@ARTICLE{2008A&A...490..279M,
       author = {{Mason}, E. and {Howell}, S.~B. and {Barman}, T. and {Szkody}, P. and {Wickramasinghe}, D.},
        title = "{VV Puppis in a low state: secondary-star irradiation or stellar activity?}",
      journal = {\aap},
         year = 2008,
        month = oct,
       volume = {490},
       number = {1},
        pages = {279-286},
          doi = {10.1051/0004-6361:20078804},
archivePrefix = {arXiv},
       eprint = {0809.5041},
 primaryClass = {astro-ph},
       adsurl = {https://ui.adsabs.harvard.edu/abs/2008A&A...490..279M}
}

@ARTICLE{1989MNRAS.236...31M,
       author = {{Meggitt}, S.~M.~A. and {Wickramasinghe}, D.~T.},
        title = "{Two-pole emission in AM Herculis-type systems.}",
      journal = {\mnras},
         year = 1989,
        month = jan,
       volume = {236},
        pages = {31-45},
          doi = {10.1093/mnras/236.1.31},
       adsurl = {https://ui.adsabs.harvard.edu/abs/1989MNRAS.236...31M}
}

@ARTICLE{2015JATIS...1a4003R,
       author = {{Ricker}, George R. and {Winn}, Joshua N. and {Vanderspek}, Roland and {Latham}, David W. and {Bakos}, G{\'a}sp{\'a}r {\'A}. and {Bean}, Jacob L. and {Berta-Thompson}, Zachory K. and {Brown}, Timothy M. and {Buchhave}, Lars and {Butler}, Nathaniel R. and {Butler}, R. Paul and {Chaplin}, William J. and {Charbonneau}, David and {Christensen-Dalsgaard}, J{\o}rgen and {Clampin}, Mark and {Deming}, Drake and {Doty}, John and {De Lee}, Nathan and {Dressing}, Courtney and {Dunham}, Edward W. and {Endl}, Michael and {Fressin}, Francois and {Ge}, Jian and {Henning}, Thomas and {Holman}, Matthew J. and {Howard}, Andrew W. and {Ida}, Shigeru and {Jenkins}, Jon M. and {Jernigan}, Garrett and {Johnson}, John Asher and {Kaltenegger}, Lisa and {Kawai}, Nobuyuki and {Kjeldsen}, Hans and {Laughlin}, Gregory and {Levine}, Alan M. and {Lin}, Douglas and {Lissauer}, Jack J. and {MacQueen}, Phillip and {Marcy}, Geoffrey and {McCullough}, Peter R. and {Morton}, Timothy D. and {Narita}, Norio and {Paegert}, Martin and {Palle}, Enric and {Pepe}, Francesco and {Pepper}, Joshua and {Quirrenbach}, Andreas and {Rinehart}, Stephen A. and {Sasselov}, Dimitar and {Sato}, Bun'ei and {Seager}, Sara and {Sozzetti}, Alessandro and {Stassun}, Keivan G. and {Sullivan}, Peter and {Szentgyorgyi}, Andrew and {Torres}, Guillermo and {Udry}, Stephane and {Villasenor}, Joel},
        title = "{Transiting Exoplanet Survey Satellite (TESS)}",
      journal = {Journal of Astronomical Telescopes, Instruments, and Systems},
         year = 2015,
        month = jan,
       volume = {1},
          eid = {014003},
        pages = {014003},
          doi = {10.1117/1.JATIS.1.1.014003},
       adsurl = {https://ui.adsabs.harvard.edu/abs/2015JATIS...1a4003R}
}

@ARTICLE{2010PASP..122..935E,
       author = {{Eastman}, Jason and {Siverd}, Robert and {Gaudi}, B. Scott},
        title = "{Achieving Better Than 1 Minute Accuracy in the Heliocentric and Barycentric Julian Dates}",
      journal = {\pasp},
         year = 2010,
        month = aug,
       volume = {122},
       number = {894},
        pages = {935},
          doi = {10.1086/655938},
archivePrefix = {arXiv},
       eprint = {1005.4415},
 primaryClass = {astro-ph.IM},
       adsurl = {https://ui.adsabs.harvard.edu/abs/2010PASP..122..935E}
}

@PHDTHESIS{2022PhDT........26K,
       author = {{Khangale}, Zwidofhelangani},
        title = "{Accretion processes in magnetic cataclysmic variables - a detailed study of UZ Fornacis}",
       school = {University of Cape Town, South Africa},
         year = 2022,
        month = jan,
       adsurl = {https://ui.adsabs.harvard.edu/abs/2022PhDT........26K}
}

@BOOK{2001cvs..book.....H,
       author = {{Hellier}, Coel},
        title = "{Cataclysmic Variable Stars}",
        publisher = {Springer},
         year = 2001,
       adsurl = {https://ui.adsabs.harvard.edu/abs/2001cvs..book.....H}
}

@ARTICLE{2025MNRAS.544..309K,
       author = {{Khangale}, Z.~N. and {Potter}, S.~B. and {Buckley}, D.~A.~H. and {Barrett}, P.~E.},
        title = "{Phase-resolved spectroscopic observations of the magnetic cataclysmic binary EF Eridani: revealing complex magnetic accretion during a high state}",
      journal = {\mnras},
         year = 2025,
        month = nov,
       volume = {544},
       number = {1},
        pages = {309-320},
          doi = {10.1093/mnras/staf1754},
archivePrefix = {arXiv},
       eprint = {2510.08266},
 primaryClass = {astro-ph.SR},
       adsurl = {https://ui.adsabs.harvard.edu/abs/2025MNRAS.544..309K}
}

@ARTICLE{2018PASP..130f4505T,
       author = {{Tonry}, J.~L. and {Denneau}, L. and {Heinze}, A.~N. and {Stalder}, B. and {Smith}, K.~W. and {Smartt}, S.~J. and {Stubbs}, C.~W. and {Weiland}, H.~J. and {Rest}, A.},
        title = "{ATLAS: A High-cadence All-sky Survey System}",
      journal = {\pasp},
         year = 2018,
        month = jun,
       volume = {130},
       number = {988},
        pages = {064505},
          doi = {10.1088/1538-3873/aabadf},
archivePrefix = {arXiv},
       eprint = {1802.00879},
 primaryClass = {astro-ph.IM},
       adsurl = {https://ui.adsabs.harvard.edu/abs/2018PASP..130f4505T}
}

@ARTICLE{2015A&A...579A..77K,
       author = {{Kotze}, E.~J. and {Potter}, S.~B. and {McBride}, V.~A.},
        title = "{Exploring inside-out Doppler tomography: non-magnetic cataclysmic variables}",
      journal = {\aap},
         year = 2015,
        month = jul,
       volume = {579},
          eid = {A77},
        pages = {A77},
          doi = {10.1051/0004-6361/201526381},
archivePrefix = {arXiv},
       eprint = {1507.05213},
 primaryClass = {astro-ph.SR},
       adsurl = {https://ui.adsabs.harvard.edu/abs/2015A&A...579A..77K}
}

@ARTICLE{2016A&A...595A..47K,
       author = {{Kotze}, E.~J. and {Potter}, S.~B. and {McBride}, V.~A.},
        title = "{Exploring inside-out Doppler tomography: magnetic cataclysmic variables}",
      journal = {\aap},
         year = 2016,
        month = oct,
       volume = {595},
          eid = {A47},
        pages = {A47},
          doi = {10.1051/0004-6361/201629120},
archivePrefix = {arXiv},
       eprint = {1610.09841},
 primaryClass = {astro-ph.SR},
       adsurl = {https://ui.adsabs.harvard.edu/abs/2016A&A...595A..47K}
}

@ARTICLE{1995A&A...293..764S,
       author = {{Schwope}, A.~D. and {Thomas}, H. -C. and {Beuermann}, K. and {Burwitz}, V. and {Jordan}, S. and {Haefner}, R.},
        title = "{Two-pole accretion in the high-field polar RXJ 1938.6-4612.}",
      journal = {\aap},
         year = 1995,
        month = jan,
       volume = {293},
        pages = {764-776},
       adsurl = {https://ui.adsabs.harvard.edu/abs/1995A&A...293..764S}
}

@ARTICLE{2001MNRAS.322..631R,
       author = {{Rosen}, S.~R. and {Rainger}, J.~F. and {Burleigh}, M.~R. and {Mittaz}, J.~P.~D. and {Buckley}, D.~A.~H. and {Sirk}, M.~M. and {Lieu}, R. and {Howell}, S.~B. and {de Martino}, D.},
        title = "{Multiwavelength monitoring of QS Tel}",
      journal = {\mnras},
         year = 2001,
        month = apr,
       volume = {322},
       number = {3},
        pages = {631-642},
          doi = {10.1046/j.1365-8711.2001.04141.x},
       adsurl = {https://ui.adsabs.harvard.edu/abs/2001MNRAS.322..631R}
}

@ARTICLE{1993MNRAS.262...93B,
       author = {{Buckley}, D.~A. and {O'Donoghue}, D. and {Hassall}, B.~J.~M. and {Kellett}, B.~J. and {Mason}, K.~O. and {Sekiguchi}, K. and {Watson}, M.~G. and {Wheatley}, P.~J. and {Chen}, A.},
        title = "{Discovery of an EUV-bright polar in the period gap from the ROSAT Wide Field Camera sky survey.}",
      journal = {\mnras},
         year = 1993,
        month = may,
       volume = {262},
        pages = {93-108},
          doi = {10.1093/mnras/262.1.93},
       adsurl = {https://ui.adsabs.harvard.edu/abs/1993MNRAS.262...93B}
}

@INPROCEEDINGS{2015gacv.workE..18B,
       author = {{Breytenbach}, H. and {Buckley}, D. and {Bonnet-Bidaud}, J.~M. and {Mouchet}, M.},
        title = "{Quasi-periodic oscillations in magnetic Cataclysmic Variables: Results for V834 Cen}",
    booktitle = {The Golden Age of Cataclysmic Variables and Related Objects - III (Golden2015)},
         year = 2015,
        month = jan,
          eid = {18},
        pages = {18},
          doi = {10.22323/1.255.0018},
       adsurl = {https://ui.adsabs.harvard.edu/abs/2015gacv.workE..18B}
}

@article{2020A&A...635A.204B,
  author = {{Bonnet-Bidaud}, J.~M. and {de Martino}, D. and {Falanga}, M. and {Mouchet}, M.},
  title = "{VV Puppis: an X-ray and optical view of a magnetic accreting white dwarf}",
  journal = {Astronomy \& Astrophysics},
  volume = {635},
  pages = {A204},
  year = {2020},
  month = {March},
  doi = {10.1051/0004-6361/201937132},
  archivePrefix = {arXiv},
  eprint = {2001.04128},
  adsurl = {https://ui.adsabs.harvard.edu/abs/2020A&A...635A.204B}
}

@ARTICLE{1990SSRv...54..195C,
       author = {{Cropper}, Mark},
        title = "{The Polars}",
      journal = {\ssr},
         year = 1990,
        month = dec,
       volume = {54},
       number = {3-4},
        pages = {195-295},
          doi = {10.1007/BF00177799},
       adsurl = {https://ui.adsabs.harvard.edu/abs/1990SSRv...54..195C}
}

@ARTICLE{1996ApJ...473..483S,
       author = {{Schmidt}, Gary D. and {Szkody}, Paula and {Smith}, Paul S. and {Silber}, Andrew and {Tovmassian}, Gaghik and {Hoard}, D.~W. and {G{\"a}nsicke}, B.~T. and {de Martino}, D.},
        title = "{AR Ursae Majoris: The First High-Field Magnetic Cataclysmic Variable}",
      journal = {\apj},
         year = 1996,
        month = dec,
       volume = {473},
        pages = {483},
          doi = {10.1086/178160},
       adsurl = {https://ui.adsabs.harvard.edu/abs/1996ApJ...473..483S}
}

@ARTICLE{1992MNRAS.256..252F,
       author = {{Ferrario}, Lilia and {Wickramasinghe}, D.~T. and {Bailey}, Jeremy and {Hough}, J.~H. and {Tuohy}, I.~R.},
        title = "{Detection of photospheric Zeeman features and cyclotron emission lines inV 834 CEN in a low state.}",
      journal = {\mnras},
         year = 1992,
        month = may,
       volume = {256},
        pages = {252-260},
          doi = {10.1093/mnras/256.2.252},
       adsurl = {https://ui.adsabs.harvard.edu/abs/1992MNRAS.256..252F}
}

@ARTICLE{1993MNRAS.265..605F,
       author = {{Ferrario}, L. and {Wickramasinghe}, D.~T.},
        title = "{A model for the optical continuum and Balmer emission lines in intermediate polars.}",
      journal = {\mnras},
         year = 1993,
        month = dec,
       volume = {265},
        pages = {605-618},
          doi = {10.1093/mnras/265.3.605},
       adsurl = {https://ui.adsabs.harvard.edu/abs/1993MNRAS.265..605F}
}

@book{Frank1992,
  author    = {Frank, J. and King, A. and Raine, D.},
  title     = {Accretion Power in Astrophysics},
  edition   = {2},
  publisher = {Cambridge University Press},
  address   = {Cambridge},
  year      = {1992},
  series    = {Cambridge Astrophysics Series},
  volume    = {21}
}

@ARTICLE{2005A&A...439..823K,
       author = {{Kalomeni}, B. and {Pek{\"u}nl{\"u}}, E.~R. and {Yakut}, K.},
        title = "{Cyclotron emission from accretion plasma columns in magnetic cataclysmic variable stars}",
      journal = {\aap},
         year = 2005,
        month = sep,
       volume = {439},
       number = {3},
        pages = {823-833},
          doi = {10.1051/0004-6361:20041780},
       adsurl = {https://ui.adsabs.harvard.edu/abs/2005A&A...439..823K}
}

@ARTICLE{2018ApJ...867..105T,
       author = {{Tonry}, J.~L. and {Denneau}, L. and {Flewelling}, H. and {Heinze}, A.~N. and {Onken}, C.~A. and {Smartt}, S.~J. and {Stalder}, B. and {Weiland}, H.~J. and {Wolf}, C.},
        title = "{The ATLAS All-Sky Stellar Reference Catalog}",
      journal = {\apj},
         year = 2018,
        month = nov,
       volume = {867},
       number = {2},
          eid = {105},
        pages = {105},
          doi = {10.3847/1538-4357/aae386},
archivePrefix = {arXiv},
       eprint = {1809.09157},
 primaryClass = {astro-ph.IM},
       adsurl = {https://ui.adsabs.harvard.edu/abs/2018ApJ...867..105T}
}

@ARTICLE{2001ApJ...553..823S,
       author = {{Schmidt}, Gary D. and {Ferrario}, Lilia and {Wickramasinghe}, D.~T. and {Smith}, Paul S.},
        title = "{The Cyclotron Fundamental Exposed in the High-Field Magnetic Variable V884 Herculis}",
      journal = {\apj},
         year = 2001,
        month = jun,
       volume = {553},
       number = {2},
        pages = {823-831},
          doi = {10.1086/320967},
archivePrefix = {arXiv},
       eprint = {astro-ph/0012473},
 primaryClass = {astro-ph},
       adsurl = {https://ui.adsabs.harvard.edu/abs/2001ApJ...553..823S}
}

@ARTICLE{2000MNRAS.314..672P,
       author = {{Potter}, S.~B.},
        title = "{Stokes imaging of the AM Her system ST LMi}",
      journal = {\mnras},
         year = 2000,
        month = jun,
       volume = {314},
       number = {4},
        pages = {672-680},
          doi = {10.1046/j.1365-8711.2000.03280.x},
       adsurl = {https://ui.adsabs.harvard.edu/abs/2000MNRAS.314..672P}
}

@ARTICLE{1994MNRAS.268..128F,
       author = {{Ferrario}, L. and {Wickramasinghe}, D.~T. and {Bailey}, J.~A. and {Buckley}, D.~A.~H.},
        title = "{The Polarization and Magnetic Field of REJ1938-461 during its Low State}",
      journal = {\mnras},
         year = 1994,
        month = may,
       volume = {268},
        pages = {128},
          doi = {10.1093/mnras/268.1.128},
       adsurl = {https://ui.adsabs.harvard.edu/abs/1994MNRAS.268..128F}
}

@ARTICLE{1998MNRAS.301..767S,
       author = {{Smith}, D.~A. and {Dhillon}, V.~S.},
        title = "{The secondary stars in cataclysmic variables and low-mass X-ray binaries}",
      journal = {\mnras},
         year = 1998,
        month = dec,
       volume = {301},
       number = {3},
        pages = {767-781},
          doi = {10.1046/j.1365-8711.1998.02065.x},
archivePrefix = {arXiv},
       eprint = {astro-ph/9808217},
 primaryClass = {astro-ph},
       adsurl = {https://ui.adsabs.harvard.edu/abs/1998MNRAS.301..767S}
}

@ARTICLE{1982ApJ...257L..71M,
       author = {{Middleditch}, J.},
        title = "{Two-second variability in AM HER binaries.}",
      journal = {\apjl},
         year = 1982,
        month = jun,
       volume = {257},
        pages = {L71-L75},
          doi = {10.1086/183811},
       adsurl = {https://ui.adsabs.harvard.edu/abs/1982ApJ...257L..71M}
}

@ARTICLE{1997ApJ...489..912M,
       author = {{Middleditch}, John and {Imamura}, James N. and {Steiman-Cameron}, Thomas Y.},
        title = "{Discovery of Quasi-periodic Oscillations in the AM Herculis Object BL Hydri}",
      journal = {\apj},
         year = 1997,
        month = nov,
       volume = {489},
       number = {2},
        pages = {912-916},
          doi = {10.1086/304834},
       adsurl = {https://ui.adsabs.harvard.edu/abs/1997ApJ...489..912M}
}

@ARTICLE{1987A&A...181L..15L,
       author = {{Larsson}, S.},
        title = "{Discovery of 2-3s quasi-periodic oscillations in EF Eri.}",
      journal = {\aap},
         year = 1987,
        month = jul,
       volume = {181},
        pages = {L15-L17},
       adsurl = {https://ui.adsabs.harvard.edu/abs/1987A&A...181L..15L}
}

@ARTICLE{1989ApJ...342L..35W,
       author = {{Wickramasinghe}, D.~T. and {Ferrario}, Lilia and {Bailey}, Jeremy},
        title = "{A 56 MG Field at the Second Pole in VV Puppis}",
      journal = {\apjl},
         year = 1989,
        month = jul,
       volume = {342},
        pages = {L35},
          doi = {10.1086/185478},
       adsurl = {https://ui.adsabs.harvard.edu/abs/1989ApJ...342L..35W}
}

@book{potter1998multiwavelenght,
  title={A multiwavelenght study of the accretion region in magnetic cataclysmic variables},
  author={Potter, Stephen Brian},
  year={1998},
  publisher={University of London, University College London (United Kingdom)}
}

@ARTICLE{2020A&A...642A.134S,
       author = {{Schwope}, A.~D. and {Worpel}, H. and {Traulsen}, I. and {Sablowski}, D.},
        title = "{The various accretion modes of AM Herculis: Clues from multi-wavelength observations in high accretion states}",
      journal = {\aap},
         year = 2020,
        month = oct,
       volume = {642},
          eid = {A134},
        pages = {A134},
          doi = {10.1051/0004-6361/202037714},
archivePrefix = {arXiv},
       eprint = {2008.08402},
 primaryClass = {astro-ph.SR},
       adsurl = {https://ui.adsabs.harvard.edu/abs/2020A&A...642A.134S}
}

@ARTICLE{2011A&A...529A.116T,
       author = {{Traulsen}, I. and {Reinsch}, K. and {Schwope}, A.~D. and {Burwitz}, V. and {Dreizler}, S. and {Schwarz}, R. and {Walter}, F.~M.},
        title = "{XMM-Newton observations of the X-ray soft polar QS Telescopii}",
      journal = {\aap},
         year = 2011,
        month = may,
       volume = {529},
          eid = {A116},
        pages = {A116},
          doi = {10.1051/0004-6361/201016352},
archivePrefix = {arXiv},
       eprint = {1103.4575},
 primaryClass = {astro-ph.SR},
       adsurl = {https://ui.adsabs.harvard.edu/abs/2011A&A...529A.116T}
}

@INPROCEEDINGS{1986SPIE..627..733T,
       author = {{Tody}, Doug},
        title = "{The IRAF Data Reduction and Analysis System}",
    booktitle = {Instrumentation in astronomy VI},
         year = 1986,
       editor = {{Crawford}, David L.},
       series = {Society of Photo-Optical Instrumentation Engineers (SPIE) Conference Series},
       volume = {627},
        month = jan,
        pages = {733},
          doi = {10.1117/12.968154},
       adsurl = {https://ui.adsabs.harvard.edu/abs/1986SPIE..627..733T}
}

@INPROCEEDINGS{1993ASPC...52..173T,
       author = {{Tody}, Doug},
        title = "{IRAF in the Nineties}",
    booktitle = {Astronomical Data Analysis Software and Systems II},
         year = 1993,
       editor = {{Hanisch}, R.~J. and {Brissenden}, R.~J.~V. and {Barnes}, J.},
       series = {Astronomical Society of the Pacific Conference Series},
       volume = {52},
        month = jan,
        pages = {173},
       adsurl = {https://ui.adsabs.harvard.edu/abs/1993ASPC...52..173T}
}

@ARTICLE{1976Ap&SS..39..447L,
       author = {{Lomb}, N.~R.},
        title = "{Least-Squares Frequency Analysis of Unequally Spaced Data}",
      journal = {\apss},
         year = 1976,
        month = feb,
       volume = {39},
       number = {2},
        pages = {447-462},
          doi = {10.1007/BF00648343},
       adsurl = {https://ui.adsabs.harvard.edu/abs/1976Ap&SS..39..447L}
}

@ARTICLE{1982ApJ...263..835S,
       author = {{Scargle}, J.~D.},
        title = "{Studies in astronomical time series analysis. II. Statistical aspects of spectral analysis of unevenly spaced data.}",
      journal = {\apj},
         year = 1982,
        month = dec,
       volume = {263},
        pages = {835-853},
          doi = {10.1086/160554},
       adsurl = {https://ui.adsabs.harvard.edu/abs/1982ApJ...263..835S}
}

@ARTICLE{2024A&A...684A.190B,
       author = {{Balman}, {\c{S}}. and {Khamitov}, I. and {Kolbin}, A. and {Aktekin {\c{C}}al{\i}{\c{s}}kan}, E. and {Bikmaev}, I. and {{\"O}zd{\"o}nmez}, A. and {Burenin}, R. and {K{\i}l{\i}{\c{c}}}, Y. and {Eseno{\u{g}}lu}, H.~H. and {Yelkenci}, K.~F. and {{\c{C}}amurdan}, D. Zengin and {Gilfanov}, M. and {Nas{\i}ro{\u{g}}lu}, I. and {Sonba{\c{s}}}, E. and {Gabdeev}, M. and {Irtuganov}, E. and {Sayga{\c{c}}}, A.~T. and {Nikolaeva}, E. and {Sakhibullin}, N. and {Er}, H. and {Sazonov}, N.~S.~S. and {Medvedev}, P. and {G{\"u}ver}, T. and {Fi{\c{s}}ek}, S.},
        title = "{Optical identification and follow-up observations of SRGA J213151.5+491400. A new magnetic cataclysmic variable discovered with the SRG observatory}",
      journal = {\aap},
         year = 2024,
        month = apr,
       volume = {684},
          eid = {A190},
        pages = {A190},
          doi = {10.1051/0004-6361/202348455},
archivePrefix = {arXiv},
       eprint = {2401.06001},
 primaryClass = {astro-ph.HE},
       adsurl = {https://ui.adsabs.harvard.edu/abs/2024A&A...684A.190B}
}

@ARTICLE{1997A&A...319..894S,
       author = {{Schwope}, A.~D. and {Mantel}, K. -H. and {Horne}, K.},
        title = "{Phase-resolved high-resolution spectrophotometry of the eclipsing polar HU Aquarii.}",
      journal = {\aap},
         year = 1997,
        month = mar,
       volume = {319},
        pages = {894-908},
          doi = {10.48550/arXiv.astro-ph/9701094},
archivePrefix = {arXiv},
       eprint = {astro-ph/9701094},
 primaryClass = {astro-ph},
       adsurl = {https://ui.adsabs.harvard.edu/abs/1997A&A...319..894S}
}

@INPROCEEDINGS{2005ASPC..330..429R,
       author = {{Romero-Colmenero}, E. and {Potter}, S.~B. and {Buckley}, D.~A.~H.},
        title = "{An eclipsing geometry for the polar QS Tel}",
    booktitle = {The Astrophysics of Cataclysmic Variables and Related Objects},
         year = 2005,
       editor = {{Hameury}, J. -M. and {Lasota}, J. -P.},
       series = {Astronomical Society of the Pacific Conference Series},
       volume = {330},
        month = aug,
        pages = {429},
       adsurl = {https://ui.adsabs.harvard.edu/abs/2005ASPC..330..429R}
}








\bsp	
\label{lastpage}
\end{document}